\documentclass[11pt]{article}
\pdfoutput=1
\usepackage[onehalfspacing]{setspace}
\usepackage[ 
	colorlinks = true , 
	urlcolor = blue , 
	linkcolor = black ,
	citecolor = blue 
	]{hyperref}

\usepackage{amsmath}
\usepackage{amsfonts}
\usepackage{amssymb}
\usepackage{amsthm}

\usepackage[authoryear,sort&compress,round]{natbib}
\usepackage[nottoc,numbib]{tocbibind} 

\usepackage[letterpaper,margin=1in]{geometry}
\usepackage{pdflscape}

\usepackage{enumerate}

\usepackage{afterpage} 
\usepackage[capposition=top,footfont=small,subfloatrowsep=none]{floatrow}

\usepackage{xcolor}
\usepackage{graphicx}
\usepackage[position=bottom]{subfig} 
\usepackage{tikz}
\usetikzlibrary{backgrounds,fit,positioning, calc, %
	shapes.geometric, shapes.multipart, shapes, %
	arrows.meta, arrows, %
	decorations.markings, decorations.pathmorphing }
\usepackage[edges]{forest}

\usepackage{booktabs}
\usepackage{multirow}
\usepackage{rotating}
\usepackage{array}
\newcolumntype{L}[1]{>{\raggedright\let\newline\\\arraybackslash\hspace{0pt}}m{#1}}
\newcolumntype{C}[1]{>{\centering}m{#1}}
\usepackage{caption} 
\usepackage{amsmath}
\usepackage{siunitx}
\usepackage{multirow}
\usepackage{tabularx}
\usepackage{makecell}

\usepackage{enumitem}

\usepackage{appendix}

\usepackage{chngcntr} 

\title{Using Household Grants to Benchmark the Cost Effectiveness of a USAID Workforce Readiness Program}
\date{This version:  September 2, 2020}

\author{ 
Craig McIntosh\thanks{\protect\normalsize University of California, San Diego, \href{mailto:ctmcintosh@ucsd.edu}{ctmcintosh@ucsd.edu}} \,  and
Andrew Zeitlin\thanks{\protect\normalsize Georgetown University, \href{mailto:andrew.zeitlin@georgetown.edu}{andrew.zeitlin@georgetown.edu}}
}

\begin{document}

\maketitle
\thispagestyle{empty}
\begin{abstract}  
We use a randomized experiment to compare a workforce  training program to cash transfers in Rwanda.  Conducted in a sample of poor and underemployed youth, this study measures the impact of the training program not only relative to a control group but relative to the counterfactual of simply disbursing the cost of the program directly to beneficiaries.  While the training program was successful in improving a number of core outcomes (productive hours, assets, savings, and subjective well-being), cost-equivalent cash transfers move all these outcomes as well as consumption, income, and wealth.  In the head-to-head costing comparison cash proves superior across a number of economic outcomes, while training outperforms cash only in the production of business knowledge.  We find little evidence of complementarity between human and physical capital interventions, and no signs of heterogeneity or spillover effects.  
\end{abstract}

\begin{flushleft}
\begin{singlespace}
\textbf{Study Information:} \quad  This study is registered with the AEA Trial Registry as Number AEARCTR-0004388, and is covered by Rwanda National Ethics Committee IRB 114/RNEC/2017, IPA-IRB:14609, and UCSD IRB 161112.  The research was paid for by USAID grant AID-0AA-A-13-00002 (SUB 00009051).  We thank the Education Development Center, GiveDirectly, and USAID for their close collaboration in executing the study, Innovations for Poverty Action for their data collection work, and USAID Rwanda, DIV, and Google.org for funding.  This study is made possible by the support of the American People through the United States Agency for International Development (USAID.) The contents of this study are the sole responsibility of the authors and do not necessarily reflect the views of USAID or the United States Government.
\end{singlespace}
\end{flushleft}

\clearpage

\pagenumbering{roman} 

\section*{Executive Summary}\label{s:exec}

The Huguka Dukore/Akazi Kanoze program (meaning `Get Trained and Let's Work/Work Well Done' in Kinyarwanda) is a five-year project (2017-2021) aimed at providing 40,000 vulnerable youth with employability skills in 19 (of 30 total) districts in Rwanda. The program targets youth ages 16-30 from poor households with less than secondary education, with an emphasis on women and youth with disabilities. Huguka Dukore includes several interventions that aim to improve workforce readiness through education, training, and on-the-job training or internship experiences.  Each of the three components of the program lasts 10 weeks, consisting of i) workforce readiness preparation; ii) individual youth entrepreneurship and microenterprise start-up; and iii) technical training for specific trades, after which trainees may be placed in apprenticeships. The program builds on lessons learned from a randomized controlled trial (RCT) of the USAID-supported Akazi Kanoze Youth Livelihoods Project, implemented by the Education Development Center. 

\bigskip

This report details the 18-month midline results from an impact evaluation that benchmarked Huguka Dukore to unconditional cash grants, provided via mobile money by the U.S. non-profit GiveDirectly. Another round of data collection is planned that will measure impacts after 36-months.

\bigskip

\subsection*{Methodology}

\textbf{A randomized controlled trial (RCT) was designed to measure the impact of the Huguka Dukore relative to cash grants of the same cost to the funder, and also to understand how any impacts compare to what would have happened in the absence of the program(s).} The evaluation was primarily interested in measuring impacts on the following outcomes: i) beneficiary employment status, ii) time use, iii) beneficiary income, iv) household consumption, and v) productive assets, but also looked at a range of secondary outcomes and intermediate mechanisms, including business knowledge, savings, subjective wellbeing, and wealth.

\bigskip

The study enrolled poor, underemployed youth who expressed willingness to enroll in a training program at baseline. Average yearly income in this population was about \$190 a year on average. Of 2,275 individuals who attended an orientation meeting and signed up for Huguka Dukore, 1,967 met the program’s eligibility criteria. A further 119 could not be located either in the village of their stated residence, leaving a total of 1,848 youth who were enrolled in the study. After a baseline survey, conducted from December 2017-February 2018, thirteen public lotteries were used to randomly assign the youth into five groups:

\bigskip

\begin{enumerate}[noitemsep]
\item	The Huguka Dukore program group
\item	A cash grant group (intended to be the same cost as Huguka Dukore)
\item	Cash grant and Huguka Dukore combined (to test if the interventions complement each other)
\item	A larger cash grant, which happened to be roughly equal to the cost of the combined arm (about \$845)
\item	Control group, in which no program was offered at the time of study
\end{enumerate}

\bigskip

Given that the total cost of the programs was not fully known before the study began, the
research team conducted a detailed costing exercise prior to, and also after, the intervention period. The costing beforehand was used to estimate the total cost of the Huguka Dukore intervention, as well as the estimated overhead costs to GiveDirectly of providing household grants in this context. 

\bigskip

This exercise arrived at a per-beneficiary cost of \$452.47 of the Huguka Dukore program. However, the program ended up costing substantially less: \$388 per person, which once we accounted for non-compliance is only \$332 per study subject. Therefore researchers use regression adjustment compared the program to a cash transfer costing the same amount, which would have delivered \$255.04 to beneficiaries (see section 2 for costing details).

\bigskip

The Huguka Dukore program was implemented for nine months, from January 2018 - November 2018 and the 
cash transfers were delivered between May 2018 - July 2018. The follow-up survey was conducted 18 months after a baseline survey, from July 2019 – August 2019, which was 18 months after the baseline survey (8-9 months after the program ended). A longer-term follow-up survey will be conducted 36 months after baseline (November 2020 – February 2021).

\bigskip

\subsection*{Findings}

\noindent \textbf{Main Findings: Huguka Dukore compared to no intervention}
The main findings of evaluation of the Huguka Dukore program, compared to the comparison group, are as follows:

\begin{itemize}[noitemsep]
\item	Youth experienced a surge in productive asset values, which rose to 154\% higher than the control group average, a large and notable impact given the program made no material transfers to the beneficiaries.
\item	The program also led to an increase in productive hours: Huguka Dukore was successful in driving a 3 hour increase to a base of 18.4 hours, an improvement of 16\%.
\item	However, youth who received the program were no more likely to be employed than the comparable youth who did not receive the program, nor did program youth experience higher incomes or consumption as a result of the program.
\item	Average savings doubled.
\item	Subjective wellbeing improved (based on a survey about happiness and life-satisfaction)
\item	Business knowledge increased: participants performed better on a test of business knowledge built against the course curriculum.
\end{itemize}

\bigskip

\noindent \textbf{Main Findings: Cash grants compared to no intervention}
The main findings of the evaluation of the central cash grant amount (on average 14 months after transfers took place), compared to the control group, are as follows:
\begin{itemize}[noitemsep]
\item	Youth in the cash group also experienced a surge in productive assets; values almost quadruple relative to control. 
\item	Youth experienced higher incomes and their household- and individual-level consumption increased. 
\item	Productive hours are non-linear in transfer amount, with the middle transfer amount leading to a significant 6.5 hour per week increase, and none of the other transfers having a significant impact.  Youth in the large cash transfer arm achieve an insignificant 1.6-hour improvement. This is the first evidence to suggesting that once transfers become sufficiently large they may reduce the incentive to work. 
\item	However, youth who received the program were no more likely to be employed than the comparable youth.
\item	Average savings more than doubled 
\item	Subjective well-being improved
\item	Net, non-land wealth increased by 90\%
\end{itemize}

\bigskip

\noindent \textbf{Main Findings: Huguka Dukore compared to cash grants}
In the head-to-head comparison, the evaluation findings can be summarized as follows:

\begin{itemize}[noitemsep]
\item  The cost-equivalent cash grant performed significantly better than Huguka Dukore at increasing monthly income, productive assets, subjective well-being, beneficiary consumption, and household livestock wealth.  
\item	Huguka Dukore was better at increasing business knowledge (the only outcome in which it outperformed cash)
\item	In sum, over the 18 month horizon, youth benefited more from cash grants than from Huguka Dukore program across a range of indicators central to beneficiary economic welfare, while Huguka Dukore was more effective at generating business knowledge.
\item  Neither Huguka Dukore or cash grants had a statistically significant impact on employment after 18 months. 
\end{itemize}

\bigskip

\noindent \textbf{Other noteworthy findings:}
\begin{itemize}[noitemsep]
\item	The evaluation did not find any complementarity between HD and cash; when they are implemented together we see the same or worse than we would expect by adding up the independent effect of the two programs. Rather, in something of a challenge to ever-more complex bundled programs, each of these interventions has a distinct set of benefits that operated independently.
\item	Nor did the evaluation find any ‘spillover’ effects on outcomes of non-beneficiaries in the same villages, though evidence suggests that take-up of HD is highest when that program is implemented with high geographic intensity. 
\item	Both interventions had a relatively consistent effect across richer and poorer, male and female, older and younger, and across local labor market conditions. 
\item	While neither program significantly improved overall employment rates during the study period, a more detailed analysis shows that youth who received cash were more likely to move from wage labor into self-employment (they became more entrepreneurial), while Huguka Dukore beneficiaries became engaged in more off-farm wage labor (their training propelled them into wage jobs). In other words, at cost-equivalent levels, cash and training have launched youth into distinct forms of employment.
\end{itemize}

\bigskip


\clearpage

\begin{singlespace}
\thispagestyle{empty}
\setcounter{tocdepth}{2}
\tableofcontents 
\end{singlespace}

\clearpage 
\pagenumbering{arabic}
\setcounter{page}{1}
\setcounter{section}{0}

\section{Introduction}\label{s:intro}
The demographic dividend in Sub-Saharan Africa is a double-edged sword. A young population provides an opportunity to benefit from the many productive years ahead while bearing a limited burden of dependency from older generations---but not if young people are unable to find productive employment \citep{fox2016youth}.  
In spite of gains in formal educational attainment, youth unemployment rates remain high; for example while 40 percent of Rwanda's population is between the ages of 14--30, 65 percent of these youth are unemployed.  This raises the prospect of both a lost generation of opportunity, and the political risks that accompany a large, unemployed, urban, young population \citep{
bongaarts2016development}.  Hence, it is critical to understand the barriers in physical and human capital that prevent youth from being fully productive.  

In spite of this pressing need, policymakers have limited access to evidence-based interventions with a track record of effectiveness.  
This is not to say that active labor-market interventions have not been studied; for example, a recent review discusses nine randomized evaluations from developing countries \citep{McK17almp}. Despite some signs of success in generating employment \citep{diaz2016impact,AlfBanBasBurRasSulVit19unpub}, the impacts of programs aimed at lifting human capital have been variable and less impressive than hoped in terms of labor and income benefits.  A systematic review by \cite{kluve2017interventions} finds labor market interventions to have positive effects on employment and income, but these impacts are small and highly variable across studies.  At the same time, the costs of relaxing capital constraints are falling due to the widespread availability of mobile money in the developing world.  A large literature finds that unconditional cash transfers are invested in durables \citep{haushofer2016short},  productive assets \citep{gertler2012investing, blattman2018long}, and microenterprises  \citep{de2012one}, suggesting that cash may be a reasonable alternative in delivering economic livelihood assistance to youth.  Given that the literature has long recognized both `money and ideas' may serve as constraints to the productivity of young entrepreneurs \citep{gine2014money}, rigorous comparative cost-effectiveness research across these different modalities is sorely needed, as well as a better understanding of potential complementarities between them.

This study addresses these challenges by undertaking an exercise in \emph{cash benchmarking}:  the direct comparison of in-kind- to cash-transfer programs in a single experimental setting. As an applied-science exercise, such a study is a form of comparative cost-effectiveness analysis; it compares the returns to alternative forms of programming on a pre-defined set of outcomes. And it can answer this counterfactual question subject to a distributional constraint, seeking to hold the value of programming per beneficiary constant across modalities.\footnote{Our companion study \cite{mcintzeit2019gikuriro} uses a similar approach to benchmark a USAID-funded child nutrition program against cash in Rwanda.}  
Such cash-benchmarking exercises also inform a basic-science question, by lifting distinct constraints to individual employment outcomes. Similar efforts include \citet{ahmed2016kinds} 
who compare BRAC's ultra-poor programming to cash, or  \citet{KarOseOseUdr14qje} who examine the comparative impact of the relaxation of credit and risk constraints in agriculture.  In the context of youth livelihoods, training programs and cash grants each move alternative potential constraints to productive employment---skills and liquidity, respectively.  One way of conceiving of the value of this benchmarking activity is that for any given outcome, our design allows us to cast the opportunity cost of skills improvement in pecuniary terms, despite the fact that these skills cannot be bought on the market.  This allows us not only to determine the benefit generated by an increment of skills improvement, but also to calculate the counterfactual cost of generating the same benefit by relaxing financial (rather than human capital) constraints.  The inclusion of a combined arm allows us to study complementarities, asking if the returns to relaxing capital constraints improve when human capital constraints have also been relaxed.\footnote{\cite{fox2017works} say that "It may be helpful to experiment more with transferable skills development initiatives\ldots combined with cash transfers to youth or access to finance."}

We study this question using an individually randomized trial with 1,848 underemployed Rwandan youth to understand how a `standard' package of training, soft skills, and networking interventions compares not only to an experimental control group but to an additional arm that receives household grants intended to be of equal cost to the donor---a \emph{cash benchmark}. The study follows poor, underemployed youth aged 15-30  who expressed interest in participating in the training program. The core program is called Huguka Dukore/ Akazi Kanoze which means `Get Trained and Let's Work/Work Well Done'  in Kinyarwanda (abbreviated henceforth as HD); it follows USAID's strategy on workforce readiness and skills training and was implemented by Education Development Center, Inc. (EDC). The benchmarking cash transfer program was implemented by GiveDirectly (GD), a US-based nonprofit that specializes in making unconditional household grants via mobile money. USAID also uses cash transfers in its programming in a number of dimensions, so this study compares two different means through which it could attempt to deliver benefits.\footnote{USAID currently uses cash mostly in the humanitarian space, but is also involved in new efforts to explore cash as a form of development assistance in countries such as Morocco and Nigeria.}  These two treatments are compared to a control group, namely a set of individuals that receive neither program, and a combined arm that receives both. Our study provides a methodology incorporating randomization of transfer amounts and ex-post regression cost adjustment that can achieve this benchmarking objective in a general way.  In a penultimate `Value for Money' section we show how our study design can be used to conduct either cost equivalence comparisons, or to do cost effectiveness analysis by comparing benefit-cost ratios across arms. 

The Huguka Dukore program is a particularly attractive candidate for a benchmarking evaluation.  It is a five-year project (2017-2021) aiming to provide 40,000 vulnerable youth with employability skills in 19 (of 30 total) districts nationwide. Targeting youth from poor households with less than secondary education, with an emphasis on women and youth with disabilities, HD offers multiple program pathways including: i) employment preparation; ii) individual and cooperative youth microenterprise start-up; iii) business development for existing microenterprises.
HD is based on a predecessor \emph{Akazi Kanoze} program, which operated in the country from 2012-2017, and which was evaluated as successful in an RCT led by EDC \citep{alcid2014randomized}.  It is a carefully designed and intensive training program, and it is backed by rigorous research heading in to this study.

The Government of Rwanda places a high priority on such programs: Priority Area 1 of the ``Economic Transformation Pillar'' in its seven-year plan for the period 2017--2024 includes the key strategic intervention to ``support and empower youth and women to create businesses through entrepreneurship and access to finance'' \citep[p. 3]{rwanda17sevenyrplan}. 
Further, training programs of this sort are widespread across the developing world: \citet{BlaRal15review} estimate that the World Bank alone spends almost a billion dollars annually on skills training programs.   
In spite of their prevalence, however, the cost-effectiveness of such programs is far from certain. Reviewing evidence on active labor market programs that operate on the supply side of the labor market, \citet{McK17almp} finds that employment and earnings impacts are modest, with costs averaging 50 times the monthly income gain.
And indeed, in its \emph{Future Drivers of Growth} report, produced jointly with the the World Bank, the Government of Rwanda raises the possibility that ``for a significant portion of the population who will continue creating their own jobs, capital-centric programs may be more effective and cheaper to implement than simple training programs'' \citep[p. 81]{rwanda19futuredrivers}. Our study seeks to resolve this uncertainty by direct comparison.

The momentum for benchmarking has built as numerous studies have shown meaningful impacts of cash transfers on important life outcomes in the short term, such as child nutrition \citep{aguero2006impact, seidenfeld2014impact}, schooling \citep{skoufias2001conditional}, mental health \citep{baird2013income, samuels2016being}, teen pregnancy and HIV \citep{baird2011cash}, microenterprise outcomes \citep{de2012one}, consumer durables \citep{haushofer2016short}, and productive assets \citep{gertler2012investing}.  The evidence on the long-term impacts of cash transfers is more mixed \citep{blattman2018long}, but some studies have found substantial impacts \citep{fernald200910, barham2014schooling, aizer2016long, hoynes2016long, balboni2019people}.\footnote{For examples of studies that find dissipating long-term benefits, see \cite{baird2019money},  \cite{araujo2017can}, and \cite{brudevold2017firm}.  Evidence from systematic reviews of cash transfers on schooling \citep{molina2016long} and child health \citep{pega2014unconditional, manley2013effective} has also been uneven, with substantial heterogeneity in findings across studies.}   The largest extant literature on benchmarking is based on the comparison of cash aid to food aid \citep{leroy2010cash, schwab2013form, hoddinott2014impact, hidrobo2014cash, ahmed2016kinds, CunGioJayXXrestud}, which has uncovered a fairly consistent result that food aid leads to a larger change in total calories while cash aid leads to an improvement in the diversity of foods consumed.  Efforts to benchmark more complex, multi-dimensional programs to cash include BRAC's Targeting the Ultra-Poor program \citep{chowdhury2016valuing}, microfranchising \citep{brudevold2017firm}, and graduation programs \citep{sedlmayr2017cash}.  
 
The randomized controlled trial proceeded in four steps. First, EDC's three local implementing partners within the study ran recruiting workshops drawing more than 2,000 eligible individuals who expressed interest in participating in HD.  From this group we then conducted baselines, checked eligibility, and recruited an experimental study sample of 1848 individuals who consented to the lottery and the baseline survey and were included in the randomization.  These individuals come from 328 villages in non-urban parts of the three districts of Rwamagana, Muhanga and Nyamagabe. The HD-imposed eligibility criteria for their training of vulnerable youth consist of (a) ages ranging from 16--30, and (b) between 6-12 years of education. Because of the conditions placed on GiveDirectly by the Rwandan government, we further strictly limited eligibility to (c) households registered in Ubudehe poverty status 1 or 2 (the poorest). In addition, in order to provide a study that has compliance rates with the HD training that are as high as possible, we further restricted eligibility to those who (d) expressed interest in participating in the employment and entrepreneurship readiness training. These individuals were recruited at a first `orientation' meeting at which the local HD implementers and the survey firm (Innovations for Poverty Action, or IPA) recorded sufficient information to enroll them and to subsequently perform baseline surveys at the household (there were no refusals to the household survey). Second, IPA collected baseline data, implementing survey instruments that collected information both at the household level and at the individual beneficiary level for study participants.  Third, IPA conducted a series of 13 public lotteries at the sector level, overseen by sector- and local-level officials, at which individuals were assigned to four main arms and a control, to be treated accordingly by implementers.  Finally, the study collects baseline and midline (18 month) indicators across a range of economic, psychological, and business-related outcomes to measure comparative impacts.  

We costed both programs in detail prior to, and after, the intervention period, following \cite{levin2017economic}. The ex-ante costing exercise was used to identify the approximate total cost of the HD intervention, as well as the estimated overhead costs to GiveDirectly of providing household grants in this context. The ex-ante costing of HD arrived at a per-beneficiary cost of \$464.25.  We then randomized transfer amounts at the individual level in the cash arm across four possible transfer amounts. These amounts were chosen to provide informative benefit/cost comparisons across two different margins: HD vs cash, and small versus large cash transfer amounts.  Incorporating GiveDirectly's operating costs, the amount actually received by households that generates the same expected cost to USAID as HD is \$410.65. The comparison between these two arms therefore provides a straightforward window on expected cost-equivalent impacts.  Because we anticipated that the exact numbers from the ex-post costing exercise would differ from the ex-ante exercise, we  randomized two bracketing cash transfer arms which transfer \$317 and \$503 to households.  Thus even our smallest transfer is providing individuals with 167\% of annual average per capita income.  In the end the HD program turned out to be less expensive than expected; the final ex-post costing figure of \$332.27 is used to  regression-adjust outcomes across transfer amounts to arrive at a comparison between HD and cash estimated at the exact cost-equivalent amount from the perspective of the donor, USAID.  Because of the low costing number, an adjustment which was intended to be an interpolation over GD transfer costs ends up being an extrapolation to a value 16\% lower than the smallest GD arm's cost.
 
The study also features a combined arm that receives both the middle GD transfer amount and HD training.  The inclusion of this arm permits a classic test for complementarities between human capital and financial interventions.  Finally, the larger cash transfer arm  was an amount chosen by the cash implementer as maximizing their own cost-effectiveness (transferring \$750, and costing \$846.71 which turns out to be almost exactly the cost of the combined arm).  The inclusion of this arm provides a statistically high-powered way of examining how benefit/cost ratios shift as the transfer amount rises.  This opens up a different type of comparative cost effectiveness question:  would the net benefit from cash transfers be maximized by concentrating large payments on a few individuals, or by spreading out smaller transfers to more people?  And if more money is to be invested over the basic transfer, should it be in the form of additional physical capital, or is training then more effective?

Our results show that at 18 months after baseline, on average 15 months after being offered HD programming and at least 3 months after any training would have ended, Huguka Dukore has delivered real benefits.  While there is no overall improvement in employment rates, the HD arm sees an increase in productive hours,  productive asset values more than double, average savings increase by 60\%, and subjective well-being is higher.  In addition, the HD arm performs a half a standard deviation better on a test of business knowledge built against the course curriculum, showing that the program clears the basic bar of having created real learning.  

The cash transfer arm, on average 14 months after transfers took place, sees improvements across a broad range of economic and psychological outcomes. These impacts prove surprisingly invariant to the transfer amount variation present in this study, suggesting that even our lower transfers clear a barrier that generates real benefit to households.  With the exception of the improvement that HD generates on business knowledge, cash improves every outcome that HD improves, generally with a greater magnitude, and in addition drives monthly income, household- and individual-level consumption, livestock value, and overall wealth, to higher levels.

Consequently, when we conduct our pre-specified comparative impact analysis, we find cost-equivalent cash to generate significantly larger benefits for income, productive and livestock assets, individual consumption, and subjective well-being, while HD is more effective at generating the human capital benefit of business knowledge.  We find no evidence of complementarity between human and physical capital interventions; if anything the combination appears to do worse than we would expect by adding up the individual impact of the two programs.  Significant negative complementarities are present for productive hours of work and for subjective well-being.  In summary, then, at the cost of around \$330 per beneficiary where the core comparison is done, cash moves a set of economic and psychological outcomes more than HD,  HD is better at improving business knowledge, and both of the efforts we engaged in to increase the amount spent above this (whether through more cash or through adding HD) had disappointing returns.  

We then look for signs of heterogeneity in impacts across gender, as well as baseline consumption, risk aversion, and local employment rates. Overall we see little meaningful heterogeneity, suggesting that both interventions have a relatively consistent effect across richer and poorer, male and female, and across local labor market conditions.  We then exploit the random variation generated by the lotteries in the intensity of treatment at the village level to look for evidence of spillovers.\footnote{This issue is particularly important given our household-level assignment and the recent evidence on spillovers both from job training programs \citep{crepon2013labor} and cash transfer programs \citep{angelucci2009indirect, egger2019general}.}  The only evidence we find for spillovers are that compliance with HD, which overall in the sample is 85\%, is improved by having more people in your village assigned to HD as well.  Looking at our primary outcomes we find no evidence that spillovers from any treatment or to any treatment group are present, suggesting that by this measure at least the study is internally valid.

These results illustrate the complexity of the comparisons created by this type of benchmarked design.  Considered relative to the control, HD has been successful in moving some of the key welfare indicators the program is geared towards, and has strongly improved the core metric of learning.  Even in a comparative sense, it is impressive that a program that made no material transfer to its beneficiaries could generate improvements in asset values half as large, and improvements in savings two thirds as large, as a program that gave them hundreds of dollars.  Nonetheless when we compare the programs in a head-to-head way it becomes clear that at least at 18 months from baseline, cash is outperforming the training program across a set of indicators likely to be central to beneficiary economic welfare (individual consumption, income, livestock wealth, and subjective well-being).  In something of a challenge to ever-more complex bundled programs, we find that each of these interventions has a distinct set of benefits that operate independently, and little is gained by providing them together.

In the remainder of the paper, we provide details of the experimental design, survey structure, and costing exercise in  Section \ref{s:research_design}, and then present the results of the study in Section \ref{s:results}.  Section \ref{s:conclusions} concludes.

\section{Experimental Design}\label{s:research_design}

\subsection{Interventions}\label{ss:Interventions}

\subsubsection*{Huguka Dukore:  Employment and entrepreneurship readiness training}

Huguka Dukore is a five-year activity providing 40,000 vulnerable youth with increased opportunities for wage and self-employment through a suite of interventions that includes market relevant work readiness training, employability skills training, work based learning, internship opportunities, links to employment and entrepreneurship training at the youth level.  The program builds on lessons learned from EDC’s prior work in this area through the Akazi Kanoze Youth Livelihoods Project (henceforth AK). 

Over the life of the project, HD will prepare 21,000 new youth for employment with Rwandan employers, with an additional 2,000 alumni receiving middle management training. It is assisting 13,000 new HD participants to start their own microenterprise, while supporting 4,000 youth (2,000 new and 2,000 AK alumni) with an existing microenterprise to grow their business, linking 15,000 youth to financial services.  Finally, HD provides support to its 30 local Implementing Partners (IPs) to improve their job placement rates.  It is important to note that the full HD program includes a number of higher-level interventions, including training front-line providers, organizational capacity building in workforce development systems, and facilitating linkages between businesses, government, and local NGOs.  Because our intervention studies the cross-individual variation within communities in which HD is working, we measure only the youth-level components of the intervention and not these more systematic dimensions.

The HD program consists of a number of separate modules which are taken serially over the course of a year.  The first of these is `Work Ready Now!', consisting of eight sub-modules (Personal Development, Interpersonal Communication, Work Habits and Conduct, Leadership, Health and Safety at Work, Worker and Employer Rights and Responsibilities, Financial Fitness, and Exploring Entrepreneurship).  This module is taken by all students as the lead-in to the HD training, and consists of 10 five-day weeks of full-day training.  

From here students choose the additional modules and the sector of work in which they receive additional training, and the curriculum splits according to the nature of formal employment opportunities in local markets.  In more urban areas students would then move on to a Technical and Vocational Training (TVET) module, Transition to Work programming, and Work Based Learning Services.  Because our study areas are almost exclusively rural, HD instead encourages students to focus on self-employment, meaning that the next module of HD would be the `Be Your Own Boss' training, which is an entrepreneurship curriculum that is tailored to the specific interests and opportunities in a specific cohort of students, and lasts another 10 weeks.  After this point HD students are typically placed in an internship or apprenticeship position with a local entrepreneur working in the selected sector.  During this interval students have regular check-ins with their trainers.  Within a year of the initiation of training students are considered `graduates' of HD.\footnote{Additional components of the broader HD curriculum include assisting students with access to finance through assistance in the formation of Savings and Internal Lending Communities and access to bank financing, and the use of a job matching resource that maintains a list of open positions and attempts to match graduates to them.  These components of HD were not operative in the study districts at the time that we ran this evaluation.}

Because the curriculum involves several components of choice (whether to pursue vocational or small business training, the sector in which to be trained), our experimental analysis will treat HD as a single intervention of which this choice is an integral component.  We provide an analysis of the determinants of participation in various components of the potential HD curriculum.  

\subsubsection*{GiveDirectly: Household grants program}

To benchmark the impact of the HD program on cash, we worked with GiveDirectly, a US-based 501(c)3 Non-Profit organization. GiveDirectly specializes in sending mobile money transfers directly to the mobile phones of beneficiary households to provide large-scale household grants in developing countries including Kenya, Uganda, and Rwanda. GiveDirectly’s typical model has involved targeting households using mass-scale proxy targeting criteria such as roof quality. GiveDirectly builds an in-country infrastructure that allows them to enroll and make transfers to households while simultaneously validating via calls from a phone bank that transfers have been received by the correct people and in a timely manner. Their typical transfers are large and lump-sum, on the order of \$1,000, and the organization provides a programatically relevant counterfactual to standard development aid programs, because it has a scalable business model that would in fact be capable of providing transfers to the tens of thousands of households that are served by the HD program.

Since eligibility did not condition on having a cellphone, during the enrollment process individuals who did not themselves own a cell phone provided a number belonging to a trusted family member or friend, and transfers were sent to them through this intermediary. The payments were made to beneficiaries in two installments two months apart, with the first payment comprising 40 percent of the total to be paid to the beneficiary, and the second payment completing the transfer. After each payment is made, staff in the GiveDirectly call center team in Kigali contact every recipient to verify that payments have been received.

In terms of implementation timing, GD orientation commenced immediately after lotteries to notify youth randomized to receive a household grant and introduce them to the program. The value of household grants was not to be disclosed until the ‘GD Treatment’ step below. GD Treatment (where transfer values will be disclosed to recipients) did not commence anywhere until the lotteries have been conducted everywhere in the district so as to avoid emphasizing the cash treatment prior to the completion of recruitment.

\subsubsection*{The Combined Arm}

 The Combined arm received both treatments (GD Middle plus HD).  Both interventions were received at the same time as others in their same sector, meaning that they typically started the HD treatment several months before they would receive the household grant from GD.

\subsection{Enrollment criteria}\label{ss:enrollment} 

A detailed timeline showing the evolution of the study is presented in Figure \ref{f:timeline}.  The study recruits youth from 13 geographic `sectors' in the districts of Rwamagana, Muhanga and Nyamagabe.\footnote{In Rwanda, the \emph{sector} is the geo-political unit below the district. There are 30 districts in Rwanda, and 416 sectors in total across those 30 districts.}
Study participants had to be eligible for Huguka Dukore, to attend an informational session about Huguka Dukore, to enroll in a lottery to determine participation in  that program following that informational setting, and to be traceable to a residence in a village in the sector where they were recruited.  Attendance in person at the public lottery was not required for program enrollment. The study enrolled in its sample all individuals who met criteria for treatment by Huguka Dukore in the study sectors.  More detail on sample recruitment and the conduct of the lotteries is provided in the Appendix.

Table \ref{t:vetting_recruitment} shows the process by which we moved from the original oversubscription universe to the final sample of 1848 individuals deemed as fully eligible who were recruited into the study and randomized.  
Of 2,275 individuals who attended an orientation meeting and signed up for HD, 1,967 were found based on administrative review to meet the eligibility criteria.  A further 119 could not be located either in the village of their stated residence, or were found to be resident outside the sector entirely, and consequently were deemed ineligible for intervention and the study. 

There were no survey refusals at baseline, so our study sample reflects the full population of individuals who were assigned to treatments. 
The final study sample therefore consists of the universe of all individuals who met the enrollment criteria for Huguka Dukore, who attended an information session; who agreed at that information session to be included in the assignment lottery; who were found resident in the relevant sector at baseline. 

Demographic and employment characteristics (the latter of which will be defined in greater detail in Section \ref{ss:outcomes} below) of the study participants are detailed in Table \ref{t:balance}.  Consistent with Huguka Dukore's `soft' targeting criteria, the sample is 59 percent female, with an average age of 23.5, (among the random sample assigned to control).  They have an average of 7.6 years of education, and typically live in households of approximately five individuals.  

Although Huguka Dukore seeks to bolster employment opportunities for underemployed youth, it does not employ a hard criterion regarding employment for eligibility. Consequently, it is not unusual for individuals to report that they are employed:  33 percent of (control-group) respondents reported being employed at baseline, using a definition that \emph{excludes} agricultural work on a farm belonging to their own household (see Section \ref{ss:outcomes} for more details).  By endline the employment rate had risen to 48\% in the control group.

Nonetheless, individuals in the study population are quite poor. 32 percent reside in households that the Government of Rwanda categorizes as Ubudehe I---its lowest socio-economic category, denoting a condition of `extreme poverty'.  Median consumption per adult equivalent is 5,879 RWF per month, which in 2018 PPP terms translates to a consumption level of USD 0.66 per day.

\subsection{Assignment protocol}\label{ss:assignment}

The allocation of these study households to treatment was undertaken on a randomized basis across eligible, interested individuals using a public lottery.  A public lottery was selected as the assignment mechanism given the very large sums of money being transferred and the desire by all parties to the research to ensure that the assignment was considered to be fair and impartial by the research subjects.  

Lotteries were conducted at the sector level in each of the 13 sectors in the study, and the proportions assigned to each treatment were fixed at each lottery. This results in a fairly standard `blocked' randomization structure across the 13 blocks in the study.  Participants drew their own treatment status as tokens of different colors from a sack, where each token corresponded to a given treatment arm and the number of tokens in the hat was determined by IPA according to the number of participants with fixed proportions assigned to each treatment.

The detailed protocol for the lottery is as follows:
 \begin{enumerate}
    \item Beneficiaries did not have to be physically present at the lottery to be included in the study.  
    \item We explicitly recognized the right of EDC/HD to eliminate from eligibility any individuals who they feel, for whatever reason, was not ‘serious’ about the program and that they did not believe will fully enroll in HD if selected.
    \item  Detailed information about GD was not provided prior to the lottery, but GD was described in detail at the lottery and every effort was made to preserve the separate identity of HD and GD so as not to provoke confusion about the broader HD program.  All information provided at the lottery was given to everyone, and there was not an attempt to separate groups and give private information.
    \item A representative of both GD and HD (or its local partner) were present at every lottery.
    \item Individuals were notified whether they have been assigned to the GD, the ‘combined’ arm or the HD arm at the time of the lottery. 
    \item Individuals assigned to GD received a variety of colors which correspond to different transfer amounts.  This means that the random assignment to GiveDirectly simultaneously randomly assigned individuals to the different transfer size amounts. The exact financial amounts were not discussed at the time of the lotteries. GD  explained that youth randomized to GD would be contacted soon after the lottery to orient them to the program, and visited at their place of residence to undertake the enrollment process. 
 \end{enumerate}

\afterpage{
\begin{landscape}
    \vspace*{\fill}
\begin{table}[!h]\centering
\caption{Study Design}\label{t:treatment_assignment}

\begin{tabular}{L{1in} *{7}{C{1in}}}
\toprule %
		&			&				& \multicolumn{4}{c}{GiveDirectly} & Combined \tabularnewline
\cmidrule(lr){4-7} \cmidrule(lr){8-8} %
Sector 	&  Control 	& Huguka Dukore & 317.16 & 410.65 & 502.96 & 750.30 & HD + 410.65 \tabularnewline 
\midrule %
Kaduha	&	63	&	60	&	21	&	21	&	22	&	22	&	26	\tabularnewline 
Kibumbwe	&	32	&	37	&	10	&	10	&	12	&	13	&	13	\tabularnewline 
Kigabiro	&	14	&	12	&	4	&	5	&	4	&	5	&	5	\tabularnewline 
Kiyumba	&	17	&	17	&	6	&	6	&	6	&	6	&	8	\tabularnewline 
Mugano	&	51	&	51	&	18	&	18	&	18	&	18	&	22	\tabularnewline 
Muhazi	&	39	&	40	&	13	&	19	&	13	&	18	&	17	\tabularnewline 
Munyaga	&	34	&	34	&	10	&	10	&	10	&	12	&	14	\tabularnewline 
Munyiginya	&	25	&	25	&	8	&	8	&	8	&	10	&	10	\tabularnewline 
Musange	&	30	&	29	&	10	&	10	&	10	&	9	&	12	\tabularnewline 
Mushishiro	&	24	&	23	&	6	&	6	&	6	&	9	&	8	\tabularnewline 
Nyakariro	&	49	&	50	&	16	&	17	&	19	&	17	&	22	\tabularnewline 
Nyarusange	&	57	&	54	&	21	&	20	&	19	&	19	&	24	\tabularnewline 
Shyogwe	&	53	&	53	&	18	&	18	&	18	&	20	&	22	\tabularnewline 
\midrule %
Total	&	488	&	485	&	161	&	168	&	165	&	178	&	203	\tabularnewline 
\bottomrule 
\end{tabular}
\floatfoot{ 
\begin{footnotesize}
Note :  This table gives the number of study individuals assigned to each treatment arm in each of the 13 sectors within which lotteries were conducted.  The lotteries were blocked so that fixed fractions of individuals are assigned to each arm.
\end{footnotesize}
}
\end{table}
\vspace*{\fill}
\end{landscape}
}

Table \ref{t:treatment_assignment} shows the outcome of the lottery process, giving the number of individuals assigned to each of the treatment arms within each lottery, as well as overall.  

\vspace{10pt} The assignment of individuals to the main study arms was as follows:
\begin{enumerate}
\item HD beneficiaries (485 individuals);
\item Recipients of unconditional household grants (672 individuals);
\item Combined arm who received both HD and the household grants intervention (203 individuals)
\item A comparison group, in which no program was offered (488 individuals).
\end{enumerate}

Household grants were randomized at the individual level over four transfer amounts. The value of the first transfer amount was made equivalent to the total cost of providing HD to each beneficiary, which is \$452.47. Less GD's own associated costs of delivery, this means that an amount of \$410.19 was actually transferred to households in this arm to make them cost-equivalent to USAID. Because we did not know the true per-capita cost of HD with certainty beforehand, we randomize GD transfer amounts to two additional values that bracket this expected cost.  The bracketing amounts are derived by supposing that the number of beneficiaries for the year two tranche of HD funding nationwide might vary between 8,000 and 12,000 beneficiaries, meaning that the per-capita cost would vary between \$377.05 and \$565.58.  Again netting out GD's costs of making transfers, that means that households in these arms actually receive \$317.34 and \$503.04, respectively (note that because we costed each GD transfer amount separately and because many of GD's costs are fixed at the individual level, the fraction of total cost that is operating cost declines as the transfer amount increases). The fourth transfer amount was designed to maximize the benefit-cost ratio of household grants, and transferred \$750 to beneficiaries.  

In the first phase of lotteries, comprising 792 study participants---we randomized purely at the individual level, as the study design did not anticipate multiple enrollees from the same household. In fact, the 792 participants in the first tranche of lotteries comprised 732 unique households.  This resulted 34 households in which individuals in the same household were assigned to different treatments (at the level of the major arms of the study). Having recognized this issue, we altered the protocol in the second phase of lotteries and assigned treatment at the household level.  To reflect this issue we cluster standard errors at the household level.

Given the public nature of the lottery assignment, the study was not blinded either to participants or to the survey firm.  The study is not a pipeline design, and to avoid expectancy biases we made it clear to the subjects at the time of the lottery that there would be no subsequent treatment by these implementers in the area.

\subsection{Program Participation}

Compliance with GiveDirectly treatment was nearly perfect. One individual in the middle GD arm was found to be ineligible by Ubudehe status and was not treated, and one individual assigned to the lower GD arm actually received the upper GD treatment.  For GD the ITT is therefore effectively the average treatment effect.

As anticipated, HD was most successful in achieving participation in its initial 10-week training (Work Ready Now).  86\% of the full HD treatment group (both HD-only and combined arms) were counted as enrolled according to the contractual definition (attending the end of the first week of WRN training).  This is the rate that the costing exercise uses since it alone determines the amount paid from USAID to the local implementing partner.  But we can use institutional data from the HD program to examine participation in more detail.  

Retention during the course of WRN is high; 79\% of the overall sample completes this 10-week training program, which focuses on general workforce readiness.\footnote{The modules of the WRN curriculum are:  personal development, interpersonal communication, work habits and conduct, leadership, health and safety at work, worker rights, financial fitness, and exploring entrepreneurship.}  69\% of the of the HD sample complete the Build Your Own Business class (which is focused on entrepreneurship and self-employment); 13 individuals who did not take WRN did then go on to enroll in BYOB.  Finally, the technical training component of the HD intervention provides focused vocational instruction in a specific sector.  48\% completed the Technical Training component of the program.  In the combined arm, participation with each of these components is about 5 pp higher than in the HD-only arm.  Again, participation in Technical Training is not strictly confined to individuals who participated in any of the previous combinations of treatments.\footnote{The formality of the sectors towards which the technical training is geared varies, ranging from formalized (hospitality) to quasi-formal (tailoring, hairdressing) to more agricultural forms of self-employment (poultry, pig rearing).  Tailoring and poultry make up almost 75\% of the trainings.}

To understand the factors that determine selection into the various components of HD, we regress participation in each component of the program on a battery of baseline characteristics, pooling the entire HD treatment group (HD only and combined).  The results are presented Table \ref{t:HD_completion}.  In general compliance is relatively similar across observed beneficiary characteristics; older individuals are slightly more likely to complete the entrepreneurship training but not technical training.  There is a modest, negative association between working more at baseline and the likelihood of completing each stage of training; each additional hour of productive time use at baseline is associated with a reduction of approximately 0.4 percentage points in completion at all stages. Conversely, higher debt stocks at baseline are associated with greater completion rates. Both of these effects appear to be primarily driven by initial compliance, with measured attributes having little ability to predict subsequent compositional changes.

\subsection{Survey data collection and processing}\label{ss:data}

\subsubsection{Instruments}

Because we were interested in understanding both individual-level and household-level impacts, we used two distinct instruments within each round of data collection.  A household survey was administered to the household head, and a beneficiary survey was administered to the beneficiary. For beneficiaries who lived on their own or who headed their own household, these instruments coincided.

We provide an overview of the contents of each instrument in Table \ref{t:modules}.  Construction of primary outcomes and hypothesized effect moderators are detailed in Section \ref{ss:outcomes} below.

In addition to this midline, we also intend to return to the field 36 months after baseline to conduct a longer-term follow-up survey, providing an eventual window into longer-term impacts. Data collected then will be used to revisit the longer-term evolution of these interventions on the lives of the beneficiaries.

\subsection{Attrition}

We attempted to follow up with all study beneficiaries 18 months after baseline. The tracking protocol for the post-treatment round was designed around the individual beneficiary, following him or her to whatever the relevant household was at that time (rather than tracking the baseline household).  The interventions studied in this trial have the possibility of inducing migration; consequently it was particularly important to have a strategy to address attrition.  Our tracking strategy proceeded in two phases.  First, we attempted to track all individuals who were still residing in any study district or in Kigali.  Once we had completed this exercise we were left with 122 baseline individuals who we had not yet found.  We then randomly sampled half of these individuals (blocking on treatment status), and began an `intensive tracking' phase that spent substantial resources to track them wherever they had gone, including migrating out of the country, and survey them.  This exercise resulted in IPA finding and surveying all 60 living beneficiaries in the intensive tracking sample (one had passed away).  Given this remarkable rate of contact, we have an unusual situation where we should be able to convincingly correct for attrition by simply giving the intensive tracking sample weights of 2.  

To verify that the data weighted in this way recovers the missing potential outcomes, we should establish whether the intensive sample that we drew was representative of the universe of early attritors.  We can analyze this by a balance test of the intensive tracking sampling across the baseline outcome for all the early attritors.  The sample for this is small (122) but in Table \ref{t:IntensiveTracking} we find no evidence of systematic problems with this sampling (2 outcomes out of 20 unbalanced with $p$-values below 0.10 prior to correction for the False Discovery Rate or FDR, and none significant once we have corrected). These two pieces of evidence---representative sampling in intensive tracking and near-perfect tracking rate---suggest that we have an endline sample that is uniquely representative of the randomized universe.

\subsection{Balance}\label{ss:outcomes}

The next step is then to establish whether the attrited and reweighted sample used for analysis was balanced at baseline.  To ask this question, we estimate a balance table using baseline outcomes but only for the attrited endline sample, and with the weights, blocking, and clustering used in the endline analysis.  This makes the balance test mimic the impact analysis we will run as closely as possible; these results are presented in Table \ref{t:balance}.  The experiment appears well balanced (note that this is also the case if we simply use the full unweighted baseline sample), with rates of rejection consistent with random noise and none of the joint F-tests of all treatments indicating imbalance.  We therefore proceed to the analysis of impacts with confidence that the study is internally valid.

\afterpage{
\clearpage 
\thispagestyle{empty}
\newgeometry{margin=0.5in}
\begin{table}[!hbp]
\caption{Descriptive statistics and balance}\label{t:balance}

\begin{footnotesize}
\begin{center}
\begin{tabular}{l *{6}{S} ScSS}
\toprule
 & & \multicolumn{4}{c}{GiveDirectly} & & \multicolumn{1}{c}{\raisebox{-1ex}[0pt]{Control}}  \\ 
\cmidrule(lr){3-6}
\multicolumn{1}{c}{\text{ }} & \multicolumn{1}{c}{\text{HD}} & \multicolumn{1}{c}{\text{Lower}} & \multicolumn{1}{c}{\text{Middle}} & \multicolumn{1}{c}{\text{Upper}} & \multicolumn{1}{c}{\text{Large}} & \multicolumn{1}{c}{\text{Combined}} & \multicolumn{1}{c}{\text{Mean}} & \multicolumn{1}{c}{\text{Obs.}} & \multicolumn{1}{c}{\text{$ R^2$}} & \multicolumn{1}{c}{\text{$ p$-value}}\\
\midrule
\multirow[t]{ 3 }{0.2\textwidth}{Ubudehe category I }  &      0.01 &      0.00 &      0.07 &      0.01 &      0.01 &     -0.03 &      0.32 &      1720 &      0.07 &      0.73 \\ 
 & (0.03)  & (0.05)  & (0.05)  & (0.04)  & (0.04)  & (0.04)  \\  & [    1.00]  & [    1.00]  & [    1.00]  & [    1.00]  & [    1.00]  & [    1.00]  \\ \addlinespace[1ex] 
\multirow[t]{ 3 }{0.2\textwidth}{Beneficiary female }  &      0.01 &     -0.02 &      0.03 &     -0.02 &      0.02 &     -0.05 &      0.60 &      1770 &      0.04 &      0.68 \\ 
 & (0.03)  & (0.05)  & (0.05)  & (0.04)  & (0.04)  & (0.04)  \\  & [    1.00]  & [    1.00]  & [    1.00]  & [    1.00]  & [    1.00]  & [    1.00]  \\ \addlinespace[1ex] 
\multirow[t]{ 3 }{0.2\textwidth}{Beneficiary age }  &     -0.21 &     -0.41 &     -0.12 &     -0.66 &      0.43 &     -0.33 &     23.58 &      1770 &      0.03 &      0.12 \\ 
 & (0.23)  & (0.31)  & (0.34)  & (0.32)  & (0.32)  & (0.31)  \\  & [    1.00]  & [    1.00]  & [    1.00]  & [    1.00]  & [    1.00]  & [    1.00]  \\ \addlinespace[1ex] 
\multirow[t]{ 3 }{0.2\textwidth}{Beneficiary years of education }  &      0.11 &      0.10 &     -0.03 &      0.07 &      0.03 &     -0.14 &      7.55 &      1770 &      0.07 &      0.91 \\ 
 & (0.15)  & (0.22)  & (0.21)  & (0.20)  & (0.21)  & (0.19)  \\  & [    1.00]  & [    1.00]  & [    1.00]  & [    1.00]  & [    1.00]  & [    1.00]  \\ \addlinespace[1ex] 
\multirow[t]{ 3 }{0.2\textwidth}{Household members }  &     -0.32 &     -0.36 &     -0.03 &      0.00 &     -0.07 &     -0.32 &      4.98 &      1766 &      0.03 &      0.26 \\ 
 & (0.16)  & (0.24)  & (0.33)  & (0.20)  & (0.22)  & (0.19)  \\  & [    1.00]  & [    1.00]  & [    1.00]  & [    1.00]  & [    1.00]  & [    1.00]  \\ \addlinespace[1ex] 
\multirow[t]{ 3 }{0.2\textwidth}{Employed }  &      0.04 &     -0.00 &     -0.03 &      0.01 &      0.04 &      0.03 &      0.33 &      1770 &      0.02 &      0.73 \\ 
 & (0.03)  & (0.04)  & (0.04)  & (0.04)  & (0.04)  & (0.04)  \\  & [    1.00]  & [    1.00]  & [    1.00]  & [    1.00]  & [    1.00]  & [    1.00]  \\ \addlinespace[1ex] 
\multirow[t]{ 3 }{0.2\textwidth}{Productive hours }  &      0.45 &     -1.17 &      0.19 &      2.17 &      0.44 &     -0.13 &     10.81 &      1770 &      0.02 &      0.88 \\ 
 & (1.28)  & (1.68)  & (1.82)  & (2.04)  & (1.65)  & (1.54)  \\  & [    1.00]  & [    1.00]  & [    1.00]  & [    1.00]  & [    1.00]  & [    1.00]  \\ \addlinespace[1ex] 
\multirow[t]{ 3 }{0.2\textwidth}{Monthly income }  &      0.07 &     -0.03 &     -0.24 &      0.08 &      0.11 &      0.18 &      4.37 &      1770 &      0.01 &      0.99 \\ 
 & (0.33)  & (0.47)  & (0.46)  & (0.46)  & (0.47)  & (0.42)  \\  & [    1.00]  & [    1.00]  & [    1.00]  & [    1.00]  & [    1.00]  & [    1.00]  \\ \addlinespace[1ex] 
\multirow[t]{ 3 }{0.2\textwidth}{Productive assets }  &     -0.56 &     -0.47 &     -0.10 &     -0.30 &     -0.43 &     -0.13 &      2.49 &      1770 &      0.03 &      0.56 \\ 
 & (0.28)  & (0.38)  & (0.39)  & (0.40)  & (0.40)  & (0.38)  \\  & [    1.00]  & [    1.00]  & [    1.00]  & [    1.00]  & [    1.00]  & [    1.00]  \\ \addlinespace[1ex] 
\multirow[t]{ 3 }{0.2\textwidth}{HH consumption per capita }  &     -0.12 &     -0.11 &     -0.08 &     -0.10 &     -0.19 &     -0.01 &      9.46 &      1766 &      0.05 &      0.37 \\ 
 & (0.07)  & (0.10)  & (0.10)  & (0.10)  & (0.09)  & (0.09)  \\  & [    1.00]  & [    1.00]  & [    1.00]  & [    1.00]  & [    1.00]  & [    1.00]  \\ \addlinespace[1ex] 
\multirow[t]{ 3 }{0.2\textwidth}{Beneficiary-specific consumption }  &     -0.07 &      0.07 &     -0.03 &     -0.17 &      0.11 &      0.01 &      7.53 &      1770 &      0.03 &      0.93 \\ 
 & (0.15)  & (0.19)  & (0.21)  & (0.22)  & (0.18)  & (0.20)  \\  & [    1.00]  & [    1.00]  & [    1.00]  & [    1.00]  & [    1.00]  & [    1.00]  \\ \addlinespace[1ex] 
\multirow[t]{ 3 }{0.2\textwidth}{HH net non-land wealth }  &     -0.05 &      0.35 &      0.15 &     -0.17 &      1.13 &     -0.22 &     10.53 &      1766 &      0.03 &      0.19 \\ 
 & (0.46)  & (0.54)  & (0.63)  & (0.70)  & (0.47)  & (0.60)  \\  & [    1.00]  & [    1.00]  & [    1.00]  & [    1.00]  & [    1.00]  & [    1.00]  \\ \addlinespace[1ex] 
\multirow[t]{ 3 }{0.2\textwidth}{Savings }  &     -0.26 &     -0.48 &     -0.34 &      0.03 &     -0.09 &      0.13 &      8.01 &      1770 &      0.04 &      0.82 \\ 
 & (0.30)  & (0.41)  & (0.44)  & (0.39)  & (0.39)  & (0.38)  \\  & [    1.00]  & [    1.00]  & [    1.00]  & [    1.00]  & [    1.00]  & [    1.00]  \\ \addlinespace[1ex] 
\multirow[t]{ 3 }{0.2\textwidth}{Debt }  &      0.12 &     -0.14 &     -0.30 &      0.03 &      0.15 &      0.75 &      7.84 &      1770 &      0.02 &      0.47 \\ 
 & (0.32)  & (0.45)  & (0.46)  & (0.44)  & (0.45)  & (0.40)  \\  & [    1.00]  & [    1.00]  & [    1.00]  & [    1.00]  & [    1.00]  & [    1.00]  \\ \addlinespace[1ex] 
\multirow[t]{ 3 }{0.2\textwidth}{HH livestock wealth }  &      0.29 &     -0.18 &      0.20 &      0.24 &     -0.26 &     -0.17 &      7.32 &      1766 &      0.03 &      0.93 \\ 
 & (0.40)  & (0.57)  & (0.58)  & (0.55)  & (0.56)  & (0.52)  \\  & [    1.00]  & [    1.00]  & [    1.00]  & [    1.00]  & [    1.00]  & [    1.00]  \\ \addlinespace[1ex] 
\multirow[t]{ 3 }{0.2\textwidth}{Business Knowledge }  &     -0.01 &      0.10 &     -0.01 &     -0.08 &     -0.03 &      0.09 &      0.00 &      1770 &      0.03 &      0.57 \\ 
 & (0.07)  & (0.09)  & (0.09)  & (0.09)  & (0.09)  & (0.09)  \\  & [    1.00]  & [    1.00]  & [    1.00]  & [    1.00]  & [    1.00]  & [    1.00]  \\ \addlinespace[1ex] 
\bottomrule
\end{tabular}

\end{center}
\end{footnotesize}

\floatfoot{
Notes: Table presents control means and standard deviations; regression coefficients and standard errors for associated comparisons, and $p$-value for a test of the hypothesis that all arms pool. Regression-based comparisons and associated hypothesis tests based on a regression with block indicators. $^{***}$, $^{**}$, and $^{*}$ denote statistical significance at the 1, 5, and 10 percent levels, respectively.  All continuous variables winsorized at top and bottom 1 percent.  Inverse hyperbolic sine transformation taken for monthly income, household consumption, beneficiary expenditure, savings, debt, and wealth variables.
}
\end{table}
\clearpage
\aftergroup\restoregeometry
}

\subsection{Cost Equivalence, Before and After the Fact}

\subsubsection{Costing at Scale}
The costing exercise in the study utilized the `ingredients method' \cite[for more discussion, see][]{levin2001cost, dhaliwal2012research, levin2017economic, walls2019costing}.
The policy question is asked from the perspective of the donor (in this case, USAID):  the policy objective is to achieve the highest benefit-cost ratio per intended beneficiary for each dollar that is spent on a program.  Operating expenditures in the implementation chain are an inherent part of these costs, and so the lower transactions costs in getting mobile money to the beneficiary play an important role in their potential attractiveness.  We conducted two different costing exercises at two moments in time.  The ex-ante exercise, which was based on projected budgets and staffing costs, was used to predict the cost at the time of the study design, and to choose the ranges over which the lower GiveDirectly transfer amounts would be randomized.  Then, a rigorous ex-post costing exercise was conducted for both programs after the fact, using actual budgets and expenditures.

Since the HD program is eventually to cover twenty-three districts (e.g. much larger than the study population only) we attempt to cost the full national program (not just the study sample), inclusive of all direct costs, all indirect in-country management costs including transport, real estate, utilities, and the staffing required to manage the program, and all international operating costs entailed in managing the HD program.  Because we do not want differences in scale to drive differential costs per beneficiary, we asked GiveDirectly to artificially scale up their operations and provide us with numbers reflecting the costs per beneficiary if they were running a national-scale program across eight districts, including 40,000 beneficiary households like HD.  This is the relevant scale for a USAID program officer contemplating commissioning a program to move the outcomes studied.  Beneficiary identification costs, incurred partly by the survey firm and partly by HD, are calculated on a per-head basis and added to the costs of both implementers equally.\footnote{This means that the operating costs for both implementers are slightly higher than they would have been absent the study-driven beneficiary identification costs, but these expenses drop out of the \textit{comparative} costing analysis.} 

We costed each GD arm separately, asking what the operating costs would have been if GD had run a national program at the scale of HD giving only transfers of that amount.  Operating costs as a percentage of the amount transferred decline with transfer amount for GD because fixed costs represent a large share of their total costs. This allows us to conduct the benefit/cost comparisons ‘at scale’, rather than having the artificial, multi-amount environment of the study contaminate the costing exercise across arms.

\subsubsection{Differential Compliance}\label{ss:DiffComply}

Given that the Intention-to-Treat is the heart of the experimental analysis, we construct the `cost per study subject' that corresponds to the spend on the sample over which the ITT is estimated.  The raw costing returns the cost per beneficiary, but less than this is spent per study subject to deliver the ITT if compliance is less than 1.  Both implementers face a relatively simple  relationship between cost and compliance.  For GD, individuals not treated cost nothing.  Similarly for HD, their rules stipulate that they pay sub-IPs a fixed amount based on enrollment at the end of the first of WRN training to then follow through and offer all appropriate subsequent classes in the curriculum.  These costs are almost exclusively based on offering the courses and do not scale sharply with class size.   Hence, we consider all costs as `averted' for non-compliers, and for each arm we calculate the ITT-comparable cost by multiplying the compliance rate times the cost per beneficiary.

\subsubsection{Final Costing Numbers}

Table \ref{t:costing} shows the evolution of the costing analysis.  As described above, the ex-ante costing exercise arrived at a figure of \$464.25 per HD beneficiary, with bracketing costs of \$377.03 and \$571.74.  GD took this number and applied their cost structure to it for a program scaled to 40,000 beneficiaries, and arrived at an ex-ante cost-equivalent transfer of \$410.65 to be actually delivered to beneficiaries.  The bracketing cash arms received \$317 and the upper arm \$503, and the `huge' arm received \$750, the amount that GD believed would maximize benefit-cost.  

Then, based on the ex-post costing exercise, we recalculate USAID costs applying the more accurate costing figures to the sums actually transferred.  These figures show that HD was less expensive than anticipated, and GD operating costs were slightly higher than anticipated.  This means that the effect USAID spend per beneficiary was only \$388.32, while the spending for the GD middle arm was \$493.96.  The inclusion of non-compliance further widens this gap, meaning that the USAID spend per study household in the HD arm was \$332.27, while in the GD arms it was \$394.93, \$490.99, \$590.41, and \$846.71, respectively.  The combined arm, incorporating compliance with both components of the combined treatment, ended up costing USAID \$840.20 per study individual, an amount similar to the GD large arm.  These are the numbers used in the Cost Equivalent table.  In sum, our study ends up with even the smallest of the GD cost-equivalent arms transferring somewhat too much to be directly comparable to HD, but the GD Large arm providing a very close cost counterfactual to the Combined arm.\footnote{The lower final costs arise primarily from two factors.  First, compliance was lower than expected given that we were working with a group who had expressed willingness to participate in HD.  Second, the ex post costing revealed a larger than expected share of costs in the early years of the HD budget being spent on curriculum development and implementer training.  Because these costs are amortized over beneficiaries for the full five years of the program, they pushed down the spend per beneficiary in this early year of the study.}  It is important to remember in looking at our results, then that the GD arms cost more than HD, and only through the linearity assumption in our cost-equivalence comparison can we recover the exact benchmarking amount.  An implication for future work is that using wider brackets may be reasonable given the considerable uncertainty we have uncovered moving from ex-ante to ex-post cost estimates.

\clearpage
\begin{landscape}
\vspace*{\fill}
\begin{table}[!hb]
\caption{Results of Costing Exercise}
\label{t:costing}
\begin{center}
\begin{tabular}{l c c c c c c}
\toprule %
Treatment             Arm:	&	Ex Ante Cost	&	Value received 	&	Ex Post Cost	&	Fraction  operating cost 	&	Compliance Rate 	&	Cost per study household	\\
Huguka Dukore	&	\$464.25	&	\$153.47	&	\$388.32	&	60.5\%	&	85.6\%	&	\$332.27	\\
GD lower	&	\$377.03	&	\$317.16	&	\$394.39	&	19.6\%	&	100\%	&	\$394.39	\\
GD mid	&	\$464.25	&	\$410.65	&	\$493.96	&	16.9\%	&	99.4\%	&	\$490.99	\\
GD upper	&	\$571.74	&	\$502.96	&	\$590.41	&	14.8\%	&	100\%	&	\$590.41	\\
GD large	&	\$828.47	&	\$750.3	&	\$846.71	&	11.3\%	&	100\%	&	\$846.71	\\
Combined	&	\$928.5	&	\$561.11	&	\$885.64	&	36.3\%	&	89.6\%(HD), 100\%(GD)	&	\$840.20	\\

\bottomrule %
\end{tabular}
\end{center}
\vskip-4ex 
\floatfoot{ 
\begin{footnotesize}
Note:  The first column shows the ex-ante costing data on which study was designed; the core number is the HD cost around which the GD actual transfer amounts in column 2 were designed.  Column 3 shows the results of the ex post costing exercise.  Column 4 provides the share of spending that did not reach the beneficiaries either in cash or in direct training and materials costs.  Column 5 shows the compliance rates, and since all costs are averted for non-compliers then the final column shows the final cost per study subject for each arm that are the basis of the cost-equivalent comparisons.
\end{footnotesize}
}
\end{table}
\vspace*{\fill}

\end{landscape}

\section{Results}\label{s:results}

\subsection{Overall ITT Impacts}\label{ss:ITT}

The data from the study are analyzed consistent with the design being a multi-arm, household-randomized program.  Let the subscript $i$ indicate the individual, $h$ the household, and $b$ the randomization block (lottery groups within which the randomization was conducted).  For outcomes observed both at baseline $(Y_{ihb0})$ and at endline $(Y_{ihb1})$, we conduct ANCOVA analysis including the baseline outcome, fixed effects for the sector-level  assignment blocks within which the randomization was conducted $\mu_b$, as well as a set of baseline control variables selected from the baseline data on the basis of their ability to predict the primary outcomes, denoted by  $X_{ihb0}$.  
Base regressions to estimate the Intention to Treat Effect include indicators for the HD treatment $T_{ihb}^{HD}$, a vector of indicators for \emph{each} of the three GD `small' treatment values, $T^{GDS1}_{ihb}$, $T^{GDS2}_{ihb}$, and $T^{GDS3}_{ihb}$, an indicator for the GD `large' treatment $T^{GDL}_{ihb}$, and an indicator for the combined arm $T_{ihb}^{COMB}$:
\begin{eqnarray}
Y_{ihb1} &= & \delta^{HD} T_{ihb}^{HD}  + \delta^{GDS1} T_{ihb}^{GDS1} + 
    \delta^{GDS2} T_{ihb}^{GDS2}+ \delta^{GDS3} T_{ihb}^{GDS3}  \nonumber\\  %
    & & + \delta^{GDL} T_{ihb}^{GDL} + \delta^{COMB} T_{ihb}^{COMB} + \beta X_{ihb0} + \rho Y_{ihb0} + \mu_b + \epsilon_{ihb1} \label{eq:PrimarySpec}
\end{eqnarray}
Block-level fixed effects,$\mu_b$, are included to account for the block randomization of the study.  Standard errors will be clustered at the household level because the second tranche of treatment was assigned at the household level.  Following the `post-double-LASSO' procedure of \citet{BelCheHan14restud}, a set of covariates were selected using a LASSO algorithm on the control data; further details of this procedure are provided in Appendix \ref{app:ModelSelection}.  
For outcomes that are collected at endline only, we cannot include the lagged outcome to run the ANCOVA regression, and so use the simple cross-sectional analog to Equation \eqref{eq:PrimarySpec}.

    \begin{sidewaystable}[!hp]
    \caption{ITT estimates, primary outcomes, separating GD transfer values}
    \label{t:itt_primary_full}
    \begin{footnotesize}
    \begin{tabular}{l *{6}{S} ScSSSS}
\toprule
 & & \multicolumn{4}{c}{GiveDirectly} & & \multicolumn{1}{c}{\raisebox{-1ex}[0pt]{Control}} & & & \multicolumn{3}{c}{$ p$-values} \\ 
\cmidrule(lr){3-6} \cmidrule(lr){11-13}
\multicolumn{1}{c}{\text{ }} & \multicolumn{1}{c}{\text{HD}} & \multicolumn{1}{c}{\text{Lower}} & \multicolumn{1}{c}{\text{Middle}} & \multicolumn{1}{c}{\text{Upper}} & \multicolumn{1}{c}{\text{Large}} & \multicolumn{1}{c}{\text{Combined}} & \multicolumn{1}{c}{\text{Mean}} & \multicolumn{1}{c}{\text{Obs.}} & \multicolumn{1}{c}{\text{$ R^2$}} & \multicolumn{1}{c}{\text{(a)}} & \multicolumn{1}{c}{\text{(b)}} & \multicolumn{1}{c}{\text{(c)}}\\
\midrule
\multirow[t]{ 3 }{0.15\textwidth}{Employed }  &      0.02 &      0.03 &      0.05 &      0.00 &      0.01 &      0.01 &      0.48 &      1770 &      0.16 &      0.95 &      0.57 &      0.94 \\ 
 & (0.03)  & (0.05)  & (0.05)  & (0.05)  & (0.05)  & (0.04)  \\  & [    0.30]  & [    0.30]  & [    0.16]  & [    0.50]  & [    0.46]  & [    0.49]  \\ \addlinespace[1ex] 
\multirow[t]{ 3 }{0.15\textwidth}{Productive hours }  &      2.79\ensuremath{^{*}} &      2.76 &      6.54\ensuremath{^{***}} &      3.56 &      1.12 &      2.31 &     18.64 &      1770 &      0.19 &      0.82 &      0.33 &      0.63 \\ 
 & (1.57)  & (2.34)  & (2.40)  & (2.52)  & (2.06)  & (2.03)  \\  & [    0.07]  & [    0.16]  & [    0.01]  & [    0.12]  & [    0.33]  & [    0.16]  \\ \addlinespace[1ex] 
\multirow[t]{ 3 }{0.15\textwidth}{Monthly income }  &      0.31 &      0.76\ensuremath{^{**}} &      1.08\ensuremath{^{***}} &      1.14\ensuremath{^{***}} &      0.73\ensuremath{^{**}} &      1.04\ensuremath{^{***}} &      8.05 &      1770 &      0.21 &      0.31 &      0.24 &      0.41 \\ 
 & (0.26)  & (0.36)  & (0.34)  & (0.35)  & (0.35)  & (0.32)  \\  & [    0.16]  & [    0.04]  & [    0.00]  & [    0.00]  & [    0.04]  & [    0.00]  \\ \addlinespace[1ex] 
\multirow[t]{ 3 }{0.15\textwidth}{Productive assets }  &      1.54\ensuremath{^{***}} &      3.94\ensuremath{^{***}} &      3.80\ensuremath{^{***}} &      3.84\ensuremath{^{***}} &      4.02\ensuremath{^{***}} &      4.42\ensuremath{^{***}} &      5.61 &      1770 &      0.26 &      0.00 &      0.00 &      0.42 \\ 
 & (0.35)  & (0.46)  & (0.50)  & (0.46)  & (0.47)  & (0.44)  \\  & [    0.00]  & [    0.00]  & [    0.00]  & [    0.00]  & [    0.00]  & [    0.00]  \\ \addlinespace[1ex] 
\multirow[t]{ 3 }{0.15\textwidth}{HH consumption per capita }  &      0.05 &      0.20\ensuremath{^{**}} &      0.27\ensuremath{^{***}} &      0.23\ensuremath{^{***}} &      0.36\ensuremath{^{***}} &      0.27\ensuremath{^{***}} &      9.46 &      1737 &      0.33 &      0.12 &      0.67 &      0.31 \\ 
 & (0.06)  & (0.08)  & (0.09)  & (0.07)  & (0.07)  & (0.07)  \\  & [    0.21]  & [    0.01]  & [    0.00]  & [    0.00]  & [    0.00]  & [    0.00]  \\ \addlinespace[1ex] 
\bottomrule
\end{tabular}

    \floatfoot{
    Note:  The six columns of the table provide the estimate on dummy variables for each of the treatment arms, compared to the control group.  The five primary outcomes are in rows.  Regressions include but do not report the lagged dependent variable, fixed effects for randomization blocks, and a set of LASSO-selected baseline covariates, and are weighted to reflect intensive tracking. Standard errors are (in soft brackets) are clustered at the household level to reflect the design effect, and p-values corrected for False Discovery Rates across all the outcomes in the table are presented in hard brackets.  Stars on coefficient estimates are derived from the FDR-corrected $p$-values, *=10\%, **=5\%, and ***=1\% significance. Reported $p$-values in final three columns derived from $F$-tests of hypotheses that cost-benefit ratios are equal between: (a) GD Lower and HD; (b) GD Lower and GD Large; and (c) GD Large and Combined treatments. Employed is a dummy variable for spending more than 10 hours per week working for a wage or as primary operator of a microenterprise.  Productive hours are measured over prior 7 days in all activities other than own-farm agriculture.  Monthly income, productive assets, and household consumption are winsorized at 1\% and 99\% and analyzed in Inverse Hyperbolic Sine, meaning that treatment effects can be interpreted as percent changes.
    }
    \end{footnotesize}
    \end{sidewaystable}


    \begin{sidewaystable}[!hp]
    \caption{ITT estimates, secondary outcomes, separating GD transfer values}
    \label{t:itt_secondary_full}
    \begin{footnotesize}
    \vskip-2ex
    \begin{tabular}{l *{6}{S} ScSSSS}
\toprule
 & & \multicolumn{4}{c}{GiveDirectly} & & \multicolumn{1}{c}{\raisebox{-1ex}[0pt]{Control}} & & & \multicolumn{3}{c}{$ p$-values} \\ 
\cmidrule(lr){3-6} \cmidrule(lr){11-13}
\multicolumn{1}{c}{\text{ }} & \multicolumn{1}{c}{\text{HD}} & \multicolumn{1}{c}{\text{Lower}} & \multicolumn{1}{c}{\text{Middle}} & \multicolumn{1}{c}{\text{Upper}} & \multicolumn{1}{c}{\text{Large}} & \multicolumn{1}{c}{\text{Combined}} & \multicolumn{1}{c}{\text{Mean}} & \multicolumn{1}{c}{\text{Obs.}} & \multicolumn{1}{c}{\text{$ R^2$}} & \multicolumn{1}{c}{\text{(a)}} & \multicolumn{1}{c}{\text{(b)}} & \multicolumn{1}{c}{\text{(c)}}\\
\midrule
\multicolumn{13}{l}{\emph{Panel A. Beneficiary welfare}}  \\ 
\addlinespace[0.5ex] \multirow[t]{ 3 }{0.15\textwidth}{Subjective well-being }  &      0.19\ensuremath{^{***}} &      0.40\ensuremath{^{***}} &      0.53\ensuremath{^{***}} &      0.48\ensuremath{^{***}} &      0.55\ensuremath{^{***}} &      0.41\ensuremath{^{***}} &      0.00 &      1770 &      0.13 &      0.08 &      0.12 &      0.17 \\ 
 & (0.07)  & (0.09)  & (0.10)  & (0.09)  & (0.09)  & (0.09)  \\  & [    0.00]  & [    0.00]  & [    0.00]  & [    0.00]  & [    0.00]  & [    0.00]  \\ \addlinespace[0.5ex] 
\multirow[t]{ 3 }{0.15\textwidth}{Mental health }  &     -0.04 &     -0.06 &      0.07 &      0.03 &      0.11 &      0.12 &      0.00 &      1770 &      0.07 &      0.89 &      0.22 &      0.93 \\ 
 & (0.07)  & (0.09)  & (0.09)  & (0.09)  & (0.10)  & (0.09)  \\  & [    0.30]  & [    0.27]  & [    0.27]  & [    0.37]  & [    0.16]  & [    0.12]  \\ \addlinespace[0.5ex] 
\multirow[t]{ 3 }{0.15\textwidth}{Beneficiary-specific consumption }  &      0.15 &      0.51\ensuremath{^{***}} &      0.61\ensuremath{^{***}} &      0.62\ensuremath{^{***}} &      0.45\ensuremath{^{***}} &      0.69\ensuremath{^{***}} &      8.27 &      1770 &      0.23 &      0.01 &      0.01 &      0.11 \\ 
 & (0.12)  & (0.12)  & (0.13)  & (0.12)  & (0.15)  & (0.12)  \\  & [    0.12]  & [    0.00]  & [    0.00]  & [    0.00]  & [    0.00]  & [    0.00]  \\ \addlinespace[0.5ex] 
\multicolumn{13}{l}{\emph{Panel B. Household wealth}}  \\ 
\addlinespace[0.5ex] \multirow[t]{ 3 }{0.15\textwidth}{HH net non-land wealth }  &     -0.18 &      0.16 &      1.20\ensuremath{^{***}} &      1.33\ensuremath{^{***}} &      1.11\ensuremath{^{***}} &      0.89\ensuremath{^{**}} &     11.28 &      1770 &      0.21 &      0.56 &      0.53 &      0.67 \\ 
 & (0.40)  & (0.59)  & (0.44)  & (0.42)  & (0.41)  & (0.48)  \\  & [    0.36]  & [    0.42]  & [    0.01]  & [    0.00]  & [    0.01]  & [    0.05]  \\ \addlinespace[0.5ex] 
\multirow[t]{ 3 }{0.15\textwidth}{HH livestock wealth }  &     -0.01 &      1.76\ensuremath{^{***}} &      1.84\ensuremath{^{***}} &      2.64\ensuremath{^{***}} &      2.17\ensuremath{^{***}} &      2.21\ensuremath{^{***}} &      7.81 &      1770 &      0.25 &      0.00 &      0.12 &      0.92 \\ 
 & (0.37)  & (0.49)  & (0.52)  & (0.45)  & (0.47)  & (0.45)  \\  & [    0.42]  & [    0.00]  & [    0.00]  & [    0.00]  & [    0.00]  & [    0.00]  \\ \addlinespace[0.5ex] 
\multirow[t]{ 3 }{0.15\textwidth}{Savings }  &      1.03\ensuremath{^{***}} &      1.04\ensuremath{^{***}} &      1.29\ensuremath{^{***}} &      1.56\ensuremath{^{***}} &      1.43\ensuremath{^{***}} &      1.69\ensuremath{^{***}} &      9.24 &      1770 &      0.20 &      0.56 &      0.22 &      0.39 \\ 
 & (0.23)  & (0.32)  & (0.34)  & (0.30)  & (0.31)  & (0.27)  \\  & [    0.00]  & [    0.00]  & [    0.00]  & [    0.00]  & [    0.00]  & [    0.00]  \\ \addlinespace[0.5ex] 
\multirow[t]{ 3 }{0.15\textwidth}{Debt }  &      0.41 &     -0.10 &     -0.23 &     -0.56 &     -0.37 &      0.00 &      8.75 &      1770 &      0.20 &      0.19 &      0.86 &      0.45 \\ 
 & (0.28)  & (0.41)  & (0.43)  & (0.45)  & (0.42)  & (0.38)  \\  & [    0.10]  & [    0.42]  & [    0.34]  & [    0.13]  & [    0.24]  & [    0.42]  \\ \addlinespace[0.5ex] 
\multicolumn{13}{l}{\emph{Panel C. Beneficiary cognitive and non-cognitive skills}}  \\ 
\addlinespace[0.5ex] \multirow[t]{ 3 }{0.15\textwidth}{Locus of control }  &      0.06 &      0.13 &      0.02 &      0.00 &      0.08 &      0.23\ensuremath{^{**}} &      0.00 &      1770 &      0.28 &      0.59 &      0.30 &      0.12 \\ 
 & (0.06)  & (0.08)  & (0.08)  & (0.08)  & (0.08)  & (0.08)  \\  & [    0.65]  & [    0.38]  & [    1.00]  & [    1.00]  & [    0.65]  & [    0.03]  \\ \addlinespace[0.5ex] 
\multirow[t]{ 3 }{0.15\textwidth}{Aspirations }  &     -0.01 &      0.08 &     -0.05 &      0.13 &      0.03 &      0.14 &      0.00 &      1770 &      0.08 &      0.41 &      0.48 &      0.26 \\ 
 & (0.07)  & (0.09)  & (0.09)  & (0.09)  & (0.09)  & (0.08)  \\  & [    1.00]  & [    0.69]  & [    1.00]  & [    0.38]  & [    1.00]  & [    0.33]  \\ \addlinespace[0.5ex] 
\multirow[t]{ 3 }{0.15\textwidth}{Big Five index }  &      0.12 &      0.08 &      0.11 &      0.02 &     -0.08 &      0.02 &      0.00 &      1770 &      0.10 &      0.55 &      0.25 &      0.35 \\ 
 & (0.07)  & (0.10)  & (0.09)  & (0.09)  & (0.09)  & (0.09)  \\  & [    0.33]  & [    0.72]  & [    0.62]  & [    1.00]  & [    0.69]  & [    1.00]  \\ \addlinespace[0.5ex] 
\multirow[t]{ 3 }{0.15\textwidth}{Business knowledge }  &      0.65\ensuremath{^{***}} &      0.09 &      0.08 &      0.06 &     -0.03 &      0.63\ensuremath{^{***}} &      0.00 &      1770 &      0.23 &      0.00 &      0.29 &      0.00 \\ 
 & (0.07)  & (0.09)  & (0.09)  & (0.09)  & (0.09)  & (0.09)  \\  & [    0.00]  & [    0.69]  & [    0.69]  & [    0.83]  & [    1.00]  & [    0.00]  \\ \addlinespace[0.5ex] 
\multirow[t]{ 3 }{0.15\textwidth}{Business attitudes }  &      0.12 &      0.19 &      0.19 &      0.10 &      0.06 &      0.15 &      0.00 &      1770 &      0.09 &      0.63 &      0.08 &      0.40 \\ 
 & (0.07)  & (0.10)  & (0.09)  & (0.09)  & (0.09)  & (0.09)  \\  & [    0.33]  & [    0.31]  & [    0.29]  & [    0.65]  & [    0.83]  & [    0.33]  \\ \addlinespace[0.5ex] 
\bottomrule
\end{tabular}

    \vskip-2ex
    \floatfoot{
    Notes:   Regressions include but do not report the lagged dependent variable, fixed effects for randomization blocks, and a set of LASSO-selected baseline covariates, and are weighted to reflect intensive tracking.  Standard errors are (in soft brackets) are clustered at the household level to reflect the design effect, and $p$-values corrected for False Discovery Rates across all the outcomes in the table are presented in hard brackets.  Stars on coefficient estimates are derived from the FDR-corrected $p$-values, *=10\%, **=5\%, and ***=1\% significance.  Reported $p$-values in final three columns derived from $F$-tests of hypotheses that cost-benefit ratios are equal between: (a) GD Lower and HD; (b) GD Lower and GD Large; and (c) GD Large and Combined treatments
    }
    \end{footnotesize}

    \end{sidewaystable}



To mitigate risks of false discovery across multiple outcomes and treatments, we report \citeauthor{anderson2008multiple}'s \citeyear{anderson2008multiple} False Discovery Rate to adjust p-values within each of the four relevant families (primary outcomes and the three families of secondary outcomes outlined in Section \ref{sss:secondary}), ensuring that the false discovery rate at the family level is controlled at five percent.  This follows the procedures described in our Pre-Analaysis Plan (PAP).

Tables \ref{t:itt_primary_full} and \ref{t:itt_secondary_full} present the results of this analysis, with the five primary outcomes in the rows of the table, and in the columns we include the core treatment arms of the study:  HD, each of the three smaller GD arms that were designed to be cost equivalent, the large cash transfer arm (GD Large), and the arm that receives both the medium GD cash transfer amount and HD (Combined).   Appendix Tables \ref{t:itt_primary_pooled} and \ref{t:itt_secondary_pooled}  present the more parsimonious specification that pools the three smaller GD transfer amounts into one arm.

Because each treatment is measured with a dummy variable the outcomes here should be interpreted as differences relative to the control group which received no intervention.  For each point estimate we present both the unadjusted standard error (in soft brackets) as well as the False Discovery Rate adjusted q-value [in hard brackets].  The stars on the coefficients are based on the adjusted q-values, with one, two, and three stars indicating significance at 10\%, 5\%, and 1\% respectively.

Beginning with Table \ref{t:itt_primary_full}, we see that none of the programs were successful in driving the core outcome of employment rates.  Relative to a control group employment rate of 48\%, neither HD, cash, nor the combination saw improvements of more than 5 percentage points.  Given that the minimum detectable effect of our study (which can be calculated by multiplying the standard error times 1.96) was between 6 and 10 pp, these modest improvements are not close to being significant.  Moving to the continuous measure of the number of productive hours worked in a week, we see more promising impacts.  Here, HD was successful in driving a 3 hour increase off a base of 18.4, an improvement of 16\%.  Over the 18 months since baseline the trend in the control group shows substantial increases in employment rates (from 33 to 48\%) and productive hours (from 11 to 19 hours per week), and treatment effects relative to this large secular improvement are modest.\footnote{This pattern is suggestive of the `Ashenfelter's dip' the literature has long understood to exist in the evaluation of job training programs, given that those interested in entering them have systematically experienced negative wage outcomes \citep{ashenfelter1985using}.  This makes RCTs of such programs particularly valuable.}

The GD transfers appear to have had a non-linear effect on productive time use; the middle transfer has a strong effect (an increase of 6.5 hours per week) but both smaller and larger transfers are insignificant.    As the transfer amount increases the productive labor increase disappears, with the Large arm achieving an insignificant 1.1 hour improvement.  This is the first evidence of an apparent non-monotonicity in the impact of cash, suggesting that moderate transfers induce a complementary increase in labor inputs, but once transfers become sufficiently large they begin to shift the opportunity cost of leisure, discouraging labor inputs. The Combined arm, far from showing signs of a complementarity between the two programs, suggests that when individuals receive both cash and training there is no corresponding improvement in labor inputs, while had they received either alone this benefit would have been observed.  In the following subsection we provide a more detailed analysis of the employment impacts of the different programs.

Next we consider the primary household outcomes of monthly income, productive assets, and consumption per capita.  All of these outcomes are measured as the inverse hyperbolic sine, meaning that impacts can be interpreted as percent changes (like a log, but capable of handling zeroes).\footnote{Note that none of the recall windows include the period in which the transfers were received, and so our income measures are not mechanically picking up the receipt of the transfers themselves.}  Beginning with the impact of HD, we see a 31\% increase in income (not significant), no meaningful change in consumption, but an impressive and highly significant surge in productive assets; these rise to be 151\% higher than the control group mean.  For a program that has made no material transfers to the beneficiaries, this improvement is substantial and impressive.  When we look at the impact of cash transfers, we replicate the results of many other studies, showing that cash can lead to substantial improvements in these indicators.  Measuring outcomes roughly 14 months after the transfers were made, we find the GD treatments leading to a doubling of monthly income, again non-monotonic with intermediate transfers appearing to be most effective.  There is a quadrupling of productive asset values, and household consumption increases by 20-36 percent.  For the Large arm (whose value received is 80\% larger than the GD Middle arm) we see outcomes that are in general very similar to the smaller arms; slightly smaller impacts on income, essentially identical impacts on productive asset values, and slightly larger effects on consumption.  So again, these results are consistent with the idea that the Large arm induces a smaller productive effect than the Main arm leading to less earned income (see penultimate column of Table \ref{t:itt_primary_full} for significance tests).  Given this, the superior consumption for the Large arm appears to be being permitted by the spending down of transferred resources.  In all cases the Combined arm has impacts that look relatively similar to the Middle arm (despite having spent almost twice as much), suggesting a lack of complementarity.

For the GD arms where compliance is 100\%, the intention to treat measures effect of actually receiving the program, namely, the treatment on the treated (ToT).  For the HD arm where the loosest measure of compliance is 85.6\%, if we are willing to assume that those not participating received no indirect effect of being included in the treatment, then we can back out the ToT by dividing by the compliance rate.  The resulting ToT estimate would 17\% larger than the ITT for each variable, with the same significance level.

Moving then to the secondary outcomes of the study, in Table  \ref{t:itt_secondary_full} we present three families of outcomes:  beneficiary welfare, household wealth, and cognitive and non-cognitive skill development.  
Beginning with beneficiary welfare, we have two ways of measuring mental health.  The first of these is a composite of the answers to two simple Likert-scaled questions about subjective well-being, one on happiness and one on life satisfaction.  For this outcome, we see every arm improving subjective well-being, with effects for any arm receiving cash that are more than twice as large as the HD arm.  Our second measure of mental health is built around reporting on a set of potential mental health issues over the past two weeks, including stress, ability to concentrate, losing sleep, confidence, and feelings of worthlessness.  Interestingly, none of the interventions improve this measure, suggesting that we are seeing improvements in general life satisfaction but no decrease in specific negative symptoms of poor mental health.  Our final measure of beneficiary welfare is the inverse hyperbolic sine of the consumption of the specific beneficiary (as opposed to the primary outcome of household consumption).  Here we see no improvements under HD, and increases of approximately 50\% over the control outcome for any arm that receives cash (again larger for the middle-sized transfers).  

Next we examine measures of household wealth.  Perhaps unsurprisingly, cash transfers move these outcomes powerfully.  Once again we see an impact that appears to be relatively homogeneous and somewhat non-monotonic in all arms that receive cash transfers, with net non-land wealth more than doubling in all arms but the smallest, livestock wealth tripling, and savings doubling.  In Figure \ref{f:saving_CDF} we plot the Cumulative Density Functions of savings across the different treatments.  All the treatment distributions first-order dominate the control, implying treatment effects for savers of all levels.  While coefficients on debt are negative for all cash arms, there is no evidence of significant pay-down of debt arising from cash transfers over this time.  HD exhibits interesting impacts, with no change in the core measures of household wealth, but an increase both in savings (103\% increase, significant at the 1\% level) and debt (41\% increase, insignificant).  

Finally, we examine cognitive and non-cognitive skills.  Using the standard Locus of Control index we find our first evidence of complementarities, in that only the Combined arm moves this outcome.  None of the arms influence measures of Aspirations or the Big 5 index.\footnote{The original aspirations survey which we borrowed from \citet{bernard2014future} used the gap between desired future income/wealth standing and the current standing to measure aspiration; we found that our interventions had substantial positive impacts on the current economic standing at the time of the survey and little effect on desired future standing.  This showed up as a negative treatment impact on aspirations; we therefore have deviated from our PAP and present the aspirations results using only the desired future economic standing.}  Our measure of business knowledge is a score on a set of questions built to reflect the HD curriculum, and hence serves as a manipulation check on their beneficiaries having learned what was intended.  The results show that any arm who received HD (HD or Combined) has about a half of a standard deviation improvement in their performance on this test, a very sizeable effect. GD does not move this score at all.  The business attitudes measure we use captures attitudes toward entrepreneurship; while none of the interventions significantly improve it they all lead to an increase of about 0.1 SD, and prior to the correction for false discovery the smaller GD arms (which saw a surge in productive hours) had an effect on this measure significant at the 10\% level.  The improvement in business knowledge for HD participants does not translate into an increase in entrepreneurship.

In Tables \ref{t:itt_primary_pooled} and \ref{t:itt_secondary_pooled} we present a more parsimonious specification that pools the three smaller GD arms into one; this analysis was pre-specified due to a concern about a potential lack of power when the smaller GD arms are analyzed separately.  In the event we have many significant results for the disaggregated analysis, indicating that power is not a concern, and so the specification pooling these arms simply provides an overview of the average effects across these smaller arms.  

\subsubsection{Detailed Analysis of Employment Effects}

Given the central importance of employment in this study, we now seek to dig deeper into the differential impact of the programs.  More nuance can be provided for these results in a number of dimensions.   First, we can examine how the treatments shift the full distribution of time use by plotting the densities of productive time use across arms.  Figure \ref{f:prod_hours_CDF} shows a modest rightward shift in the CDFs (implying an increase in hours worked) across most of the distribution.  None of the interventions moved hours for those who would have worked least, the Combined arm is most effective around the middle of the distribution, and the HD Main arm is most effective for those working the most hours.  This speaks to the enabling effect of additional capital being particularly important for those already heavily engaged in productive work.

In Appendix Table \ref{t:employment_details} we present a more detailed analysis using the five disaggregated categories in which we asked the underlying time use questions.  These report the hours beneficiaries worked in the previous week in  agricultural wage labor, non-agricultural wage labor, non-agricultural enterprises, agricultural self-employment, or agricultural processing and trading.  Interestingly, the HD treatment induces an increase in non-agricultural wage labor of an estimated 6 percentage points (against a control-arm counterfactual of 30 percent), as well as a modest increase in agriculture-related self-employment, while the impacts of cash transfers are concentrated in non-agricultural enterprise and self-employment labor, at the expense of agricultural (and, at high transfer values, other forms of) wage labor.

Next, we delve into the source of the dissonance between our binary measure of employment, which shows no impact, and the continuous measure of productive hours, which does.  The obvious suspect here is the specific threshold used to define `employed'.  To examine this, we vary this threshold continuously from 5 hours per week to 40 hours per week, and present a visualization of employment rates across thresholds and across treatment arms.  We use the disaggregated labor categories analyzed in Table \ref{t:employment_details}, which are presented in the Appendix in decreasing order of their importance to total employment.  First, Figure \ref{f:hours_noagric} presents effects on non-agricultural wage labor (the category with the most overall time use). In general, the treatments effects represent similar vertical shifts across all the employment thresholds, implying that the binary estimate of employment effects would not be sensitive to threshold.  HD exerts a positive effect, and GD Large a negative effect, on wage employment.  Non-agricultural self employment, as shown in Figure \ref{f:hours_enterp}, responds little to HD, and to cash amounts in a relatively monotonic way (bigger transfers mean bigger impacts).  Agricultural wage employment, which is both arduous and low-paid, appears to demonstrate an income effect in being depressed by any kind of cash transfer (Figure \ref{f:hours_employed_farm}).\footnote{Using our own data to divide the amount earned by the hours worked for each of the five labor sectors, we find agricultural wage labor having by far the lowest pay, 10 cents per hour.  The pay in non-agricultural wage labor is almost three times as high (26 cents/hour).  The hourly wage rates we record in self-employment are even higher (36 cents for ag self-employment, 56 for non-ag self-employment, and 71 cents for ag processing), but these should be interpreted with some caution due to the complexity of telling net from gross income in self employment.}  Agricultural self-employment is increased by all the treatments, and it is here that the non-monotonicity in transfer amounts is strongest, with GD Main seeing a larger effect than GD Large (Figure \ref{f:hours_semploy}).  Finally, agricultural processing and trading appears to be a sector that HD encourages individuals to \textit{leave}, helping to explain smaller total productive hours effects despite the surge seen in non-ag wage employment and agricultural self-employment (Figure
\ref{f:hours_agroprocess}).  Given the general invariance of the impacts to the thresholds used, this suggests that the dissonance arises from the fact that the treatment effects on productive hours are occurring among those who would anyways have been counted as employed.

Finally, we can exploit a different source of data, which is a series of phone calls made by IPA to all of the study participants over the period from November 2017 (around the end of the HD intervention) to May 2019 (a few months before the midline).  The core purpose of these phone calls was to acquire tracking information for the in-person survey, but we block randomized the month in which we called each individual, and during the call asked the same time use questions that would be included in the midline.  This survey was successful in reaching 1,797 of our study subjects (97\%), and given the randomization by month provides a clean way of estimating the time path of treatment effects.  In Figure \ref{f:phone_survey_productive} we show coefficient plots of ITT regressions run separately within each month's sample of surveys.  The HD Main treatment has a consistent positive effect on time use across every monthly sub-sample, while the coefficients on the other treatments are more unstable across time.  A core purpose of this phone survey was to see what happened to employment as the apprenticeship phase of HD came to an end.  In Figure \ref{f:phone_survey_apprenticeship} we can indeed see the elevated rate of apprenticeships for HD participants as program participation came to an end (late 2018), which then contextualizes the uptick in productive hours visible in the prior table for HD beneficiaries in the early months of 2019. 

Summarizing this analysis, we see reasonable labor market effects of the HD program in a number of dimensions.  We would not expect productive time use to improve \textit{during} the intervention, and our midline survey appears to come about six months after the labor market impacts emerge.  HD focuses on job training and on helping beneficiaries start agricultural enterprises, and it is successful in both these endeavors.  Because most of the individuals entering this work are leaving agricultural wage employment or agricultural trading (which we also count as employment), the headline employment rate does not shift substantially.  Cash transfers, on the other hand, particularly strongly drive individuals away from agricultural wage work and into self-employment (both agricultural and non-agricultural).  So these two interventions have the effect of pushing participants down somewhat different paths towards income generation.

\subsection{Cost-Equivalent Benchmarking}\label{ss:CE}

We pre-specified a regression adjustment strategy for benchmarking HD at an exactly cost-equivalent level using the ex-post costing data from both programs.   First, begin with the total GD donor cost per study subject within each transfer amount arm, denoted by $t_c$ (this is the final column in Table \ref{t:costing}).  Subtract from this number the benchmarked HD cost per household $C$ from the same column, and denote the difference $t_c-C=\tau_c$; this is the deviation (positive or negative) of each GD arm from the benchmarked HD cost.  Set $\tau_c$ to zero in the control and HD arms.  We can then re-run regression \eqref{eq:PrimarySpec} above omitting the combined arm, and controlling for a linear term in $\tau_c$, a dummy for either treatment, and a dummy for receiving HD:
\begin{equation}\label{eq:CostEquiv}
Y_{ihb1}= \delta^T T_{ihb} + \delta^{HD} T_{ibh}^{HD} + \beta X_{ihb0} + \rho Y_{ihb0} + \gamma_1 \tau_c + \mu_b + + \epsilon_{ihb1} 
\end{equation}

In this specification $T_{ihb}$ is a dummy variable indicating that individual $i$ in household $h$ of randomization block $b$ was assigned to any treatment (HD or GD).  Subject to the assumption of linear transfer amount effects, the slope coefficient $\tau_c$ captures impacts arising from deviations in GD cost from HD cost, the coefficient $\delta$ effectively gives the impact of GD at the cost of HD, and the dummy variable $\delta^{HD}$  provides a direct benchmarking test:  the differential impact of HD over GD at the same cost per eligible.  We impose the simple linear functional form to preserve as much statistical power as possible for the core cost-equivalent benchmarking comparison, although it is straightforward to make this more flexible.

 Table \ref{t:cost_equiv_primary} shows the cost-equivalent benchmarking results for primary outcomes, and Table \ref{t:cost_equiv_secondary} for secondaries.  Beginning with the third column we linearize across transfer arms and test the marginal effect of receiving an additional \$100 from GD. Remarkably, across all 17 outcomes we do not have a single case where we can detect a significant slope across transfer amounts; indeed more than half of the transfer slope coefficients are negative.\footnote{This results stands in stark contrast to our previous benchmarking study \citep{mcintzeit2019gikuriro}, in which the GD Main cash transfer amounts were much smaller (\$84 on average) and a sharp differentiation was visible between these small transfers and the Large arm (\$532).}  The takeaway seems to be that this study features transfer amounts large enough to have cleared an impact threshold beyond which additional amounts of money do not lead to meaningfully better outcomes.  For the purposes of cost adjustment, this result also indicates that the pre-specified way of using a linear transfer amount control to adjust the GD impacts will not have a strong effect on the predicted cost-equivalent outcome.

The second column, \emph{Cost-equivalent GD impact}, reports the coefficient on an indicator for receiving any treatment.  This estimates the effect of GD at the cost-equivalent level; here we see as significant the outcomes already shown to be moved by GD (hours worked, income, productive assets, consumption, subjective well-being, livestock wealth, and savings).  In this pooled specification and with regression adjustment we now get a small effect of cost-equivalent cash on business attitudes, significant at the 10\% level.  

The first column, \emph{Differential impact of HD}, presents the core of the comparative cost effectiveness exercise that this study was built to conduct.  Using the linearized cost adjustment it gives the differential impact of HD relative to a cash transfer of precisely equal amount.  The results indicate that HD leads to somewhat lower monthly income and substantially fewer productive assets than a cost-equivalent cash transfer.  In terms of secondary outcomes, HD also does significantly worse at driving subjective well-being, beneficiary consumption, and livestock wealth, and of all the outcomes in the study only for the Business Knowledge index does HD do significantly better than cash.

Appendix Figure \ref{f:CE_secondary} provides graphical interpretation for our cost equivalent comparison, using subjective well-being and business knowledge as outcomes.  In this figure the grey circle plots the Control in outcome/cost space, the four black circles represent the GD arms, and the black diamond gives the actual HD outcome.  The cost adjustment strategy first fits a linear regression of the outcome by transfer amount in the GD arms (solid black line), then extrapolates this line to the predicted value that would have obtained had the GD transfers exactly equalled the ex-post HD cost (hollow circle).  Mapping the picture to the regression estimands, then, the third column gives the slope of the line, the second column gives the vertical differential between the predicted hollow circle and the control, and the core cost equivalent test measures the vertical difference between the hollow circle and the black diamond.  This last quantity is the core purpose of our comparative exercise, the difference between what we actually observed in HD and what we predict we would have observed in GD at the exact same cost.  For subjective well-being, despite an improvement in HD over the control we see the predicted cash outcome being more than twice as a far above the control.  For business knowledge, it is clear that no amount of money generates an effect resembling HD.  Similar figures for all primary outcomes, along with a side-by-side comparison to a more standard cost-effectiveness approach, are included in Figure \ref{f:CE_CEff_Comparison} and discussed in Section \ref{s:value_for_money}.

These results illustrate the value of the double counterfactual created by the cash benchmarking design.  Compared to what would have happened in the absence of the program HD is successful, leading to meaningful improvements in a number of the core outcomes it was designed to move.  Considering the cost of the program, however, and comparing to what would have happened if this cost had been distributed directly to beneficiaries, the picture is less rosy.  Given that the direct distribution of these costs would have led to a surge in consumption and investment, the hurdle for success is raised, and we find the HD program falling short across most outcomes.  Unless policymakers had a strong preference for the specific human capital built by HD and measured in our Business knowledge index, over the 18 month time horizon the benefits of cash would dominate.  It is nonetheless impressive that HD managed to generate meaningful improvements in productive assets, savings, and subjective well-being without having made any material transfers to beneficiaries.

\begin{table}[!hbtp]
\caption{Cost-equivalent analysis:  Primary outcomes}
\label{t:cost_equiv_primary}
\begin{footnotesize}
\begin{center}
\begin{tabular}{l *{3}{S} ScS}
\toprule
\multicolumn{1}{c}{\text{ }} & \multicolumn{1}{c}{\text{\makecell[b]{Differential impact \\ of HD}}} & \multicolumn{1}{c}{\text{\makecell[b]{Cost-equivalent\\GD impact}}} & \multicolumn{1}{c}{\text{\makecell[b]{Transfer\\Value}}} & \multicolumn{1}{c}{\text{\makecell[b]{Control\\Mean}}} & \multicolumn{1}{c}{\text{Obs.}} & \multicolumn{1}{c}{\text{$ R^2$}}\\
\midrule
\multirow[t]{ 3 }{0.2\textwidth}{Employed }  &     -0.01 &      0.04 &     -0.00 &      0.48 &      1578 &      0.17 \\ 
 & (0.04)  & (0.04)  & (0.01)  \\  & [    0.60]  & [    0.33]  & [    0.60]  \\ \addlinespace[1ex] 
\multirow[t]{ 3 }{0.2\textwidth}{Productive hours }  &     -2.25 &      5.05\ensuremath{^{**}} &     -0.66 &     18.64 &      1578 &      0.19 \\ 
 & (2.19)  & (2.14)  & (0.56)  \\  & [    0.29]  & [    0.05]  & [    0.27]  \\ \addlinespace[1ex] 
\multirow[t]{ 3 }{0.2\textwidth}{Monthly income }  &     -0.71\ensuremath{^{*}} &      0.99\ensuremath{^{**}} &     -0.02 &      8.05 &      1578 &      0.22 \\ 
 & (0.32)  & (0.33)  & (0.09)  \\  & [    0.05]  & [    0.01]  & [    0.60]  \\ \addlinespace[1ex] 
\multirow[t]{ 3 }{0.2\textwidth}{Productive assets }  &     -2.27\ensuremath{^{***}} &      3.83\ensuremath{^{***}} &      0.03 &      5.61 &      1578 &      0.27 \\ 
 & (0.43)  & (0.44)  & (0.12)  \\  & [    0.00]  & [    0.00]  & [    0.60]  \\ \addlinespace[1ex] 
\multirow[t]{ 3 }{0.2\textwidth}{HH consumption per capita }  &     -0.13 &      0.19\ensuremath{^{**}} &      0.03 &      9.46 &      1548 &      0.31 \\ 
 & (0.08)  & (0.08)  & (0.02)  \\  & [    0.12]  & [    0.04]  & [    0.16]  \\ \addlinespace[1ex] 
\bottomrule
\end{tabular}

\end{center}
\floatfoot{
Note:  This table uses a linear adjustment of primary outcomes for program cost to compare HD and GD at exactly equivalent costs.  The \emph{Transfer value} column estimates the marginal effect of spending an extra \$100 through cash transfers.  The \emph{Cost-equivalent GD impact} column is estimated as a dummy for either HD or GD treatment, and estimates the impact of cash at the exact cost of HD.  The \emph{Differential impact of HD} column then estimates the differential effect of HD above cash at this benchmarked cost. Regressions include but do not report the lagged dependent variable, fixed effects for randomization blocks, and a set of LASSO-selected baseline covariates, and are weighted to reflect intensive tracking.  Standard errors are (in soft brackets) are clustered at the household level to reflect the design effect, and p-values corrected for False Discovery Rates across all the outcomes in the table are presented in hard brackets.  Stars on coefficient estimates are derived from the FDR-corrected p-values, *=10\%, **=5\%, and ***=1\% significance.  Employed is a dummy variable for spending more than 10 hours per week working for a wage or as primary operator of a microenterprise.  Productive hours are measured over prior 7 days in all activities other than own-farm agriculture.  Monthly income, productive assets, and household consumption are winsorized at 1\% and 99\% and analyzed in Inverse Hyperbolic Sine, meaning that treatment effects can be interpreted as percent changes. 
}
\end{footnotesize}
\end{table}

\clearpage 
\begin{table}[!hbtp]
\caption{Cost-equivalent analysis:  Secondary outcomes}
\label{t:cost_equiv_secondary}
\begin{footnotesize}
\begin{center}
\begin{tabular}{l *{3}{S} ScS}
\toprule
\multicolumn{1}{c}{\text{ }} & \multicolumn{1}{c}{\text{\makecell[b]{Differential impact \\ of HD}}} & \multicolumn{1}{c}{\text{\makecell[b]{Cost-equivalent\\GD impact}}} & \multicolumn{1}{c}{\text{\makecell[b]{Transfer\\Value}}} & \multicolumn{1}{c}{\text{\makecell[b]{Control\\Mean}}} & \multicolumn{1}{c}{\text{Obs.}} & \multicolumn{1}{c}{\text{$ R^2$}}\\
\midrule
\multicolumn{7}{l}{\emph{Panel A. Beneficiary welfare}}  \\ 
\addlinespace[1ex] \multirow[t]{ 3 }{0.2\textwidth}{Subjective well-being }  &     -0.23\ensuremath{^{**}} &      0.42\ensuremath{^{***}} &      0.03 &      0.00 &      1578 &      0.16 \\ 
 & (0.08)  & (0.09)  & (0.02)  \\  & [    0.01]  & [    0.00]  & [    0.23]  \\ \addlinespace[1ex] 
\multirow[t]{ 3 }{0.2\textwidth}{Mental health }  &     -0.00 &     -0.04 &      0.03 &      0.00 &      1578 &      0.08 \\ 
 & (0.09)  & (0.09)  & (0.02)  \\  & [    0.50]  & [    0.50]  & [    0.23]  \\ \addlinespace[1ex] 
\multirow[t]{ 3 }{0.2\textwidth}{Beneficiary-specific consumption }  &     -0.44\ensuremath{^{***}} &      0.61\ensuremath{^{***}} &     -0.02 &      8.27 &      1578 &      0.24 \\ 
 & (0.12)  & (0.12)  & (0.04)  \\  & [    0.00]  & [    0.00]  & [    0.50]  \\ \addlinespace[1ex] 
\multicolumn{7}{l}{\emph{Panel B. Household wealth}}  \\ 
\addlinespace[1ex] \multirow[t]{ 3 }{0.2\textwidth}{HH net non-land wealth }  &     -0.80 &      0.62 &      0.15 &     11.28 &      1578 &      0.21 \\ 
 & (0.49)  & (0.48)  & (0.12)  \\  & [    0.30]  & [    0.35]  & [    0.35]  \\ \addlinespace[1ex] 
\multirow[t]{ 3 }{0.2\textwidth}{HH livestock wealth }  &     -1.92\ensuremath{^{***}} &      1.90\ensuremath{^{***}} &      0.09 &      7.81 &      1578 &      0.27 \\ 
 & (0.45)  & (0.46)  & (0.12)  \\  & [    0.00]  & [    0.00]  & [    0.52]  \\ \addlinespace[1ex] 
\multirow[t]{ 3 }{0.2\textwidth}{Savings }  &     -0.10 &      1.13\ensuremath{^{***}} &      0.08 &      9.24 &      1578 &      0.21 \\ 
 & (0.28)  & (0.29)  & (0.08)  \\  & [    0.52]  & [    0.00]  & [    0.46]  \\ \addlinespace[1ex] 
\multirow[t]{ 3 }{0.2\textwidth}{Debt }  &      0.57 &     -0.16 &     -0.06 &      8.75 &      1578 &      0.21 \\ 
 & (0.39)  & (0.39)  & (0.11)  \\  & [    0.35]  & [    0.52]  & [    0.52]  \\ \addlinespace[1ex] 
\multicolumn{7}{l}{\emph{Panel C. Beneficiary cognitive and non-cognitive skills}}  \\ 
\addlinespace[1ex] \multirow[t]{ 3 }{0.2\textwidth}{Locus of control }  &     -0.00 &      0.07 &     -0.00 &      0.00 &      1578 &      0.29 \\ 
 & (0.08)  & (0.08)  & (0.02)  \\  & [    1.00]  & [    0.99]  & [    1.00]  \\ \addlinespace[1ex] 
\multirow[t]{ 3 }{0.2\textwidth}{Aspirations }  &     -0.06 &      0.06 &     -0.00 &      0.00 &      1578 &      0.08 \\ 
 & (0.09)  & (0.08)  & (0.02)  \\  & [    0.99]  & [    0.99]  & [    1.00]  \\ \addlinespace[1ex] 
\multirow[t]{ 3 }{0.2\textwidth}{Big Five index }  &     -0.01 &      0.13 &     -0.04 &      0.00 &      1578 &      0.11 \\ 
 & (0.09)  & (0.09)  & (0.02)  \\  & [    1.00]  & [    0.76]  & [    0.76]  \\ \addlinespace[1ex] 
\multirow[t]{ 3 }{0.2\textwidth}{Business knowledge }  &      0.54\ensuremath{^{***}} &      0.11 &     -0.03 &      0.00 &      1578 &      0.23 \\ 
 & (0.09)  & (0.09)  & (0.02)  \\  & [    0.00]  & [    0.76]  & [    0.82]  \\ \addlinespace[1ex] 
\multirow[t]{ 3 }{0.2\textwidth}{Business attitudes }  &     -0.09 &      0.21\ensuremath{^{*}} &     -0.03 &      0.00 &      1578 &      0.09 \\ 
 & (0.08)  & (0.08)  & (0.02)  \\  & [    0.82]  & [    0.09]  & [    0.76]  \\ \addlinespace[1ex] 
\bottomrule
\end{tabular}

\end{center}
\vskip-2ex 
\floatfoot{
Note:  This table uses a linear adjustment of secondary outcomes for program cost to compare HD and GD at exactly equivalent costs.  The \emph{Transfer value} column estimates the marginal effect of spending an extra \$100 through cash transfers.  The \emph{Cost-equivalent GD impact} column is estimated as a dummy for either HD or GD treatment, and estimates the impact of cash at the exact cost of HD.  The \emph{Differential impact of HD} column then estimates the differential effect of HD above cash at this benchmarked cost.  Regressions include but do not report the lagged dependent variable, fixed effects for randomization blocks, and a set of LASSO-selected baseline covariates, and are weighted to reflect intensive tracking.  Standard errors are (in soft brackets) are clustered at the household level to reflect the design effect, and p-values corrected for False Discovery Rates across all outcomes within each family are presented in hard brackets.  Stars on coefficient estimates are derived from the FDR-corrected p-values, *=10\%, **=5\%, and ***=1\% significance.  
}
\end{footnotesize}
\end{table}

\cleardoublepage
\subsection{Complementarities}\label{ss:complementarities}

Next, we present a canonical statistical analysis of complementarities.  To do this, we use only the control, the HD arm, the middle GD transfer, and the combined arm (who received HD and the middle GD transfer).  This sets up a standard 2x2 design that cross-cuts the two treatments.  Secondly, we redefine the treatment dummies so that we include one control for `any HD' (HD or combined), one for `any GD' (GD or combined), and a dummy for the combined arm. Using this approach, the `combined' arm dummy now measures not the difference relative to the control but instead whether there is an additional impact from the combination of HD and GD that is greater than what would be expected by adding together the independent HD and GD effects.  It is therefore a direct test of complementarities; whether the whole of the combined arm is something different than the sum of the two parts.

Across the board, the evidence in Tables \ref{t:complementarities_1} and \ref{t:complementarities_2} suggests that far from finding positive complementarities, the whole appears to be less than the sum of the parts.  The sign on the core test (column 3) is negative for all primary outcomes and for two thirds of secondary outcomes.  The complementarity is significantly negative for productive hours and subjective well-being.  

Complex, multi-dimensional programs are often justified on the grounds that poverty presents a range of constraints, meaning that individuals are unable to benefit unless more than one constraint is relaxed at once.  Here we find that the effort to reduce human capital and physical capital barriers simultaneously generates no additional benefit.  Outcomes that are driven by cash are not further helped by HD, outcomes driven by HD are not further helped by cash, and for two of our key outcomes the combination is actually worse than what we would expect from the independent effect of the two programs.  On the other hand, by chance we have a circumstance where the cost of the Combined arm is very similar to the GD Large arm, and we also do not find evidence that the increase in cash transfer amounts from the Main to the Large transfer amounts is worthwhile.  The combined message is then that both efforts to increase expenditure per household from ~\$300 to ~\$750 were not justified, and more moderate expenditure amounts would dominate in a standard cost-benefit sense.

\begin{table}[!hbtp]
\caption{Complementarities: Primary outcomes}
\label{t:complementarities_1}
\begin{footnotesize}
\begin{center}
\begin{tabular}{l *{3}{S} ScS}
\toprule
\multicolumn{1}{c}{\text{ }} & \multicolumn{1}{c}{\text{HD}} & \multicolumn{1}{c}{\text{GD}} & \multicolumn{1}{c}{\text{Complementarity}} & \multicolumn{1}{c}{\text{\makecell[b]{Control\\Mean}}} & \multicolumn{1}{c}{\text{Obs.}} & \multicolumn{1}{c}{\text{$ R^2$}}\\
\midrule
\multirow[t]{ 3 }{0.2\textwidth}{Employed }  &      0.03 &      0.06 &     -0.08 &      0.48 &      1289 &      0.18 \\ 
 & (0.03)  & (0.05)  & (0.06)  \\  & [    0.27]  & [    0.18]  & [    0.18]  \\ \addlinespace[1ex] 
\multirow[t]{ 3 }{0.2\textwidth}{Productive hours }  &      3.41\ensuremath{^{**}} &      7.23\ensuremath{^{***}} &     -8.18\ensuremath{^{**}} &     18.64 &      1289 &      0.22 \\ 
 & (1.58)  & (2.37)  & (3.13)  \\  & [    0.04]  & [    0.01]  & [    0.02]  \\ \addlinespace[1ex] 
\multirow[t]{ 3 }{0.2\textwidth}{Monthly income }  &      0.35 &      1.15\ensuremath{^{***}} &     -0.44 &      8.05 &      1289 &      0.22 \\ 
 & (0.26)  & (0.34)  & (0.45)  \\  & [    0.18]  & [    0.00]  & [    0.26]  \\ \addlinespace[1ex] 
\multirow[t]{ 3 }{0.2\textwidth}{Productive assets }  &      1.57\ensuremath{^{***}} &      3.79\ensuremath{^{***}} &     -0.87 &      5.61 &      1289 &      0.26 \\ 
 & (0.36)  & (0.51)  & (0.68)  \\  & [    0.00]  & [    0.00]  & [    0.18]  \\ \addlinespace[1ex] 
\multirow[t]{ 3 }{0.2\textwidth}{HH consumption per capita }  &      0.06 &      0.28\ensuremath{^{***}} &     -0.09 &      9.46 &      1260 &      0.36 \\ 
 & (0.06)  & (0.09)  & (0.11)  \\  & [    0.26]  & [    0.01]  & [    0.31]  \\ \addlinespace[1ex] 
\bottomrule
\end{tabular}

\end{center}
\floatfoot{ 
Notes:  HD, GD indicators defined as taking value of one if individual is in \emph{either} the corresponding arm \emph{or} the combined arm.  \emph{Complementarity} column reports the differential effect of being in the combined arm, compared to the sum of HD and GD impacts.  Among individuals assigned to GD, only the mid-sized transfer arm is included in this analysis. Regressions include but do not report the lagged dependent variable, fixed effects for randomization blocks, and a set of LASSO-selected baseline covariates, and are weighted to reflect intensive tracking.  Standard errors are (in soft brackets) are clustered at the household level to reflect the design effect, and p-values corrected for False Discovery Rates across all the outcomes in the table are presented in hard brackets.  Stars on coefficient estimates are derived from the FDR-corrected p-values, *=10\%, **=5\%, and ***=1\% significance.
}
\end{footnotesize}
\end{table}

\begin{table}[!hbtp]
\caption{Complementarities: Secondary outcomes}
\label{t:complementarities_2}
\begin{footnotesize}
\begin{center}
\begin{tabular}{l *{3}{S} ScS}
\toprule
\multicolumn{1}{c}{\text{ }} & \multicolumn{1}{c}{\text{HD}} & \multicolumn{1}{c}{\text{GD}} & \multicolumn{1}{c}{\text{Complementarity}} & \multicolumn{1}{c}{\text{\makecell[b]{Control\\Mean}}} & \multicolumn{1}{c}{\text{Obs.}} & \multicolumn{1}{c}{\text{$ R^2$}}\\
\midrule
\multicolumn{7}{l}{\emph{Panel A. Beneficiary welfare}}  \\ 
\addlinespace[1ex] \multirow[t]{ 3 }{0.2\textwidth}{Subjective well-being }  &      0.20\ensuremath{^{***}} &      0.53\ensuremath{^{***}} &     -0.32\ensuremath{^{**}} &      0.00 &      1289 &      0.16 \\ 
 & (0.07)  & (0.10)  & (0.13)  \\  & [    0.01]  & [    0.00]  & [    0.02]  \\ \addlinespace[1ex] 
\multirow[t]{ 3 }{0.2\textwidth}{Mental health }  &     -0.03 &      0.08 &      0.07 &      0.00 &      1289 &      0.09 \\ 
 & (0.07)  & (0.10)  & (0.14)  \\  & [    0.43]  & [    0.35]  & [    0.43]  \\ \addlinespace[1ex] 
\multirow[t]{ 3 }{0.2\textwidth}{Beneficiary-specific consumption }  &      0.17 &      0.64\ensuremath{^{***}} &     -0.08 &      8.27 &      1289 &      0.25 \\ 
 & (0.12)  & (0.14)  & (0.18)  \\  & [    0.17]  & [    0.00]  & [    0.43]  \\ \addlinespace[1ex] 
\multicolumn{7}{l}{\emph{Panel B. Household wealth}}  \\ 
\addlinespace[1ex] \multirow[t]{ 3 }{0.2\textwidth}{HH net non-land wealth }  &     -0.20 &      1.13\ensuremath{^{**}} &     -0.06 &     11.28 &      1289 &      0.21 \\ 
 & (0.40)  & (0.45)  & (0.71)  \\  & [    0.70]  & [    0.03]  & [    0.87]  \\ \addlinespace[1ex] 
\multirow[t]{ 3 }{0.2\textwidth}{HH livestock wealth }  &     -0.03 &      1.92\ensuremath{^{***}} &      0.33 &      7.81 &      1289 &      0.26 \\ 
 & (0.37)  & (0.52)  & (0.70)  \\  & [    0.87]  & [    0.00]  & [    0.70]  \\ \addlinespace[1ex] 
\multirow[t]{ 3 }{0.2\textwidth}{Savings }  &      1.05\ensuremath{^{***}} &      1.27\ensuremath{^{***}} &     -0.62 &      9.24 &      1289 &      0.23 \\ 
 & (0.24)  & (0.34)  & (0.43)  \\  & [    0.00]  & [    0.00]  & [    0.25]  \\ \addlinespace[1ex] 
\multirow[t]{ 3 }{0.2\textwidth}{Debt }  &      0.45 &     -0.24 &     -0.23 &      8.75 &      1289 &      0.24 \\ 
 & (0.28)  & (0.42)  & (0.57)  \\  & [    0.21]  & [    0.70]  & [    0.70]  \\ \addlinespace[1ex] 
\multicolumn{7}{l}{\emph{Panel C. Beneficiary cognitive and non-cognitive skills}}  \\ 
\addlinespace[1ex] \multirow[t]{ 3 }{0.2\textwidth}{Locus of control }  &      0.05 &      0.01 &      0.16 &      0.00 &      1289 &      0.29 \\ 
 & (0.06)  & (0.08)  & (0.11)  \\  & [    0.51]  & [    0.64]  & [    0.43]  \\ \addlinespace[1ex] 
\multirow[t]{ 3 }{0.2\textwidth}{Aspirations }  &     -0.00 &     -0.04 &      0.19 &      0.00 &      1289 &      0.12 \\ 
 & (0.07)  & (0.10)  & (0.13)  \\  & [    0.64]  & [    0.63]  & [    0.43]  \\ \addlinespace[1ex] 
\multirow[t]{ 3 }{0.2\textwidth}{Big Five index }  &      0.12 &      0.11 &     -0.20 &      0.00 &      1289 &      0.12 \\ 
 & (0.07)  & (0.09)  & (0.13)  \\  & [    0.39]  & [    0.43]  & [    0.43]  \\ \addlinespace[1ex] 
\multirow[t]{ 3 }{0.2\textwidth}{Business knowledge }  &      0.66\ensuremath{^{***}} &      0.07 &     -0.11 &      0.00 &      1289 &      0.24 \\ 
 & (0.08)  & (0.10)  & (0.14)  \\  & [    0.00]  & [    0.51]  & [    0.51]  \\ \addlinespace[1ex] 
\multirow[t]{ 3 }{0.2\textwidth}{Business attitudes }  &      0.12 &      0.16 &     -0.15 &      0.00 &      1289 &      0.10 \\ 
 & (0.07)  & (0.09)  & (0.12)  \\  & [    0.39]  & [    0.39]  & [    0.43]  \\ \addlinespace[1ex] 
\bottomrule
\end{tabular}

\end{center}
\vskip-2ex 
\floatfoot{ 
Notes:  HD, GD indicators defined as taking value of one if individual is in \emph{either} the corresponding arm \emph{or} the combined arm. \emph{Complementarity} column reports the differential effect of being in the combined arm, compared to the sum of HD and GD impacts.  Among individuals assigned to GD, only the mid-sized transfer arm is included in this analysis.  Regressions include but do not report the lagged dependent variable, fixed effects for randomization blocks, and a set of LASSO-selected baseline covariates, and are weighted to reflect intensive tracking.  Standard errors are (in soft brackets) are clustered at the household level to reflect the design effect, and p-values corrected for False Discovery Rates across all the outcomes in the table are presented in hard brackets.  Stars on coefficient estimates are derived from the FDR-corrected p-values, *=10\%, **=5\%, and ***=1\% significance.
}
\end{footnotesize}
\end{table}

\cleardoublepage 
\subsection{Analysis of Heterogeneity}\label{ss:heterogeneity}

Our pre-analysis plan indicates four baseline dimensions over which we would look for signs of heterogeneity.   These are gender, household consumption per capita, risk aversion (measured by the choices in a Binswanger--Eckel--Grossman lottery) and local labor market conditions (the employment rate within each of the 67 `cells' of the study).\footnote{We filed our initial PAP with the AEA and then subsequently submitted to the \textit{Journal of Development Economics'} pre-registry facility.  In the time between these two submissions we came to realize that the data from the Convex Time Budget exercise we conducted at baseline \citep{andreoni2012estimating} had not produced meaningful results.  As a result we dropped the analysis of heterogeneity using discount rates and hyperbolicity derived from the CTB, and present here only the heterogeneity tests included in the later PAP filed with the JDE.}
Following the PAP, we use a standard interaction between treatment indicators and baseline characteristics to study heterogeneity across these baseline covariates, and examine only primary outcomes.  The covariates are demeaned prior to interaction so that the uninteracted coefficient should be interpreted as impact at the mean of the interaction variable.  To avoid interpretation issues arising from co-linearity, we omit both the baseline outcome (ANCOVA) and also the LASSO-selected covariates that are included but not reported for all the prior regressions.

Results of subgroup analyses by gender, risk, consumption, and cell-level employment shares are presented in Appendix Tables \ref{t:het_gender}, \ref{t:het_risk}, \ref{t:het_consumption}, and \ref{t:het_employ}, respectively.  Overall, we find very little evidence of meaningful heterogeneity in the impact of the program across these four dimensions.  Gender itself has a huge effect (female beneficiaries are less likely to be employed, put in fewer productive hours, and have lower levels of income and assets), but none of the interventions affect women in a manner significantly different than men.  It is important to recognize, however, that power starts to become a greater concern as we split the study into smaller cells.  Reading point estimates, we see that in fact HD does lead to a 9 pp increase in employment rates among men, while women see a 3 pp \textit{deterioration} in employment rates during HD.  These effects are not significant, however, because the minimum detectable effect on HD among men is 10 pp, and on the gender interaction is 14 pp.  Table \ref{t:het_risk} shows the risk averse being somewhat better at translating the interventions into income, with few other differences.  Encouragingly, Table \ref{t:het_consumption} shows impacts that are relatively invariant to baseline consumption, indicating that both cash and HD are equally effective for the very poor (the GD Large arm has slightly larger benefits, and the Combined arm slightly larger benefits, for those with higher consumption at baseline).  Given that many related programs have had an easier time creating benefits for the non-poor, this suggests that both of the interventions studied here should be considered good candidates for heavily poverty-targeted programs.  Baseline employment rates, studied in Table \ref{t:het_employ}, are not only not driving the impacts of the program but appear to be completely uncorrelated with outcomes overall.  Finally, based on feedback from the Rwanda USAID mission, we included age as a dimension of heterogeneity, and again in Table \ref{t:het_older} find no evidence of differences.  The bottom row of all of these tables shows the p-value on F-tests of joint significance across the four interaction terms (note these are  based on the non-FDR-adjusted significance rates); we present 25 such omnibus tests and find two of them to be significant at the 10\% level, fewer than we would expect by random chance.   Hence, taken together the analysis of heterogeneity suggests programs that are having consistent and similar effects across different types of beneficiaries and across local labor market conditions.

\subsection{Spillovers}\label{ss:spillovers}

Spillovers are of central interest in this project for several reasons.  First, for both of the programs being studied here recent literatures suggest that we should be concerned with impacts on non-beneficiaries.   \cite{crepon2013labor} show that most of the benefits of job training programs in France come from diverting a fixed set of job opportunities towards treated individuals and away from untreated ones.  Cash transfer programs appear to have complex spillovers on non-beneficiaries, with \cite{angelucci2009indirect} and \cite{egger2019general} finding potential \textit{positive} spillovers through family or labor market mechanisms, with other studies suggesting negative spillovers  to non-beneficiaries \citep{haushofer2016short, mcintzeit2019gikuriro}, particularly in terms of mental health \citep{baird2013income}.    Because our study uses an individually randomized design these spillovers are a direct threat to internal validity, and so this test is critical.

We look for spillovers both on program participation and on primary outcomes.  While in principle there may be externalities of each program at several levels of contact, we focus on spillovers that are \emph{local}, in the sense that they occur between individuals who reside in the same village at baseline.\footnote{ We conducted a social network survey measuring connections to other individuals in the study at baseline that we had intended to use for this analysis, but we found that a) networks within villages are typically completely connected, and b) we were unable to collect this data for a small subset of beneficiaries.  Because the simple treatment saturation in the village maps almost perfectly to the saturation in the social network and is universally observable, we use it for our analysis.}  The reason for doing so is both substantive---this is plausibly the level at which such interactions are most salient---as well as practical: since the randomization is blocked at the sector level, and provides no variation in treatment saturation at that or higher levels.

 Let $T_{ivb}^w$ denote the assignment of individual $i$ in \emph{village} $v$ to treatment $w\in\left\{\text{GDM}, \text{GDL}, \text{HD}\right\}$, where we pool the three smaller cash-transfer values into a single arm, $w=\text{GDM}$, as distinct from the larger transfer value, $w=\text{GDL}$.   Individuals in the combined arm have $T_{ivb}^{GDS} = T_{ivb}^{HD} =1$. Define $T_{ivb} = [T_{ivb}^{GDM},\, T_{ivb}^{GDL},\, T_{ivb}^{HD}]$ as the vector denoting individual $i$'s treatment status.  Finally, we let the vector $\bar{T}_{-i,vb}=[\bar{T}_{-i,vb}^{GDM},\, \bar{T}_{-i,vb}^{GDL},\, \bar{T}_{-i,vb}^{HD}]$ denote the average treatment status of study individuals \emph{other than} individual $i$ in village $v$ for each of the three treatments (that is, the saturation of each treatment among others in the village), and we adopt the convention that $\bar{T}_{-i,vb}=0$ if individual $i$ is the only study participant in village $v$.  This vector of village-level saturations is randomly assigned through the household-level lottery, and is independent of the own-treatment terms because we calculate saturations among others in the village.  The densities of the treatment saturations are plotted in Figure \ref{f:saturation_densities}.

\subsubsection{Spillovers on Compliance}

Using this notation, we can represent the three types of spillover analysis conducted, in increasing order of complexity.  First, we analyze whether there are spillover effects on the rate at which individuals choose to participate in Huguka Dukore (this question is not interesting for GD because compliance is so close to universal).  A major concern during the study design phase was that the assignment of one's peers to cash would discourage participation with HD. To do this, we use only the HD arm and estimate the following linear probability model:
    \begin{equation}\label{eq:TakeupExternalities}
    \mathrm{E}[P_{ivb}^w|\bar{T}_{-i,vb}] = \mu_b^w + \phi^w  \bar{T}_{-i,vb} 
    \end{equation}
where $P_{ivb}^W$ is a measure of individual $i$'s participation in treatment $w\in\{HD,Combined\}$.  

  Table \ref{t:compliance_saturation} illustrates that the density of GD treatment in a village does not drive compliance.  The first column analyzes compliance within the standalone HD arm, and the second column within the Combined arm.  The point estimates on the GD saturation rates are zero or positive, and never significant.  Instead, this table shows that HD compliance is driven by the \textit{Huguka Dukore} treatment saturation; the point estimate in the standalone HD arm is significant at the 5\% level and suggests that as an HD participant goes from having no-one else in the village treated to everyone else in the village treated, we can expect compliance to increase by 27 pp.  Since HD in the absence our of experiment would naturally attempt to treat 100\% of the willing and eligible individuals considered in this study, that means that our study features a compliance rate that may be slightly too low relative to the rate that would naturally occur. However, since the costing estimate is multiplied times the observed compliance rate and the ITT is similarly a function of observed compliance in our study, compliance falls out of the benefit/cost comparisons that we make (it is in both the numerator and the denominator).  This suggests that our estimates are still likely to be meaningful, absent substantial Essential Heterogeneity in the sense of \cite{heckman2005structural}.

\begin{table}[!h]\caption{Spillovers on Program Compliance}
\label{t:compliance_saturation}
\begin{center}
\begin{tabular}{l *{2}{S}}
\toprule
\multicolumn{1}{c}{\text{ }} & \multicolumn{1}{c}{\text{HD}} & \multicolumn{1}{c}{\text{Combined}}\\
\midrule
\multirow[t]{ 2}{0.2\textwidth}{HD Saturation} &      0.27\ensuremath{^{**}} &      0.09 \\ 
 & (0.09)  & (0.17)  \\ 
 & [    0.01]  & [    0.85]  \\ 
\addlinespace[1ex] \multirow[t]{ 2}{0.2\textwidth}{GD Main Saturation} &      0.13 &     -0.01 \\ 
 & (0.08)  & (0.14)  \\ 
 & [    0.45]  & [    1.00]  \\ 
\addlinespace[1ex] \multirow[t]{ 2}{0.2\textwidth}{GD Large Saturation} &      0.12 &      0.15 \\ 
 & (0.14)  & (0.26)  \\ 
 & [    0.85]  & [    0.85]  \\ 
\addlinespace[1ex] Average compliance  &      0.86 &      0.90 \\ 
Observations  & \multicolumn{1}{c}{      466}  & \multicolumn{1}{c}{      192}  \\ 
$ R^2$  &      0.32 &      0.55 \\ 
$ p$-value  &      0.01 &      0.85 \\ 
\addlinespace[1ex] 
\bottomrule
\end{tabular}

\end{center}
\begin{footnotesize}
\vskip-2ex
\floatfoot{ 
Notes:  Table uses a Linear Probability Model to examine the likelihood that an individual assigned to that arm participates in Huguka Dukore (Column 1) or the HD component of the Combined arm (Column 2), as a function of the saturation of each of the three treatments among other members of the same village.  Regressions include but do not report the lagged dependent variable, fixed effects for randomization blocks, and a set of LASSO-selected baseline covariates, and are weighted to reflect intensive tracking.  Standard errors are (in soft brackets) are clustered at the household level to reflect the design effect, and p-values corrected for False Discovery Rates across all the outcomes in the table are presented in hard brackets.  Stars on coefficient estimates are derived from the FDR-corrected p-values, *=10\%, **=5\%, and ***=1\% significance.  Bottom row is the p-value on an F-test of the joint significance of the three saturation terms.
}
\end{footnotesize}
\end{table}

Next, we can take our study of spillovers to the primary outcomes used in the study.  We first use a more parsimonious specification that looks for average spillover effects \textit{from} each type of treatment saturation to the other members of the village.  Table \ref{t:saturation_levels} conducts this analysis and uncovers  no evidence of spillovers in outcomes; not only are none of the saturation rates for any of the three treatments significant, but the sign of the coefficients alternates signs across outcomes for all three treatment saturations.  

For outcome $Y_{ivb1}$, we modify the specification of equation used to estimate ITT effects (equation \ref{eq:PrimarySpec}) as follows:
    \begin{equation}\label{eq:Externalities_basic}
    Y_{ivb1} = \delta_1 T_{ivb} + \delta_2 \bar{T}_{-i,vb} + \beta X_{ivb0} + \rho Y_{ivb0} + \mu_b + \varepsilon_{ivb1}.
    \end{equation}

\begin{table}[!h]\caption{Spillover effects:  levels model}
\label{t:saturation_levels}
\begin{footnotesize}
\begin{center}
\begin{tabular}{l *{5}{S}}
\toprule
\multicolumn{1}{c}{\text{ }} & \multicolumn{1}{c}{\text{Employed}} & \multicolumn{1}{c}{\text{\makecell[b]{Productive\\Hours}}} & \multicolumn{1}{c}{\text{\makecell[b]{Monthly\\Income}}} & \multicolumn{1}{c}{\text{\makecell[b]{Productive\\Assets}}} & \multicolumn{1}{c}{\text{Consumption}}\\
\midrule
\multirow[t]{ 2}{0.2\textwidth}{HD} &      0.01 &      0.91 &      0.22 &      1.16\ensuremath{^{***}} &      0.05 \\ 
 & (0.03)  & (1.25)  & (0.20)  & (0.30)  & (0.05)  \\ 
 & [    1.00]  & [    1.00]  & [    0.93]  & [    0.00]  & [    1.00]  \\ 
\addlinespace[1ex] \multirow[t]{ 2}{0.2\textwidth}{GD main} &      0.01 &      2.73 &      0.90\ensuremath{^{***}} &      3.49\ensuremath{^{***}} &      0.23\ensuremath{^{***}} \\ 
 & (0.03)  & (1.28)  & (0.19)  & (0.26)  & (0.04)  \\ 
 & [    1.00]  & [    0.13]  & [    0.00]  & [    0.00]  & [    0.00]  \\ 
\addlinespace[1ex] \multirow[t]{ 2}{0.2\textwidth}{GD Huge treatment} &      0.00 &      0.18 &      0.70 &      3.81\ensuremath{^{***}} &      0.35\ensuremath{^{***}} \\ 
 & (0.04)  & (1.92)  & (0.34)  & (0.46)  & (0.07)  \\ 
 & [    1.00]  & [    1.00]  & [    0.13]  & [    0.00]  & [    0.00]  \\ 
\addlinespace[1ex] \multirow[t]{ 2}{0.2\textwidth}{HD Saturation} &      0.01 &     -1.80 &     -0.08 &     -0.01 &      0.02 \\ 
 & (0.06)  & (2.75)  & (0.45)  & (0.66)  & (0.10)  \\ 
 & [    1.00]  & [    1.00]  & [    1.00]  & [    1.00]  & [    1.00]  \\ 
\addlinespace[1ex] \multirow[t]{ 2}{0.2\textwidth}{GD Main Saturation} &     -0.00 &      0.41 &     -0.39 &      0.75 &     -0.01 \\ 
 & (0.06)  & (2.77)  & (0.42)  & (0.59)  & (0.09)  \\ 
 & [    1.00]  & [    1.00]  & [    1.00]  & [    0.77]  & [    1.00]  \\ 
\addlinespace[1ex] \multirow[t]{ 2}{0.2\textwidth}{GD Large Saturation} &     -0.11 &     -9.89 &      0.15 &      0.12 &     -0.07 \\ 
 & (0.10)  & (4.85)  & (0.72)  & (1.04)  & (0.16)  \\ 
 & [    0.93]  & [    0.13]  & [    1.00]  & [    1.00]  & [    1.00]  \\ 
\addlinespace[1ex] Control mean  &      0.48 &     18.64 &      8.05 &      5.61 &      9.46 \\ 
Observations  & \multicolumn{1}{c}{     1770}  & \multicolumn{1}{c}{     1770}  & \multicolumn{1}{c}{     1770}  & \multicolumn{1}{c}{     1770}  & \multicolumn{1}{c}{     1737}  \\ 
$ R^2$  &      0.17 &      0.19 &      0.21 &      0.26 &      0.33 \\ 
$ p$-value  &      0.69 &      0.23 &      0.79 &      0.64 &      0.96 \\ 
\addlinespace[1ex] 
\bottomrule
\end{tabular}

\end{center}
\floatfoot{ 
Notes:  Table analyzes spillover effects of the three main treatments (HD, GD Main, and GD Large) on the five primary outcomes.  The first three rows are dummy variables for own treatment status, and the next three are the saturation rates for the three treatments among others in the village, so measure the marginal effect of going from no-one else treated to everyone else treated.  Regressions include but do not report the lagged dependent variable, fixed effects for randomization blocks, and a set of LASSO-selected baseline covariates, and are weighted to reflect intensive tracking.  Standard errors are (in soft brackets) are clustered at the household level to reflect the design effect, and p-values corrected for False Discovery Rates across all the outcomes in the table are presented in hard brackets.  Stars on coefficient estimates are derived from the FDR-corrected p-values, *=10\%, **=5\%, and ***=1\% significance.  Bottom row is the p-value on an F-test of the joint significance of the three saturation terms.
}
\end{footnotesize}
\end{table}

Finally, we use the full model from our PAP that allows for the estimation of spillovers both \textit{from} each treatment arm, and \textit{on to} each treatment arm:
    \begin{equation}\label{eq:Externalities}
    Y_{ivb1} = \delta_1 T_{ivb} + \delta_2 \bar{T}_{-i,vb} + \delta_3 T_{ivb} \bar{T}_{-i,vb} + \beta X_{ivb0} + \rho Y_{ivb0} + \mu_b + \varepsilon_{ivb1}.
    \end{equation}
In equation \eqref{eq:Externalities}, the three coefficients in $\delta_2$ provide a test of the spillover effects of each of the three treatments onto control individuals, and the nine coefficients in the vector $\delta_3$ test for whether the spillover effects \textit{from} the saturations of any of the three treatments \textit{on to} individuals directly receiving each of the three treatments differ from the control.\footnote{We report cluster-robust standard errors for each of the coefficients, clustering at the village level.  In addition, given the large number of hypotheses tested in these regressions (sixteen) we correct the p-values in these regressions using Anderson's (2008) False Discovery Rate correction across all coefficients within each regression.}

The five Appendix Tables \ref{t:interference_bn_employed}, \ref{t:interference_bn_productive_hrs}, \ref{t:interference_bn_monthly_income}, \ref{t:interference_bn_tot_prod_assetval}, and \ref{t:interference_hh_month_consumption_pc} present this analysis  Because of the large number of hypotheses being tested we present the analysis for each outcome in a separate table.  Each of these analyses shows results from a single regression with the three different sets of treatment by saturation interactions stacked as adjacent columns.  Again, this analysis is remarkably clear and consistent in showing an absence of spillover effects.  Using significance levels derived from within-regression false discovery corrections we do not have a single significant spillover effect in the control, or differential effect for any treatment, across any of the 60 spillover tests performed here.  Using unadjusted p-values we find 4 of these comparisons significant at the 10\% level and none at the 5\% level, in line with random chance.  

\begin{figure}[!p]
\caption{Expected outcomes by treatment arm, under alternative saturation rates}
\label{tab:interference_combined}
\begin{center}
\includegraphics[width=0.95\linewidth]{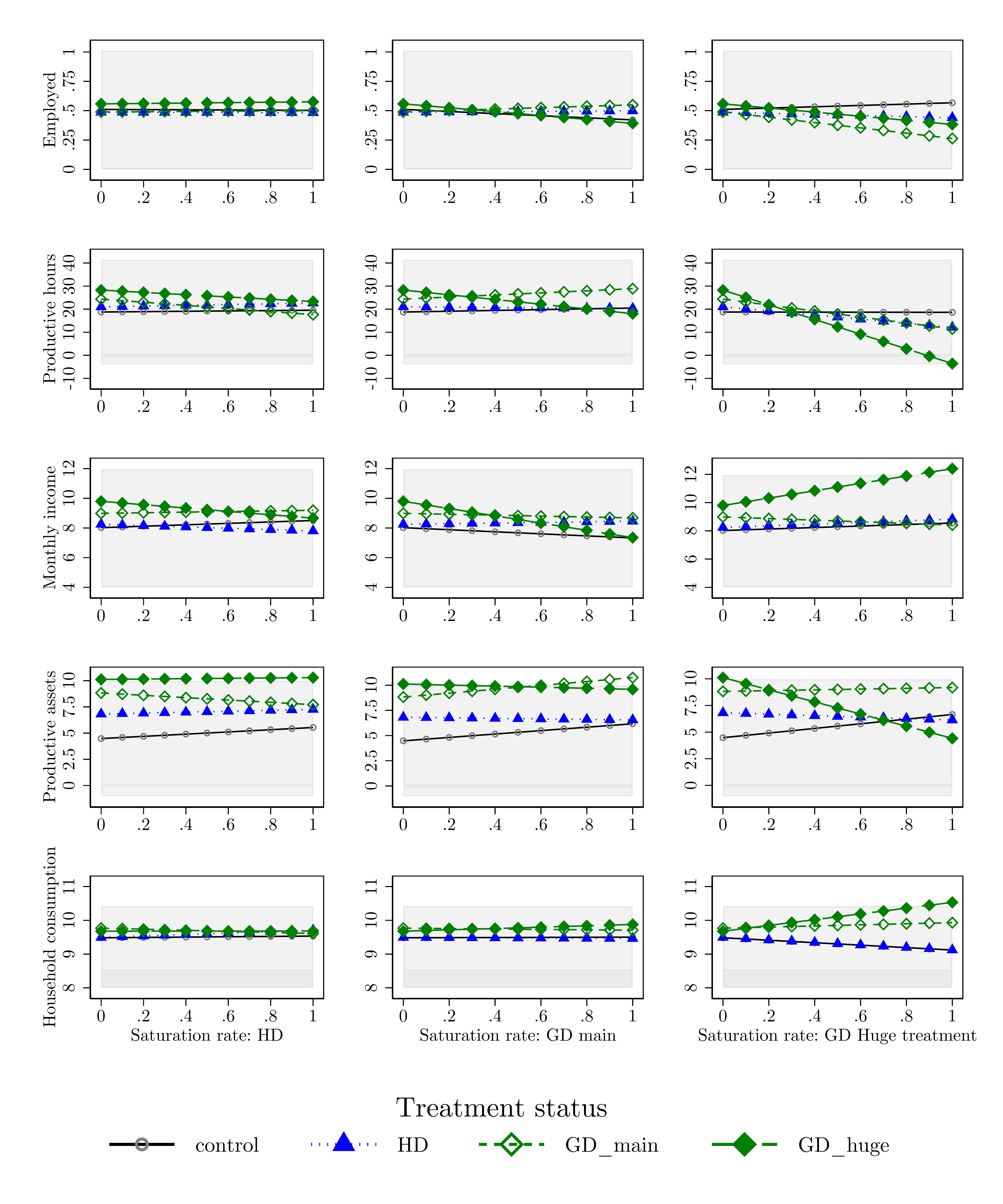}
\end{center}

\floatfoot{ 
\begin{footnotesize}
Notes:  Each panel presents predicted outcomes under each of the four main treatment arms (Control, HD, GD Main, and GD Large), as the saturation level of a specific active treatment arm changes.  Rows correspond to the outcomes of employment, productive hours, and the inverse hyperbolic sine of monthly incomes, productive assets, and household consumption per adult-equivalent, respectively. Horizontal shaded bands highlight one standard deviation above and below the control mean.  Columns illustrate effects of variation in saturation rates in HD, GD-main, and GD-large, respectively.  All predicted outcomes evaluated at means of covariates used in Equation \ref{eq:Externalities}.
\end{footnotesize}
}
\end{figure}

Figure \ref{tab:interference_combined} provides a graphical take on this analysis, showing the predicted outcome for each treatment group (in rows) and primary outcome(in columns).  Using the observed treatment saturations and the estimated saturation slopes, this exercise predicts the outcome we would expect to see within each arm and outcome as the local intensity of treatment changes, with the specific treatment saturations generating the spillovers plotted as different colored fitted lines in each graph.  The only visual signs of spillovers are restricted to the GD Large arm (which in reality has a limited variation in the saturations), while the HD and GD Main arms which are the core of the benchmarking exercise appear completely invariant to local intensity of treatment.

It therefore appears that we can quite simply conclude that our study did not generate detectable within-village spillover effects on outcomes.  While our blocked lotteries do not allow us to study the kinds of cross-village General Equilibrium (GE) effects generated in \cite{egger2019general}, the size of our transfers are tiny relative to local GDP and it appears unlikely that we would have generated meaningful local GE effects.  We do not see the kind of zero-sum diversionary treatment effects uncovered by job-training programs in more formalized labor markets \citep{crepon2013labor}, presumably because a) our overall impact on employment was small, and b) the informal labor markets in this context do not demonstrate a hard capacity constraint in the same way.  The invariance of outcomes to HD saturations suggests that the response of HD compliance to HD saturations is not generating a study ITT that is driven by the lower treatment saturations induced by the lottery.  Comfortingly, the overall takeaway is that our analysis remains internally valid and there is no need to attempt to correct for spillovers.

\subsection{Tracing cash flows}\label{s:moneyflows}

In this question, we ask a question specific to the cash-transfer arms of our study:  \emph{How have participants used the funding they received?}  Tracing these cash flows facilitates understanding in several ways.  First, variation in the use of funds by the amount of cash received, and by the provision of HD training alongside cash in the combined arm, may help us to understand the mechanisms by which these arms deliver distinct impacts.  Second, the extent to which we can account for the full values of transfers received sheds light on whether there might be `missing' dimensions of impact not accounted for in our evaluation. And third, the extent to which observed patterns of spending are consistent with a simple spend-down of cash grants---and not a set of investments likely to deliver future income gains---is indicative of the sustainability of the impacts of cash. 

Accounting for cash flows requires us to model both the inflows and the outflows induced by each active treatment arm.  Clearly, the core of the induced inflow is the value of cash itself received by the beneficiary.  But we also need to account for other income gains that each intervention can cause.  To do so, we estimate each arm's impacts on beneficiary income and on transfers \emph{received} by the beneficiary household, and we include these as inflows.  We then compare these impacts on inflows with estimated impacts on expenditures, where our expenditure measures are a mixture of flows---household consumption, and transfers and loans made to other households---as well as stocks---savings, debt, livestock, and other productive assets.  
Flow measures vary in their recall periods, with income and consumption measured over comparatively short recall periods (as analyzed in Section \ref{ss:ITT}, these are constructed as estimates for the month prior to the follow-up survey) and transfers measured over a 12-month recall period. 

We value impacts on each of these dimensions by multiplying estimated \emph{proportional} impacts of each program, from the specification in equation \eqref{eq:PrimarySpec}, by the average \emph{level} of the corresponding outcome at follow-up in the control group.  We emulate the results of Section \ref{ss:ITT} for household transfers in Appendix Table \ref{t:itt_transfers}.  Together with the ITT results on primary and secondary outcomes, and combined with estimates of control means, this allows us provide the estimated financial impact of each program on both inflows and outflows.

\begin{table}[!htb]
    \caption{Accounting for cash transfers}
    \label{t:accounting}

    \begin{center}
    \begin{tabular}{l *{6}{S}}
\toprule
&  &\multicolumn{5}{c}{Treatment effect} \\ 
\cmidrule(lr){3-7} 
\multicolumn{1}{c}{\text{ }} & \multicolumn{1}{c}{\text{Control mean}} & \multicolumn{1}{c}{\text{Lower}} & \multicolumn{1}{c}{\text{Middle}} & \multicolumn{1}{c}{\text{Upper}} & \multicolumn{1}{c}{\text{Large}} & \multicolumn{1}{c}{\text{Combined}}\\
\midrule
\multicolumn{7}{l}{\emph{Panel A.  Inflows}}  \\ 
\addlinespace[1ex] Cash received  &      0.00 &    317.16 &    410.65 &    502.96 &    750.30 &    410.65 \\ 
\addlinespace[1ex] 
Beneficiary income  &    209.36 &    158.39 &    226.36 &    239.48 &    152.48 &    218.09 \\ 
\addlinespace[1ex] 
Transfers received  &     23.38 &     40.93 &     62.50 &     61.16 &     72.74 &     57.63 \\ 
\addlinespace[1ex] 
\emph{Total inflows}  &         . &    516.47 &   699.51 &    803.60 &    975.53 &    686.37 \\ 
\addlinespace[2ex] 
\multicolumn{7}{l}{\emph{Panel B.  Outflows}}  \\ 
\addlinespace[1ex] Household consumption  &    625.85 &    124.50 &    166.53 &    146.10 &    223.08 &    169.13 \\ 
\addlinespace[1ex] 
Livestock  &    118.64 &    208.40 &    218.03 &    313.31 &    257.79 &    262.49 \\ 
\addlinespace[1ex] 
Productive assets  &     49.89 &    196.35 &    189.78 &    191.37 &    200.70 &    220.60 \\ 
\addlinespace[1ex] 
Savings  &     51.99 &     54.11 &     66.88 &     81.22 &     74.34 &     88.11 \\ 
\addlinespace[1ex] 
Debt  &     61.93 &     -5.93 &    -14.04 &    -34.77 &    -22.79 &      0.14 \\ 
\addlinespace[1ex] 
Loans made  &      3.83 &      3.46 &      1.74 &      5.41 &      2.61 &      4.74 \\ 
\addlinespace[1ex] 
Transfers made  &      4.53 &      3.08 &      8.48 &      2.56 &      1.46 &      3.65 \\ 
\addlinespace[1ex] 
\emph{Total outflows}  &         . &    595.82 &    665.47 &  774.73 &    782.78 &    748.60 \\ 
\addlinespace[2ex] 
\multicolumn{7}{l}{\emph{Panel C.  Totals}}  \\ 
\addlinespace[1ex] Share accounted  &         . &    \multicolumn{1}{c}{115\%} &      \multicolumn{1}{c}{95\%}  &      \multicolumn{1}{c}{96\%}   &      \multicolumn{1}{c}{80\%}   &      \multicolumn{1}{c}{109\%}  \\ 
\addlinespace[1ex] 
\bottomrule
\end{tabular}

    \end{center}
    
    \vskip-4ex
    \floatfoot{
        \par Note.  Table presents control means and estimated impacts on financial values, in dollars.  Beneficiary income and household consumption are estimated 12-month totals, assuming constant flows over the period between delivery of cash transfers and follow-up. Inter-household transfers and loans are 12-month recall variables.  All other variables are stocks measured at follow-up. \emph{Total inflows} are the sum of cash received, beneficiary income, and transfers received.  \emph{Total outflows} are the sum of household consumption, livestock values, other productive asset values, savings values, the negative of debt values, loans made, and transfers made.   \emph{Share accounted} is the ratio between total outflows and total inflows.  
    }

\end{table}

The results of this exercise are presented in Table \ref{t:accounting}. In Panel A of that table, we estimate impacts of each cash-transfer arm on households' total income.  This comprises not only the direct value of transfers received, but also induced increases in beneficiary income and in transfers received from other households.  Since beneficiary income is measured over a short recall, we need to make an assumption about its time path, and so we extrapolate over the 12 months since the cash transfer assuming a constant impact in all months. Whether this under- or over-states true financial inflows will depend, among other things, on whether beneficiary income measures represent true income effects or the spend-down of business stocks.  We compare these impacts on cash inflows with financial outflows, measured in Panel B.  Those outlays include measures of flows including household consumption, impacts on which are extrapolated as constant over the period since the cash transfer, as well as loans and transfers made by the household, which are measured with a 12-month recall, and so do not need to be extrapolated.  They also include values of stocks at follow-up of livestock, other productive assets, savings, and debt.  Our total outflow measure is constructed as the sum of the financial value of impacts on each of these categories, with impacts on debt entering negatively.  We then construct the share of cash inflows that can be accounted for by these measured dimensions as the ratio of total treatment-induced outflows to total treatment-induced inflows for each arm.

We draw three basic conclusions from this exercise. 
First, the fact that induced outflows constitute a large share of induced inflows in general suggests that our measures of financial assets and expenditures are relatively complete, at least as far as is relevant to the impacts of these transfers. 
Second, expenditure patterns do exhibit some differences across arms.  For example,  investments in livestock and in other productive assets do not rise proportionally with the value of transfers.  Further, comparison between expenditure patterns in the Middle and Combined arms---for which cash transfer values are equal---suggests that there are modest differences in the application of cash induced by HD training.  The Combined treatment seems to divert the flow of cash to livestock and productive assets to a greater extent than the Middle arm, at the expense, in part, of debt repayment.  While these investments have not delivered increases in income at the time of follow-up, it suggests that there remains a possibility that the Combined arm will cause differences in incomes as these investments deliver returns over the long term.\footnote{The differences in investments induced appear too small to have created meaningful differences in income over the period studied.  For example, if the \emph{differential} livestock investment in the Combined arm versus the Middle arm, of \$262-\$218=\$44, was undertaken immediately after the transfer and paid an annuity value of 5 percent, the resulting difference in incomes would have been just over \$2 in total over the year since transfers occurred.} 
Third, while our ability to draw inferences about the sustainability of cash-transfer impacts beyond the follow-up period is necessarily speculative, there are two features of this exercise that point to sustainability.  First, the consumption impacts of the cash transfers are smaller than their income impacts in absolute terms.  And to the extent that a rapidly declining consumption path over the period prior to follow-up period would have been indicative of a lack of sustainability, we note that under the current (generous) assumption about total impacts on inflows, there is little unaccounted-for expenditure that could have been part of a downward-sloping consumption pattern over time.

\section{Value for Money}\label{s:value_for_money}

There are multiple ways that one can pose that most basic question in cost effectiveness:  how can policy spending achieve the greatest effect?  Our study is designed to emphasize one comparison, namely the \emph{cost equivalent} one:  if a comparable amount of money is to be spent per beneficiary across programs, which achieves the greatest benefit?  This approach holds both the beneficiary pool and the spend per beneficiary fixed, and asks about comparative effectiveness.  

A related but different question can be asked if one is willing to concentrate spending on a subset of the beneficiary pool.  This is
\emph{comparative cost effectiveness}: how can money be spent to create the largest total benefit across the pool for a fixed overall budget?  In Tables \ref{t:itt_primary_full} and \ref{t:itt_secondary_full} when we test for differences in the ratio of the effect sizes to the cost of the arm, this is the question we are asking. 

\begin{figure}[!hptb]
\caption{Cost Equivalence versus Cost Effectiveness}
\label{f:CE_CEff_Comparison}
\begin{center}
\includegraphics[width=0.8\linewidth]{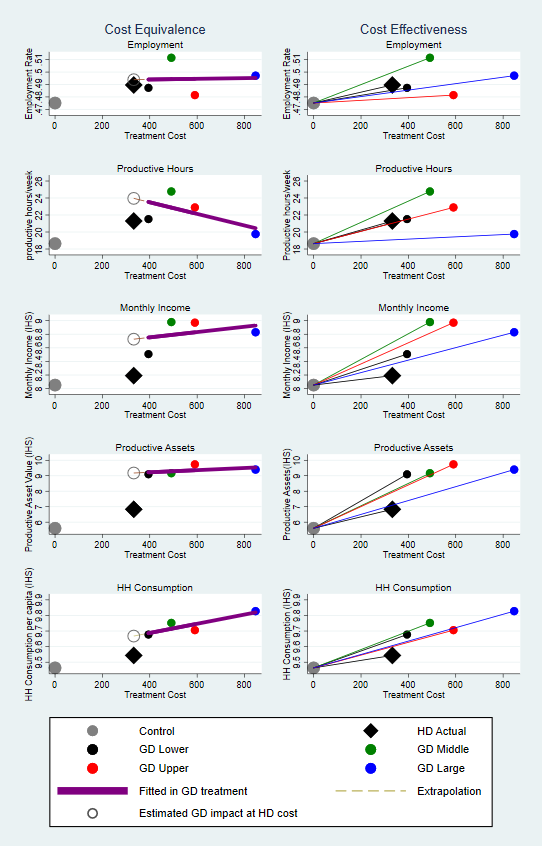}
\end{center}
\end{figure}

A visual comparison between the cost equivalence and the cost effectiveness approaches is provided in Figure \ref{f:CE_CEff_Comparison}.  In the left-hand column of figures, we illustrate the cost-equivalent comparison for the five primary outcomes.  Here, the question is focused on a specific point on the x-axis, namely the cost of HD (the black diamond), and the GD arms are pooled to estimate one counterfactual, which is the predicted value at the HD cost.  The Control need not even be included to execute this comparison.  

The right-hand column of Figure \ref{f:CE_CEff_Comparison} plots the identical outcomes by arm, but uses them to instead represent the cost effectiveness question in a visual way.  The term \emph{value for money} can be thought of as the relationship between the vertical axis in these figures (the value generated in terms of improved outcomes) over the horizontal axis (the money spent).  Hence, the slope of the line between the Control outcome and the outcome in each active treatment arm in this space is a direct representation of benefit/cost across arms.  The steepest slope has the highest value for money, assuming that the donor is willing to be flexible regarding how the number of people treated and therefore the cost per person. \footnote{In the absence of general equilibrium effects, all points along these lines connecting average outcomes in active treatment arms with the control group are achievable by mixing that treatment with a fraction of individuals left untreated.  In this sense, there is a lesser degree of extrapolation required for comparisons based on cost effectiveness than for those based on cost equivalence. }

In our study, the cost-equivalent question can be posed at one cost by design (the ex-post cost of HD, compared to the extrapolated benefit of GD at that cost), and at another cost by accident (because the Combined and HD Large arms turn out to have virtually identical costs).  The comparative cost-effectiveness question, on the other hand, is easily posed across costs since it essentially compares the gradient of additional money at each modality and total cost.  Comparative cost effectiveness is therefore an attractive way to horse-race the GD arms against each other. 

Given these two potential approaches, what does our study say about value for money?  Beginning with the central cost-equivalent comparison between HD and cash, the main results of the paper suggest that for the \$332 price tag of HD, one could produce significantly better outcomes via cash transfers across productive assets, beneficiary consumption, and livestock wealth, and for this amount of money HD is superior only at producing business knowledge.  This result can help policymakers with different objectives think about which type of intervention will best produce the outcomes that they want to see.

 Appendix Tables \ref{t:benefit_cost_ratios_primary} and \ref{t:benefit_cost_ratios_secondary} provide estimates of cost effectiveness for primary and secondary outcomes, respectively, by dividing the ITT estimates by the cost of the arm in hundreds of dollars, and testing for the differences across arms (this is the slope coefficient visually represented in the right-hand panel of Figure \ref{f:CE_CEff_Comparison}).   Seen in comparative cost effectiveness terms, the GD Middle transfer appears to be superior overall.  It has the highest benefit/cost ratio across four of the five primary outcomes, losing only to the GD Small arm on productive asset value.  Among secondary outcomes, GD Middle wins or ties in terms of cost effectiveness for subjective well-being, beneficiary consumption, and net household wealth.  HD has higher cost effectiveness in driving savings, and for business knowledge.  
 
 When we make the cost-equivalent comparison between the Combined arm and the GD Large arm, we are asking the following question:  given that we have already spent the \$494 dollars to deliver the GD Main arm, is it then better to spend another \$350 to deliver HD on top of that cash, or should that additional spending be used to amplify the cash transfer?  Here, the answers look quite different.  There is now no outcome for which the GD Large arm is significantly better than Combined, and the Combined arm continues to demonstrate HD's advantage at producing business knowledge.  Why these divergent results?  The answer appears to be evidence of diminishing marginal benefits of cash; because HD is effective at driving many of the economic outcomes and additional cash is having a weaker effect, we find these two interventions to have similar cost-equivalent benefits even for outcomes on which cash `beat' HD starting from zero. Overall, the fact that in general the less expensive interventions produce superior outcomes in cost-effectiveness terms says that we uncover no evidence for the idea that concentrating spending per person above the levels seen in the GD Middle arm is justified.

\section{Conclusions}\label{s:conclusions}
This study undertakes an exercise increasingly called for in recent years:   namely, a direct comparison of in-kind development aid to cash, in a study design that uses variation in cash transfer values to identify cost-equivalent comparisons.  The comparison program, Huguka Dukore, is a well-established workforce readiness intervention, and the beneficiary group is one in which both interventions have straightforward pathways to generate long-term improvements in welfare.  

Our findings on the impacts of Huguka Dukore contribute to the evidence base on supply-side active labor market programs, of which it is typical in design.  Huguka Dukore appears successful in increasing beneficiaries' productive time use, as well as the stock of assets productive that they accumulate.  18 months after program assignment, its participants report higher levels of subjective well being, demonstrate improved business knowledge, and have increased their stocks of savings considerably.   HD has no significant effects on employment rates or beneficiary incomes.  Given a control-group endline average income of approximately USD 17.45 per month, our point estimates on HD's impacts on beneficiary income imply that approximately 61 months of earnings at this rate would be required to pay back the realized costs to beneficiaries, consistent with supply-side labor market policies elsewhere \citep{McK17almp}.  But looking beneath the surface, we do see exploratory evidence of shifts to microenterprise and even non-agricultural wage labor, which may portend long-term labor-market impacts.  
            
On the other hand, cash transfers to this population appear to have moved a wide range of outcomes, including productive hours, incomes, productive assets, and household consumption.  These impacts are substantial for beneficiaries and provide a meaningful return on the costs of intervention:  for example, the cost to USAID of the middle transfer would be recuperated in beneficiary income impacts alone after approximately 26 months.  Secondary measures including subjective well-being, household wealth, and savings are all meaningfully moved by these transfers. Several lessons about the design mix of transfer programming are evident:  there is no evidence of complementarities between these cash-transfer impacts and the skills provided by HD, and increases in transfer size above the middle transfer amount of \$411 included in our study do not appear justified in cost-effectiveness terms.
            
It is worth noting a number of study limitations.  First, the final costing number is substantially lower than anticipated, meaning that the cost equivalent analysis must extrapolate to a cost lower than anything observed in the GD transfer arms.  Second, in order to achieve the benchmarking we confine both implementers to somewhat unnatural sample selection rules.  An organization providing cash transfers would never have a reason to target only individuals who express interest in a training program.  For HD, the study constrained them to study only poor individuals (they do not normally use Ubudehe status as a targeting criterion).   We examine only a first phase of implementation of HD, and we miss any environmental benefits to the employment landscape caused by HD's capacity building and job placement work.  Also, because we induced HD to treat at a lower intensity than they normally would (they would typically have treated \textit{all} the individuals in our study), we may not have captured the effect of a program running at greater intensity.  Nonetheless, the internal variation in our sample suggests that these issues would generate limited bias in our study:  outcomes are flat across transfer amount in GD (costing error), impacts are homogeneous across a range of beneficiary characteristics (differential targeting rules), and we find no evidence of saturation effects (HD treatment intensity).  It therefore appears likely that our study has reasonable external validity.   
   
The comparative evaluation of cash and in-kind modalities in this study---together with experimentally induced variation in cash transfer sizes---allows us to speak to the cash-benchmarking question  increasingly called for in recent years.  How would beneficiary outcomes change if a standard and widespread development intervention  were simply distributed to the beneficiaries in the form of a mobile money transfer? Proponents of cash transfers have suggested that they should be considered the `index funds' of international development, meaning a benchmark to which other programs are compared \citep{blattman2014show}.  
We estimate that at transfer values that are cost-equivalent from the donor perspective, the impacts of cash transfers exceeds that of HD by a statistically significant margin on two of five primary outcomes:  monthly income and productive assets.  On the other hand, given the substantially superior performance of HD at producing the human capital measured by the business skills index, we can provide an exact exchange rate that quantifies the tradeoffs policymakers must be willing to make.  Focusing on the comparative impacts on productive assets versus business skills, to prefer HD at 18 months policymakers must be willing to forego a 200\% increase in beneficiary productive assets in order to obtain a 0.5 SD increase in human capital.
        
This approach to cash benchmarking holds donor expenditure per beneficiary constant---in effect, restricting comparisons of in-kind programs to cash transfers that could reach an equivalent number of beneficiaries with equal-sized transfers.  This need not be the optimal intensity of transfers.  A donor who seeks to maximize an additive social welfare function in any of the outcomes considered, for example, would prefer to choose programming that maximizes the benefit-cost ratio.  Summarizing across outcomes, the middle of the cash transfer sizes considered in this study seems to perform best by this metric. 
    
One of the most surprising results of our study is the lack of complementarities between GD and HD.  It is worth noting, however, that we designed the Combined arm to provide the cleanest test of structural complementarities between physical and human capital from the beneficiary side, which required that we implemented each arm precisely as in the standalone case.  In reality, however, these two programs could be interwoven in a number of deeper ways.  At the very least, regular programming intended to add capital over training would typically only do so at the end of the training, while our Combined arm received their cash midway through HD.  More fundamentally, one might think about making the cash component conditional on participation in HD, which we did not do.  So our results should not be taken to mean that it is impossible to design cash and training programs in a complementary manner, but rather than simply providing them together does not automatically generate a whole greater than the sum of the parts.  Given the many complex, expensive, and  multi-dimensional programs pushed by the donor community, this result is worth paying attention to.  

The lessons of this study add to the evidence base not just on specific programs, but on the relative importance of the underlying constraints they seek to address.  HD does appear to have been effective in alleviating a human capital constraint:  it improves business knowledge, though non-cognitive dimensions such as aspirations and beneficiaries' self-efficacy appear to have been harder to move.  In this context, the returns to raising business knowledge appear not to have been as impactful as the returns to alleviating liquidity constraints through cash transfers. And the liquidity constraints faced by these individuals appear not to be exceedingly large, given that the returns to increases in cash-transfer size are modest.  While our prior cash-benchmarking study in Rwanda \citep{mcintzeit2019gikuriro} targeted households with malnourished children, it also focused on Rwandans within Ubudehe 1 and 2, and so provides an interesting point of comparison to the results here.  That study provided evidence that transfer sizes greater than USD 150 are required to induce changes in productive outcomes in this setting, but here we find transfers larger than USD 400 have limited additional value.   This helps to identify the `sweet spot' for cash in this context.   Moreover, this study shows that the impact of human-capital and liquidity improvements flow through different channels, in particular as the HD program does move beneficiaries in the direction of wage employment, whereas cash transfers support self-employment.   Importantly, these results are at 18-months only. The long-run impact of the mechanisms induced by each of these programs remains an important question.

\clearpage
\bibliographystyle{aer}
\bibliography{HugukaDukore}

\cleardoublepage
\appendix
\noappendicestocpagenum \addappheadtotoc
\makeatletter
\def\@seccntformat#1{Appendix\ \csname the#1\endcsname\quad}
\def\@subseccntformat#1{\csname the#1\endcsname\quad}
\makeatother
\renewcommand\thetable{\Alph{section}.\arabic{table}}
\renewcommand\thefigure{\Alph{section}.\arabic{figure}}

\counterwithin{figure}{section}
\counterwithin{table}{section}

\clearpage
\section{Supplementary tables}\label{app:MoreTables}

\begin{table}[!h]
\caption{Process of identifying the eligible sample}
\label{t:vetting_recruitment}
    \begin{center}

\begin{tabular}{lrrr}
\toprule %
Sector & Orientation sign-ups & Verified eligible & Baseline completed \tabularnewline 
\midrule %
Kaduha		&	273	 &	261	&	235	\tabularnewline 
Kibumbwe	&	144	 &	139	&	127	\tabularnewline 
Kigabiro	&	66	 &	52	&	49	\tabularnewline 
Kiyumba		&	102	 &	70	&	66	\tabularnewline 
Mugano		&	244	 &	198	&	196	\tabularnewline 
Muhazi		&	192	 &	170	&	159	\tabularnewline 
Munyaga		&	157	 &	137	&	124	\tabularnewline 
Munyiginya	&	115	 &	102	&	94	\tabularnewline 
Musange		&	170	 &	115	&	110	\tabularnewline 
Mushishiro	&	88	 &	87	&	82	\tabularnewline 
Nyakariro	&	227	 &	200	&	190	\tabularnewline 
Nyarusange	&	245	 &	226	&	214	\tabularnewline 
Shyogwe		&	252	 &	210	&	202	\tabularnewline 
\midrule %
Total		&	2275 &	1967	&	1848	\tabularnewline 
\bottomrule 
\end{tabular}
    \end{center}
    \begin{footnotesize}
    \begin{flushleft} 
Notes:  Table gives the number of individuals participating in each of three phases of study recruitment, for each of the 13 sectors in which recruitment took place.
\end{flushleft}
\end{footnotesize}

\end{table}

\begin{table}[!hbtp]
\caption{Survey modules by instrument and round}
\label{t:modules}
\begin{center}

\begin{tabular}{l c c}
\toprule
Module & Baseline instrument & Endline instrument \\ 
\midrule 
Identification & Both	& Both \\
Social network & Beneficiary & --- \\
Firm creation and employment history &  Beneficiary & --- \\ 
Wage employment &  Beneficiary & Beneficiary \\  
Microenterprise activities and assets &	Both	& Both \\ 
Time use 		&	Beneficiary & Beneficiary \\ 
Income			&   Both & Beneficiary \\ 
Savings			& 	Both		& Both \\
Borrowing		& 	Both		& Both \\
Lending			&   Both		& Both \\ 
Business contacts & Beneficiary	& --- 	\\ 
Private consumption & Beneficiary & Beneficiary \\ 
Private assets	& Beneficiary & Beneficiary \\  
Psychometrics &  Beneficiary & Beneficiary \\ 
Raven's test & Beneficiary & --- \\ 
Digit-span recall & Beneficiary & --- \\ 
Numeracy & Beneficiary & --- \\ 
Lottery choice & Beneficiary & --- \\ 
Convex time budget & Beneficiary & --- \\ 
Locus of control & Beneficiary & Beneficiary \\ 
Big Five &	---	& Beneficiary \\ 
Aspirations & ---  & Beneficiary \\ 
Mental health	& --- & Beneficiary \\ 
Business knowledge & --- & Beneficiary \\ 
Business attitudes & --- & Beneficiary \\ 
Program participation & --- & Beneficiary \\ 
Gender empowerment & --- & Beneficiary \\ 
Household roster & Household	& Household \\ 
Dwelling characteristics & Household & Household \\ 
Land use and ownership		& Household & Household \\ 
Inter-household transfers 	& Household & Household \\ 
Consumption 				& Household & Household \\ 
Dietary diversity			& Household & Household \\
Household assets			& Household & Household \\ 
\bottomrule
\end{tabular}
\end{center}
\end{table}

\begin{table}[!hbp]
\caption{Correlates of HD Participation}\label{t:HD_completion}
\begin{footnotesize}
\begin{center}
\begin{tabular}{l *{4}{S[table-format=04.4]}}
\toprule
 & \multicolumn{4}{c}{Huguka Dukore stage completed\ldots} \\ 
\cmidrule(lr){2-5}
\multicolumn{1}{c}{\text{ }} & \multicolumn{1}{c}{\text{Complier}} & \multicolumn{1}{c}{\text{Work Ready Now}} & \multicolumn{1}{c}{\text{Be Your Own Boss}} & \multicolumn{1}{c}{\text{Technical Training}}\\
\midrule
\multirow[t]{ 2}{0.2\textwidth}{Ubudehe category I} &    0.0106 &    0.0146 &    0.0128 &   -0.0199 \\ 
 & (   0.0281)  & (   0.0319)  & (   0.0356)  & (   0.0327)  \\ 
\addlinespace[1ex] \multirow[t]{ 2}{0.2\textwidth}{Beneficiary female} &    0.0142 &    0.0146 &    0.0284 &   -0.0187 \\ 
 & (   0.0285)  & (   0.0317)  & (   0.0363)  & (   0.0336)  \\ 
\addlinespace[1ex] \multirow[t]{ 2}{0.2\textwidth}{Beneficiary age} &    0.0120\ensuremath{^{***}} &    0.0100\ensuremath{^{**}} &    0.0085\ensuremath{^{*}} &    0.0067 \\ 
 & (   0.0041)  & (   0.0048)  & (   0.0052)  & (   0.0042)  \\ 
\addlinespace[1ex] \multirow[t]{ 2}{0.2\textwidth}{Beneficiary years of education} &    0.0037 &    0.0051 &    0.0050 &    0.0073 \\ 
 & (   0.0061)  & (   0.0071)  & (   0.0083)  & (   0.0073)  \\ 
\addlinespace[1ex] \multirow[t]{ 2}{0.2\textwidth}{Household members} &    0.0025 &    0.0061 &    0.0083 &    0.0146\ensuremath{^{*}} \\ 
 & (   0.0074)  & (   0.0078)  & (   0.0085)  & (   0.0079)  \\ 
\addlinespace[1ex] \multirow[t]{ 2}{0.2\textwidth}{Employed} &    0.0930\ensuremath{^{*}} &    0.0757 &    0.0846 &    0.0450 \\ 
 & (   0.0504)  & (   0.0555)  & (   0.0615)  & (   0.0590)  \\ 
\addlinespace[1ex] \multirow[t]{ 2}{0.2\textwidth}{Productive hours} &   -0.0038\ensuremath{^{***}} &   -0.0041\ensuremath{^{***}} &   -0.0047\ensuremath{^{***}} &   -0.0044\ensuremath{^{***}} \\ 
 & (   0.0014)  & (   0.0015)  & (   0.0015)  & (   0.0014)  \\ 
\addlinespace[1ex] \multirow[t]{ 2}{0.2\textwidth}{Monthly income} &   -0.0019 &   -0.0008 &    0.0004 &   -0.0026 \\ 
 & (   0.0045)  & (   0.0051)  & (   0.0055)  & (   0.0050)  \\ 
\addlinespace[1ex] \multirow[t]{ 2}{0.2\textwidth}{Productive assets} &   -0.0001 &    0.0004 &   -0.0016 &   -0.0027 \\ 
 & (   0.0030)  & (   0.0037)  & (   0.0041)  & (   0.0040)  \\ 
\addlinespace[1ex] \multirow[t]{ 2}{0.2\textwidth}{HH consumption per capita} &    0.0027 &    0.0042 &    0.0103 &    0.0157 \\ 
 & (   0.0134)  & (   0.0152)  & (   0.0179)  & (   0.0154)  \\ 
\addlinespace[1ex] \multirow[t]{ 2}{0.2\textwidth}{Beneficiary-specific consumption} &   -0.0092\ensuremath{^{*}} &   -0.0139\ensuremath{^{**}} &   -0.0109\ensuremath{^{*}} &   -0.0021 \\ 
 & (   0.0053)  & (   0.0059)  & (   0.0064)  & (   0.0058)  \\ 
\addlinespace[1ex] \multirow[t]{ 2}{0.2\textwidth}{HH net non-land wealth} &    0.0030 &    0.0040\ensuremath{^{*}} &    0.0050\ensuremath{^{*}} &    0.0026 \\ 
 & (   0.0023)  & (   0.0024)  & (   0.0028)  & (   0.0024)  \\ 
\addlinespace[1ex] \multirow[t]{ 2}{0.2\textwidth}{Savings} &   -0.0018 &   -0.0028 &   -0.0023 &   -0.0030 \\ 
 & (   0.0033)  & (   0.0036)  & (   0.0041)  & (   0.0036)  \\ 
\addlinespace[1ex] \multirow[t]{ 2}{0.2\textwidth}{Debt} &    0.0080\ensuremath{^{**}} &    0.0099\ensuremath{^{***}} &    0.0096\ensuremath{^{**}} &    0.0086\ensuremath{^{**}} \\ 
 & (   0.0034)  & (   0.0038)  & (   0.0042)  & (   0.0038)  \\ 
\addlinespace[1ex] \multirow[t]{ 2}{0.2\textwidth}{HH livestock wealth} &    0.0025 &    0.0025 &   -0.0003 &   -0.0034 \\ 
 & (   0.0026)  & (   0.0030)  & (   0.0033)  & (   0.0030)  \\ 
\addlinespace[1ex] \multirow[t]{ 2}{0.2\textwidth}{Business Knowledge} &   -0.0129 &    0.0056 &    0.0019 &   -0.0151 \\ 
 & (   0.0135)  & (   0.0154)  & (   0.0179)  & (   0.0156)  \\ 
\addlinespace[1ex] Average completion rate  &      0.86 &      0.79 &      0.69 &      0.48 \\ 
Observations  & \multicolumn{1}{c}{      668}  & \multicolumn{1}{c}{      668}  & \multicolumn{1}{c}{      668}  & \multicolumn{1}{c}{      668}  \\ 
$ R^2$  &      0.12 &      0.16 &      0.18 &      0.47 \\ 
$ p$-value  &      0.00 &      0.00 &      0.01 &      0.00 \\ 
\addlinespace[1ex] 
\bottomrule
\end{tabular}

\end{center}
\floatfoot{
Notes:  Table estimates Linear Probability Model for four measures of progression through the Huguka Dukore program:  attending the first week of the WRN coursework and hence triggering payment to the implementer, then completing each of the three subsequent components of the program.  Rows are the baseline covariates over which we look for heterogeneity in these compliance rates.  p-value in the final row is from F-test on the joint significance of all of the covariates.  Standard errors in parentheses;  *=10\%, **=5\%, and ***=1\% significance
}
\end{footnotesize}
\end{table}

\begin{table}[!hbp]
\caption{Sampling of Attritors for Intensive Tracking Exercise}\label{t:IntensiveTracking}
\vskip-1ex
\begin{footnotesize}
\begin{center}
\begin{tabular}{l *{1}{S} ScSS}
\toprule
\multicolumn{1}{c}{\text{ }} & \multicolumn{1}{c}{\text{Intensive tracking}} & \multicolumn{1}{c}{\text{Control mean}} & \multicolumn{1}{c}{\text{Observations}} & \multicolumn{1}{c}{\text{$ R^2$}}\\
\midrule
\multirow[t]{ 3 }{0.2\textwidth}{Ubudehe category I }  &     -0.10 &      0.38 &       120 &      0.01 \\ 
 & (0.09)  \\  & [    1.00]  \\ \addlinespace[1ex] 
\multirow[t]{ 3 }{0.2\textwidth}{Beneficiary female }  &     -0.00 &      0.62 &       122 &      0.00 \\ 
 & (0.09)  \\  & [    1.00]  \\ \addlinespace[1ex] 
\multirow[t]{ 3 }{0.2\textwidth}{Beneficiary age }  &      1.66 &     22.64 &       122 &      0.06 \\ 
 & (0.61)  \\  & [    0.13]  \\ \addlinespace[1ex] 
\multirow[t]{ 3 }{0.2\textwidth}{Beneficiary years of education }  &      0.07 &      8.10 &       122 &      0.00 \\ 
 & (0.45)  \\  & [    1.00]  \\ \addlinespace[1ex] 
\multirow[t]{ 3 }{0.2\textwidth}{Household members }  &     -0.25 &      4.66 &       122 &      0.00 \\ 
 & (0.37)  \\  & [    1.00]  \\ \addlinespace[1ex] 
\multirow[t]{ 3 }{0.2\textwidth}{Employed }  &     -0.00 &      0.31 &       122 &      0.00 \\ 
 & (0.09)  \\  & [    1.00]  \\ \addlinespace[1ex] 
\multirow[t]{ 3 }{0.2\textwidth}{Productive hours }  &      2.48 &     11.75 &       122 &      0.00 \\ 
 & (4.09)  \\  & [    1.00]  \\ \addlinespace[1ex] 
\multirow[t]{ 3 }{0.2\textwidth}{Monthly income }  &     -0.41 &      4.15 &       122 &      0.00 \\ 
 & (0.91)  \\  & [    1.00]  \\ \addlinespace[1ex] 
\multirow[t]{ 3 }{0.2\textwidth}{Productive assets }  &      0.65 &      2.59 &       122 &      0.00 \\ 
 & (0.85)  \\  & [    1.00]  \\ \addlinespace[1ex] 
\multirow[t]{ 3 }{0.2\textwidth}{HH consumption per capita }  &      0.14 &      9.44 &       122 &      0.00 \\ 
 & (0.20)  \\  & [    1.00]  \\ \addlinespace[1ex] 
\multirow[t]{ 3 }{0.2\textwidth}{Beneficiary-specific consumption }  &      0.08 &      7.61 &       122 &      0.00 \\ 
 & (0.36)  \\  & [    1.00]  \\ \addlinespace[1ex] 
\multirow[t]{ 3 }{0.2\textwidth}{HH net non-land wealth }  &      0.95 &     10.48 &       122 &      0.01 \\ 
 & (1.02)  \\  & [    1.00]  \\ \addlinespace[1ex] 
\multirow[t]{ 3 }{0.2\textwidth}{Savings }  &      0.42 &      7.67 &       122 &      0.00 \\ 
 & (0.83)  \\  & [    1.00]  \\ \addlinespace[1ex] 
\multirow[t]{ 3 }{0.2\textwidth}{Debt }  &     -0.21 &      7.60 &       122 &      0.00 \\ 
 & (0.86)  \\  & [    1.00]  \\ \addlinespace[1ex] 
\multirow[t]{ 3 }{0.2\textwidth}{HH livestock wealth }  &      2.22 &      5.26 &       122 &      0.04 \\ 
 & (1.09)  \\  & [    0.37]  \\ \addlinespace[1ex] 
\multirow[t]{ 3 }{0.2\textwidth}{Business Knowledge }  &      0.34 &     -0.02 &       122 &      0.03 \\ 
 & (0.18)  \\  & [    0.37]  \\ \addlinespace[1ex] 
\bottomrule
\end{tabular}

\end{center}
\vskip-5ex
\floatfoot{
Notes:  Table regresses a sequence of covariates on a dummy variable for having been sampled for intensive tracking, within the sample of original attritors.  Standard errors in parentheses, p-values corrected for False Discovery Rate across whole table in hard brackets; stars are based on FDR-adjusted values,  *=10\%, **=5\%, and ***=1\% significance.
}
\end{footnotesize}
\end{table}

\cleardoublepage 

    \global\pdfpageattr\expandafter{\the\pdfpageattr/Rotate 90}
    \newgeometry{margin=0.4in}
    \begin{landscape}
    
        \vspace*{\fill}

    \begin{table}[!hbtp]
\caption{Intent-to-treat analysis: Primary outcomes, aggregated specification}
\label{t:itt_primary_pooled}

\begin{footnotesize}
\begin{center}
\begin{tabular}{l *{4}{S} ScSSSS}
\toprule
 & & \multicolumn{2}{c}{GiveDirectly} & & \multicolumn{1}{c}{\raisebox{-1ex}[0pt]{Control}} & & & \multicolumn{3}{c}{$ p$-values} \\ 
\cmidrule(lr){3-4} \cmidrule(lr){9-11}
\multicolumn{1}{c}{\text{ }} & \multicolumn{1}{c}{\text{HD}} & \multicolumn{1}{c}{\text{Main}} & \multicolumn{1}{c}{\text{Large}} & \multicolumn{1}{c}{\text{Combined}} & \multicolumn{1}{c}{\text{Mean}} & \multicolumn{1}{c}{\text{Obs.}} & \multicolumn{1}{c}{\text{$ R^2$}} & \multicolumn{1}{c}{\text{(a)}} & \multicolumn{1}{c}{\text{(b)}} & \multicolumn{1}{c}{\text{(c)}}\\
\midrule
\multirow[t]{ 3 }{0.15\textwidth}{Employed }  &      0.02 &      0.03 &      0.01 &      0.01 &      0.48 &      1770 &      0.16 &      0.89 &      0.52 &      0.46 \\ 
 & (0.03)  & (0.03)  & (0.05)  & (0.04)  \\  & [    0.29]  & [    0.24]  & [    0.51]  & [    0.51]  \\ \addlinespace[1ex] 
\multirow[t]{ 3 }{0.15\textwidth}{Productive hours }  &      2.80\ensuremath{^{*}} &      4.34\ensuremath{^{***}} &      1.12 &      2.33 &     18.64 &      1770 &      0.18 &      0.93 &      0.03 &      0.07 \\ 
 & (1.57)  & (1.65)  & (2.06)  & (2.03)  \\  & [    0.07]  & [    0.01]  & [    0.35]  & [    0.17]  \\ \addlinespace[1ex] 
\multirow[t]{ 3 }{0.15\textwidth}{Monthly income }  &      0.32 &      1.00\ensuremath{^{***}} &      0.73\ensuremath{^{**}} &      1.04\ensuremath{^{***}} &      8.05 &      1770 &      0.21 &      0.11 &      0.02 &      0.10 \\ 
 & (0.26)  & (0.25)  & (0.35)  & (0.32)  \\  & [    0.16]  & [    0.00]  & [    0.03]  & [    0.00]  \\ \addlinespace[1ex] 
\multirow[t]{ 3 }{0.15\textwidth}{Productive assets }  &      1.54\ensuremath{^{***}} &      3.86\ensuremath{^{***}} &      4.02\ensuremath{^{***}} &      4.42\ensuremath{^{***}} &      5.61 &      1770 &      0.26 &      0.00 &      0.00 &      0.00 \\ 
 & (0.35)  & (0.34)  & (0.47)  & (0.44)  \\  & [    0.00]  & [    0.00]  & [    0.00]  & [    0.00]  \\ \addlinespace[1ex] 
\multirow[t]{ 3 }{0.15\textwidth}{HH consumption per capita }  &      0.05 &      0.23\ensuremath{^{***}} &      0.36\ensuremath{^{***}} &      0.27\ensuremath{^{***}} &      9.46 &      1737 &      0.33 &      0.05 &      0.64 &      0.17 \\ 
 & (0.06)  & (0.06)  & (0.07)  & (0.07)  \\  & [    0.23]  & [    0.00]  & [    0.00]  & [    0.00]  \\ \addlinespace[1ex] 
\bottomrule
\end{tabular}

\end{center}
\vskip-1ex
\floatfoot{
 Note:  Intention to treat pooling the three smaller cash transfer amounts into a single arm, GD Main.  Regressions include but do not report the lagged dependent variable, fixed effects for randomization blocks, and a set of LASSO-selected baseline covariates, and are weighted to reflect intensive tracking. Standard errors are (in soft brackets) are clustered at the household level to reflect the design effect, and $p$-values corrected for False Discovery Rates across all the outcomes in the table are presented in hard brackets.  Stars on coefficient estimates are derived from the FDR-corrected $p$-values, *=10\%, **=5\%, and ***=1\% significance. $p$-values in final three columns derived from $F$-tests of hypotheses that benefit-cost ratios are equal between (a) GD Main and HD; (b) GD Main and GD Large; and (c) GD Main and Combined. Employed is a dummy variable for spending more than 10 hours per week working for a wage or as primary operator of a microenterprise.  Productive hours are measured over prior 7 days in all activities other than own-farm agriculture.  Monthly income, productive assets, and household consumption are winsorized at 1\% and 99\% and analyzed in Inverse Hyperbolic Sine, meaning that treatment effects can be interpreted as percent changes.  
}
\end{footnotesize}
\end{table}

\vspace*{\fill}

\begin{table}[!h]
\caption{Intent-to-treat analysis: Secondary outcomes, aggregated specification}
\label{t:itt_secondary_pooled}
\vskip-2ex
\begin{footnotesize}
\begin{center}
\begin{tabular}{l *{4}{S} ScSSSS}
\toprule
 & & \multicolumn{2}{c}{GiveDirectly} & & \multicolumn{1}{c}{\raisebox{-1ex}[0pt]{Control}} & & & \multicolumn{3}{c}{$ p$-values} \\ 
\cmidrule(lr){3-4} \cmidrule(lr){9-11}
\multicolumn{1}{c}{\text{ }} & \multicolumn{1}{c}{\text{HD}} & \multicolumn{1}{c}{\text{Main}} & \multicolumn{1}{c}{\text{Large}} & \multicolumn{1}{c}{\text{Combined}} & \multicolumn{1}{c}{\text{Mean}} & \multicolumn{1}{c}{\text{Obs.}} & \multicolumn{1}{c}{\text{$ R^2$}} & \multicolumn{1}{c}{\text{(a)}} & \multicolumn{1}{c}{\text{(b)}} & \multicolumn{1}{c}{\text{(c)}}\\
\midrule
\multicolumn{11}{l}{\emph{Panel A. Beneficiary welfare}}  \\ 
\addlinespace[1ex] \multirow[t]{ 3 }{0.15\textwidth}{Subjective well-being }  &      0.19\ensuremath{^{***}} &      0.47\ensuremath{^{***}} &      0.55\ensuremath{^{***}} &      0.41\ensuremath{^{***}} &      0.00 &      1770 &      0.13 &      0.03 &      0.03 &      0.00 \\ 
 & (0.07)  & (0.07)  & (0.09)  & (0.09)  \\  & [    0.00]  & [    0.00]  & [    0.00]  & [    0.00]  \\ \addlinespace[1ex] 
\multirow[t]{ 3 }{0.15\textwidth}{Mental health }  &     -0.04 &      0.01 &      0.11 &      0.12 &      0.00 &      1770 &      0.07 &      0.43 &      0.48 &      0.40 \\ 
 & (0.07)  & (0.07)  & (0.10)  & (0.09)  \\  & [    0.34]  & [    0.34]  & [    0.16]  & [    0.12]  \\ \addlinespace[1ex] 
\multirow[t]{ 3 }{0.15\textwidth}{Beneficiary-specific consumption }  &      0.15 &      0.58\ensuremath{^{***}} &      0.45\ensuremath{^{***}} &      0.69\ensuremath{^{***}} &      8.27 &      1770 &      0.23 &      0.01 &      0.00 &      0.02 \\ 
 & (0.11)  & (0.10)  & (0.15)  & (0.12)  \\  & [    0.12]  & [    0.00]  & [    0.00]  & [    0.00]  \\ \addlinespace[1ex] 
\multicolumn{11}{l}{\emph{Panel B. Household wealth}}  \\ 
\addlinespace[1ex] \multirow[t]{ 3 }{0.15\textwidth}{HH net non-land wealth }  &     -0.17 &      0.91\ensuremath{^{***}} &      1.11\ensuremath{^{***}} &      0.90\ensuremath{^{**}} &     11.28 &      1770 &      0.20 &      0.02 &      0.41 &      0.28 \\ 
 & (0.40)  & (0.35)  & (0.41)  & (0.48)  \\  & [    0.31]  & [    0.01]  & [    0.01]  & [    0.05]  \\ \addlinespace[1ex] 
\multirow[t]{ 3 }{0.15\textwidth}{HH livestock wealth }  &     -0.00 &      2.08\ensuremath{^{***}} &      2.17\ensuremath{^{***}} &      2.22\ensuremath{^{***}} &      7.81 &      1770 &      0.25 &      0.00 &      0.02 &      0.02 \\ 
 & (0.37)  & (0.35)  & (0.47)  & (0.45)  \\  & [    0.46]  & [    0.00]  & [    0.00]  & [    0.00]  \\ \addlinespace[1ex] 
\multirow[t]{ 3 }{0.15\textwidth}{Savings }  &      1.04\ensuremath{^{***}} &      1.30\ensuremath{^{***}} &      1.43\ensuremath{^{***}} &      1.70\ensuremath{^{***}} &      9.24 &      1770 &      0.20 &      0.40 &      0.04 &      0.15 \\ 
 & (0.23)  & (0.23)  & (0.31)  & (0.27)  \\  & [    0.00]  & [    0.00]  & [    0.00]  & [    0.00]  \\ \addlinespace[1ex] 
\multirow[t]{ 3 }{0.15\textwidth}{Debt }  &      0.40 &     -0.29 &     -0.37 &      0.00 &      8.75 &      1770 &      0.20 &      0.02 &      0.80 &      0.34 \\ 
 & (0.28)  & (0.30)  & (0.42)  & (0.38)  \\  & [    0.10]  & [    0.19]  & [    0.22]  & [    0.46]  \\ \addlinespace[1ex] 
\multicolumn{11}{l}{\emph{Panel C. Beneficiary cognitive and non-cognitive skills}}  \\ 
\addlinespace[1ex] \multirow[t]{ 3 }{0.15\textwidth}{Locus of control }  &      0.06 &      0.05 &      0.08 &      0.23\ensuremath{^{**}} &      0.00 &      1770 &      0.27 &      0.58 &      0.98 &      0.14 \\ 
 & (0.06)  & (0.06)  & (0.08)  & (0.08)  \\  & [    0.45]  & [    0.53]  & [    0.45]  & [    0.02]  \\ \addlinespace[1ex] 
\multirow[t]{ 3 }{0.15\textwidth}{Aspirations }  &     -0.01 &      0.05 &      0.03 &      0.14 &      0.00 &      1770 &      0.08 &      0.52 &      0.60 &      0.63 \\ 
 & (0.07)  & (0.07)  & (0.09)  & (0.08)  \\  & [    1.00]  & [    0.54]  & [    1.00]  & [    0.20]  \\ \addlinespace[1ex] 
\multirow[t]{ 3 }{0.15\textwidth}{Big Five index }  &      0.12 &      0.07 &     -0.08 &      0.02 &      0.00 &      1770 &      0.10 &      0.23 &      0.10 &      0.40 \\ 
 & (0.07)  & (0.07)  & (0.09)  & (0.09)  \\  & [    0.20]  & [    0.45]  & [    0.53]  & [    1.00]  \\ \addlinespace[1ex] 
\multirow[t]{ 3 }{0.15\textwidth}{Business knowledge }  &      0.65\ensuremath{^{***}} &      0.08 &     -0.03 &      0.63\ensuremath{^{***}} &      0.00 &      1770 &      0.23 &      0.00 &      0.19 &      0.00 \\ 
 & (0.07)  & (0.07)  & (0.09)  & (0.09)  \\  & [    0.00]  & [    0.45]  & [    1.00]  & [    0.00]  \\ \addlinespace[1ex] 
\multirow[t]{ 3 }{0.15\textwidth}{Business attitudes }  &      0.12 &      0.16\ensuremath{^{*}} &      0.06 &      0.15 &      0.00 &      1770 &      0.09 &      0.81 &      0.06 &      0.26 \\ 
 & (0.07)  & (0.06)  & (0.09)  & (0.09)  \\  & [    0.20]  & [    0.06]  & [    0.65]  & [    0.20]  \\ \addlinespace[1ex] 
\bottomrule
\end{tabular}

\end{center}
\vskip-6ex
\floatfoot{ 
Notes:   Regressions include but do not report the lagged dependent variable, fixed effects for randomization blocks, and a set of LASSO-selected baseline covariates, and are weighted to reflect intensive tracking. Standard errors are (in soft brackets) are clustered at the household level to reflect the design effect, and p-values corrected for False Discovery Rates across all the outcomes in the table are presented in hard brackets.  Stars on coefficient estimates are derived from the FDR-corrected p-values, *=10\%, **=5\%, and ***=1\% significance.
}
\end{footnotesize}
\end{table}


    
    \begin{table}[!h]
    \caption{Breakdown of employment impacts}
    \label{t:employment_details}
    \begin{footnotesize}
    \begin{center}
    \begin{tabular}{l *{6}{S} ScSS}
\toprule
 & & \multicolumn{4}{c}{GiveDirectly} & & \multicolumn{1}{c}{\raisebox{-1ex}[0pt]{Control}} \\ 
\cmidrule(lr){3-6}
\multicolumn{1}{c}{\text{ }} & \multicolumn{1}{c}{\text{HD}} & \multicolumn{1}{c}{\text{Lower}} & \multicolumn{1}{c}{\text{Middle}} & \multicolumn{1}{c}{\text{Upper}} & \multicolumn{1}{c}{\text{Large}} & \multicolumn{1}{c}{\text{Combined}} & \multicolumn{1}{c}{\text{Mean}} & \multicolumn{1}{c}{\text{Obs.}} & \multicolumn{1}{c}{\text{$ R^2$}} & \multicolumn{1}{c}{\text{$ p$-value}}\\
\midrule
\multicolumn{11}{l}{\emph{Panel A. Employment composition}}  \\ 
\addlinespace[1ex] \multirow[t]{ 3 }{0.2\textwidth}{Non-agricultural microenterprise }  &      0.05 &      0.09\ensuremath{^{*}} &      0.08\ensuremath{^{*}} &      0.14\ensuremath{^{**}} &      0.17\ensuremath{^{***}} &      0.12\ensuremath{^{**}} &      0.22 &      1770 &      0.12 &      0.00 \\ 
 & (0.03)  & (0.04)  & (0.04)  & (0.05)  & (0.04)  & (0.04)  \\  & [    0.11]  & [    0.06]  & [    0.08]  & [    0.02]  & [    0.00]  & [    0.02]  \\ \addlinespace[1ex] 
\multirow[t]{ 3 }{0.2\textwidth}{Other microenterprise or self-employment }  &      0.04\ensuremath{^{*}} &      0.04 &      0.11\ensuremath{^{**}} &      0.07\ensuremath{^{*}} &      0.05\ensuremath{^{*}} &      0.04 &      0.07 &      1770 &      0.09 &      0.02 \\ 
 & (0.02)  & (0.03)  & (0.03)  & (0.03)  & (0.03)  & (0.03)  \\  & [    0.06]  & [    0.11]  & [    0.02]  & [    0.06]  & [    0.09]  & [    0.11]  \\ \addlinespace[1ex] 
\multirow[t]{ 3 }{0.2\textwidth}{Agricultural processing or trading }  &      0.01 &      0.13\ensuremath{^{**}} &      0.01 &      0.07\ensuremath{^{*}} &      0.06 &      0.02 &      0.17 &      1770 &      0.08 &      0.05 \\ 
 & (0.03)  & (0.04)  & (0.04)  & (0.04)  & (0.04)  & (0.04)  \\  & [    0.24]  & [    0.02]  & [    0.24]  & [    0.08]  & [    0.11]  & [    0.24]  \\ \addlinespace[1ex] 
\multirow[t]{ 3 }{0.2\textwidth}{Agricultural wage labor }  &     -0.02 &     -0.08\ensuremath{^{**}} &     -0.07\ensuremath{^{*}} &     -0.08\ensuremath{^{**}} &     -0.11\ensuremath{^{**}} &     -0.09\ensuremath{^{**}} &      0.22 &      1770 &      0.15 &      0.00 \\ 
 & (0.03)  & (0.03)  & (0.03)  & (0.03)  & (0.03)  & (0.03)  \\  & [    0.24]  & [    0.03]  & [    0.06]  & [    0.03]  & [    0.02]  & [    0.03]  \\ \addlinespace[1ex] 
\multirow[t]{ 3 }{0.2\textwidth}{Non-agricultural wage labor }  &      0.06\ensuremath{^{*}} &     -0.04 &      0.01 &     -0.06 &     -0.07\ensuremath{^{*}} &      0.01 &      0.30 &      1770 &      0.20 &      0.00 \\ 
 & (0.03)  & (0.04)  & (0.04)  & (0.04)  & (0.04)  & (0.04)  \\  & [    0.06]  & [    0.16]  & [    0.24]  & [    0.11]  & [    0.09]  & [    0.24]  \\ \addlinespace[1ex] 
\multicolumn{11}{l}{\emph{Panel B. Alternative hours thresholds}}  \\ 
\addlinespace[1ex] \multirow[t]{ 3 }{0.2\textwidth}{Employed (0 hr) }  &      0.05 &      0.05 &      0.07 &      0.02 &      0.06 &      0.06 &      0.70 &      1770 &      0.12 &      0.46 \\ 
 & (0.03)  & (0.04)  & (0.04)  & (0.04)  & (0.04)  & (0.04)  \\  & [    1.00]  & [    1.00]  & [    1.00]  & [    1.00]  & [    1.00]  & [    1.00]  \\ \addlinespace[1ex] 
\multirow[t]{ 3 }{0.2\textwidth}{Employed (10 hr) }  &      0.02 &      0.03 &      0.05 &      0.00 &      0.01 &      0.01 &      0.48 &      1770 &      0.16 &      0.95 \\ 
 & (0.03)  & (0.05)  & (0.05)  & (0.05)  & (0.05)  & (0.04)  \\  & [    1.00]  & [    1.00]  & [    1.00]  & [    1.00]  & [    1.00]  & [    1.00]  \\ \addlinespace[1ex] 
\multirow[t]{ 3 }{0.2\textwidth}{Employed (20 hr) }  &      0.04 &      0.04 &      0.08 &      0.02 &      0.01 &      0.04 &      0.29 &      1770 &      0.17 &      0.62 \\ 
 & (0.03)  & (0.04)  & (0.04)  & (0.04)  & (0.04)  & (0.04)  \\  & [    1.00]  & [    1.00]  & [    1.00]  & [    1.00]  & [    1.00]  & [    1.00]  \\ \addlinespace[1ex] 
\multirow[t]{ 3 }{0.2\textwidth}{Employed (30 hr) }  &      0.02 &      0.03 &      0.09 &      0.04 &      0.00 &      0.06 &      0.19 &      1770 &      0.17 &      0.37 \\ 
 & (0.03)  & (0.04)  & (0.04)  & (0.04)  & (0.04)  & (0.04)  \\  & [    1.00]  & [    1.00]  & [    1.00]  & [    1.00]  & [    1.00]  & [    1.00]  \\ \addlinespace[1ex] 
\multirow[t]{ 3 }{0.2\textwidth}{Employed (40 hr) }  &      0.03 &      0.03 &      0.09 &     -0.01 &      0.02 &      0.04 &      0.13 &      1770 &      0.17 &      0.26 \\ 
 & (0.02)  & (0.03)  & (0.04)  & (0.03)  & (0.03)  & (0.03)  \\  & [    1.00]  & [    1.00]  & [    0.70]  & [    1.00]  & [    1.00]  & [    1.00]  \\ \addlinespace[1ex] 
\bottomrule
\end{tabular}

    \end{center}
    \vskip-2ex
    \floatfoot{
        Notes: Panel A presents impacts on indicators for employment of any hours in the corresponding activity type in the preceding week.  Panel B presents impacts on an indicator for overall employment, using the reported threshold for minimum hours. Regressions include but do not report an indicator for lagged employment status, fixed effects for randomization blocks, and a set of LASSO-selected baseline covariates, and are weighted to reflect intensive tracking. Standard errors are (in soft brackets) are clustered at the household level to reflect the design effect, and p-values corrected for False Discovery Rates across outcomes in each panel are presented in hard brackets.  Stars on coefficient estimates are derived from the FDR-corrected p-values, *=10\%, **=5\%, and ***=1\% significance.
    }
    \end{footnotesize}
    \end{table}
    \end{landscape}
    \clearpage
    \global\pdfpageattr\expandafter{\the\pdfpageattr/Rotate 0}
    \restoregeometry

\begin{table}[!h]\caption{Heterogeneity: Gender}\label{t:het_gender}
\begin{footnotesize}
\begin{center}
\begin{tabular}{l *{5}{S}}
\toprule
\multicolumn{1}{c}{\text{ }} & \multicolumn{1}{c}{\text{Employed}} & \multicolumn{1}{c}{\text{\makecell[b]{Productive\\Hours}}} & \multicolumn{1}{c}{\text{\makecell[b]{Monthly\\Income}}} & \multicolumn{1}{c}{\text{\makecell[b]{Productive\\Assets}}} & \multicolumn{1}{c}{\text{Consumption}}\\
\midrule
\multirow[t]{ 2}{0.2\textwidth}{HD} &      0.09 &      1.43 &      0.14 &      1.03 &      0.01 \\ 
 & (0.05)  & (2.78)  & (0.37)  & (0.61)  & (0.10)  \\ 
 & [    0.30]  & [    0.96]  & [    0.96]  & [    0.30]  & [    1.00]  \\ 
\addlinespace[1ex] \multirow[t]{ 2}{0.2\textwidth}{GD main} &      0.05 &      4.44 &      0.79\ensuremath{^{*}} &      2.97\ensuremath{^{***}} &      0.29\ensuremath{^{**}} \\ 
 & (0.05)  & (2.98)  & (0.33)  & (0.59)  & (0.10)  \\ 
 & [    0.71]  & [    0.41]  & [    0.06]  & [    0.00]  & [    0.02]  \\ 
\addlinespace[1ex] \multirow[t]{ 2}{0.2\textwidth}{GD large} &      0.04 &     -0.08 &      0.62 &      3.18\ensuremath{^{***}} &      0.36\ensuremath{^{**}} \\ 
 & (0.07)  & (3.62)  & (0.51)  & (0.77)  & (0.12)  \\ 
 & [    0.88]  & [    1.00]  & [    0.52]  & [    0.00]  & [    0.01]  \\ 
\addlinespace[1ex] \multirow[t]{ 2}{0.2\textwidth}{Combined} &      0.05 &      3.55 &      0.59 &      3.50\ensuremath{^{***}} &      0.37\ensuremath{^{**}} \\ 
 & (0.06)  & (3.67)  & (0.43)  & (0.71)  & (0.13)  \\ 
 & [    0.75]  & [    0.71]  & [    0.44]  & [    0.00]  & [    0.02]  \\ 
\addlinespace[1ex] \multirow[t]{ 2}{0.2\textwidth}{HD $\times$ Female} &     -0.12 &      2.16 &      0.02 &      0.32 &      0.11 \\ 
 & (0.07)  & (3.35)  & (0.50)  & (0.75)  & (0.13)  \\ 
 & [    0.30]  & [    0.88]  & [    1.00]  & [    0.96]  & [    0.71]  \\ 
\addlinespace[1ex] \multirow[t]{ 2}{0.2\textwidth}{GD main $\times$ Female} &     -0.05 &     -0.32 &     -0.07 &      1.17 &     -0.06 \\ 
 & (0.07)  & (3.53)  & (0.48)  & (0.73)  & (0.12)  \\ 
 & [    0.75]  & [    1.00]  & [    1.00]  & [    0.33]  & [    0.96]  \\ 
\addlinespace[1ex] \multirow[t]{ 2}{0.2\textwidth}{GD large $\times$ Female} &     -0.04 &      1.95 &      0.28 &      1.02 &     -0.01 \\ 
 & (0.09)  & (4.29)  & (0.68)  & (0.99)  & (0.15)  \\ 
 & [    0.96]  & [    0.96]  & [    0.96]  & [    0.70]  & [    1.00]  \\ 
\addlinespace[1ex] \multirow[t]{ 2}{0.2\textwidth}{Combined $\times$ Female} &     -0.08 &     -2.65 &      0.61 &      1.25 &     -0.21 \\ 
 & (0.09)  & (4.30)  & (0.60)  & (0.89)  & (0.16)  \\ 
 & [    0.71]  & [    0.88]  & [    0.70]  & [    0.44]  & [    0.49]  \\ 
\addlinespace[1ex] \multirow[t]{ 2}{0.2\textwidth}{Female} &     -0.16\ensuremath{^{***}} &    -13.87\ensuremath{^{***}} &     -1.80\ensuremath{^{***}} &     -1.32\ensuremath{^{*}} &      0.03 \\ 
 & (0.05)  & (2.26)  & (0.35)  & (0.53)  & (0.09)  \\ 
 & [    0.01]  & [    0.00]  & [    0.00]  & [    0.05]  & [    0.96]  \\ 
\addlinespace[1ex] Control mean  &      0.48 &     18.64 &      8.05 &      5.61 &      9.46 \\ 
Observations  & \multicolumn{1}{c}{     1770}  & \multicolumn{1}{c}{     1770}  & \multicolumn{1}{c}{     1770}  & \multicolumn{1}{c}{     1770}  & \multicolumn{1}{c}{     1737}  \\ 
$ R^2$  &      0.06 &      0.10 &      0.07 &      0.11 &      0.10 \\ 
$ p$-value  &      0.53 &      0.84 &      0.81 &      0.43 &      0.38 \\ 
\addlinespace[1ex] 
\bottomrule
\end{tabular}

\end{center}
\floatfoot{
Notes: Table presents tests for heterogeneity of treatment effects by Gender.  Uninteracted coefficients in the first four rows give the treatment effect of the program on men, and the next four rows test for the differential effect between women and men. Standard errors are (in soft brackets) are clustered at the household level to reflect the design effect, and p-values corrected for False Discovery Rates across all the outcomes in the table are presented in hard brackets.  Stars on coefficient estimates are derived from the FDR-corrected p-values, *=10\%, **=5\%, and ***=1\% significance.  $p$-value in the last row from an F-test on whether treatments have a jointly differential effect by gender.  
}
\end{footnotesize}
\end{table}

\clearpage 

\begin{table}[!h]\caption{Heterogeneity: Risk aversion}\label{t:het_risk}
\begin{footnotesize}
\begin{center}
\begin{tabular}{l *{5}{S}}
\toprule
\multicolumn{1}{c}{\text{ }} & \multicolumn{1}{c}{\text{Employed}} & \multicolumn{1}{c}{\text{\makecell[b]{Productive\\Hours}}} & \multicolumn{1}{c}{\text{\makecell[b]{Monthly\\Income}}} & \multicolumn{1}{c}{\text{\makecell[b]{Productive\\Assets}}} & \multicolumn{1}{c}{\text{Consumption}}\\
\midrule
\multirow[t]{ 2}{0.2\textwidth}{HD} &      0.01 &      2.64 &      0.12 &      1.21\ensuremath{^{***}} &      0.08 \\ 
 & (0.03)  & (1.62)  & (0.27)  & (0.37)  & (0.06)  \\ 
 & [    0.84]  & [    0.23]  & [    0.84]  & [    0.01]  & [    0.40]  \\ 
\addlinespace[1ex] \multirow[t]{ 2}{0.2\textwidth}{GD main} &      0.01 &      4.25\ensuremath{^{**}} &      0.74\ensuremath{^{**}} &      3.68\ensuremath{^{***}} &      0.25\ensuremath{^{***}} \\ 
 & (0.03)  & (1.70)  & (0.26)  & (0.36)  & (0.06)  \\ 
 & [    0.84]  & [    0.05]  & [    0.02]  & [    0.00]  & [    0.00]  \\ 
\addlinespace[1ex] \multirow[t]{ 2}{0.2\textwidth}{GD large} &      0.02 &      0.78 &      0.74 &      3.83\ensuremath{^{***}} &      0.35\ensuremath{^{***}} \\ 
 & (0.05)  & (2.10)  & (0.37)  & (0.49)  & (0.08)  \\ 
 & [    0.84]  & [    0.84]  & [    0.13]  & [    0.00]  & [    0.00]  \\ 
\addlinespace[1ex] \multirow[t]{ 2}{0.2\textwidth}{Combined} &      0.01 &      2.66 &      1.01\ensuremath{^{***}} &      4.26\ensuremath{^{***}} &      0.26\ensuremath{^{***}} \\ 
 & (0.04)  & (2.11)  & (0.31)  & (0.43)  & (0.08)  \\ 
 & [    0.84]  & [    0.38]  & [    0.01]  & [    0.00]  & [    0.01]  \\ 
\addlinespace[1ex] \multirow[t]{ 2}{0.2\textwidth}{HD $\times$ Baseline risk aversion} &      0.03 &      1.68 &      0.50 &     -0.11 &      0.04 \\ 
 & (0.03)  & (1.61)  & (0.26)  & (0.37)  & (0.06)  \\ 
 & [    0.67]  & [    0.50]  & [    0.15]  & [    0.84]  & [    0.84]  \\ 
\addlinespace[1ex] \multirow[t]{ 2}{0.2\textwidth}{GD main $\times$ Baseline risk aversion} &      0.01 &     -0.51 &      0.34 &     -0.06 &      0.00 \\ 
 & (0.03)  & (1.71)  & (0.25)  & (0.35)  & (0.06)  \\ 
 & [    0.84]  & [    0.84]  & [    0.38]  & [    0.84]  & [    0.94]  \\ 
\addlinespace[1ex] \multirow[t]{ 2}{0.2\textwidth}{GD large $\times$ Baseline risk aversion} &      0.05 &      1.96 &      0.55 &     -0.68 &     -0.01 \\ 
 & (0.05)  & (2.07)  & (0.37)  & (0.49)  & (0.08)  \\ 
 & [    0.40]  & [    0.56]  & [    0.30]  & [    0.36]  & [    0.84]  \\ 
\addlinespace[1ex] \multirow[t]{ 2}{0.2\textwidth}{Combined $\times$ Baseline risk aversion} &      0.07 &      3.75 &      0.75\ensuremath{^{*}} &      0.12 &     -0.03 \\ 
 & (0.04)  & (2.07)  & (0.31)  & (0.43)  & (0.08)  \\ 
 & [    0.23]  & [    0.18]  & [    0.05]  & [    0.84]  & [    0.84]  \\ 
\addlinespace[1ex] \multirow[t]{ 2}{0.2\textwidth}{Baseline risk aversion} &     -0.01 &     -0.38 &     -0.43\ensuremath{^{*}} &      0.10 &     -0.05 \\ 
 & (0.02)  & (1.10)  & (0.19)  & (0.26)  & (0.04)  \\ 
 & [    0.84]  & [    0.84]  & [    0.06]  & [    0.84]  & [    0.50]  \\ 
\addlinespace[1ex] Control mean  &      0.48 &     18.64 &      8.05 &      5.61 &      9.46 \\ 
Observations  & \multicolumn{1}{c}{     1770}  & \multicolumn{1}{c}{     1770}  & \multicolumn{1}{c}{     1770}  & \multicolumn{1}{c}{     1770}  & \multicolumn{1}{c}{     1737}  \\ 
$ R^2$  &      0.02 &      0.03 &      0.03 &      0.11 &      0.10 \\ 
$ p$-value  &      0.42 &      0.27 &      0.13 &      0.65 &      0.95 \\ 
\addlinespace[1ex] 
\bottomrule
\end{tabular}

\end{center}
\floatfoot{
Notes: Table presents tests for heterogeneity of treatment effects by Risk Aversion.  Risk Aversion demeaned before interaction so first four rows give effect of treatment at average value, and next four rows test for differential treatment effect by risk aversion. Standard errors are (in soft brackets) are clustered at the household level to reflect the design effect, and p-values corrected for False Discovery Rates across all the outcomes in the table are presented in hard brackets.  Stars on coefficient estimates are derived from the FDR-corrected p-values, *=10\%, **=5\%, and ***=1\% significance.  $p$-value in the last row from an F-test on whether treatments have a jointly differential effect by gender.  
}
\end{footnotesize}
\end{table}

\clearpage 

\begin{table}[!h]\caption{Heterogeneity: Baseline household consumption}\label{t:het_consumption}
\begin{footnotesize}
\begin{center}
\begin{tabular}{l *{5}{S}}
\toprule
\multicolumn{1}{c}{\text{ }} & \multicolumn{1}{c}{\text{Employed}} & \multicolumn{1}{c}{\text{\makecell[b]{Productive\\Hours}}} & \multicolumn{1}{c}{\text{\makecell[b]{Monthly\\Income}}} & \multicolumn{1}{c}{\text{\makecell[b]{Productive\\Assets}}} & \multicolumn{1}{c}{\text{Consumption}}\\
\midrule
\multirow[t]{ 2}{0.2\textwidth}{HD} &      0.01 &      2.60 &      0.14 &      1.27\ensuremath{^{***}} &      0.13\ensuremath{^{*}} \\ 
 & (0.03)  & (1.62)  & (0.27)  & (0.37)  & (0.06)  \\ 
 & [    0.64]  & [    0.18]  & [    0.63]  & [    0.00]  & [    0.09]  \\ 
\addlinespace[1ex] \multirow[t]{ 2}{0.2\textwidth}{GD main} &      0.02 &      4.31\ensuremath{^{**}} &      0.76\ensuremath{^{**}} &      3.72\ensuremath{^{***}} &      0.28\ensuremath{^{***}} \\ 
 & (0.03)  & (1.70)  & (0.26)  & (0.36)  & (0.06)  \\ 
 & [    0.63]  & [    0.04]  & [    0.01]  & [    0.00]  & [    0.00]  \\ 
\addlinespace[1ex] \multirow[t]{ 2}{0.2\textwidth}{GD large} &      0.03 &      1.17 &      0.84\ensuremath{^{*}} &      3.93\ensuremath{^{***}} &      0.40\ensuremath{^{***}} \\ 
 & (0.05)  & (2.08)  & (0.36)  & (0.48)  & (0.08)  \\ 
 & [    0.63]  & [    0.63]  & [    0.06]  & [    0.00]  & [    0.00]  \\ 
\addlinespace[1ex] \multirow[t]{ 2}{0.2\textwidth}{Combined} &      0.02 &      2.90 &      1.08\ensuremath{^{***}} &      4.32\ensuremath{^{***}} &      0.26\ensuremath{^{***}} \\ 
 & (0.04)  & (2.14)  & (0.30)  & (0.42)  & (0.08)  \\ 
 & [    0.64]  & [    0.27]  & [    0.00]  & [    0.00]  & [    0.00]  \\ 
\addlinespace[1ex] \multirow[t]{ 2}{0.2\textwidth}{HD $\times$ Baseline HH consumption per AE} &     -0.04 &     -1.09 &      0.03 &     -0.19 &     -0.02 \\ 
 & (0.03)  & (1.72)  & (0.25)  & (0.38)  & (0.07)  \\ 
 & [    0.31]  & [    0.63]  & [    0.80]  & [    0.63]  & [    0.64]  \\ 
\addlinespace[1ex] \multirow[t]{ 2}{0.2\textwidth}{GD main $\times$ Baseline HH consumption per AE} &      0.01 &      0.83 &      0.29 &     -0.07 &     -0.11 \\ 
 & (0.03)  & (1.63)  & (0.27)  & (0.34)  & (0.06)  \\ 
 & [    0.64]  & [    0.63]  & [    0.37]  & [    0.76]  & [    0.13]  \\ 
\addlinespace[1ex] \multirow[t]{ 2}{0.2\textwidth}{GD large $\times$ Baseline HH consumption per AE} &      0.09\ensuremath{^{*}} &      2.51 &      0.70 &      0.46 &     -0.13 \\ 
 & (0.04)  & (1.95)  & (0.39)  & (0.48)  & (0.08)  \\ 
 & [    0.09]  & [    0.29]  & [    0.14]  & [    0.45]  & [    0.18]  \\ 
\addlinespace[1ex] \multirow[t]{ 2}{0.2\textwidth}{Combined $\times$ Baseline HH consumption per AE} &     -0.02 &     -1.44 &     -0.69\ensuremath{^{*}} &     -0.68 &      0.00 \\ 
 & (0.04)  & (2.15)  & (0.30)  & (0.41)  & (0.11)  \\ 
 & [    0.63]  & [    0.63]  & [    0.06]  & [    0.17]  & [    0.82]  \\ 
\addlinespace[1ex] \multirow[t]{ 2}{0.2\textwidth}{Baseline HH consumption per AE} &      0.02 &      0.43 &      0.10 &      0.51 &      0.32\ensuremath{^{***}} \\ 
 & (0.02)  & (1.05)  & (0.18)  & (0.27)  & (0.04)  \\ 
 & [    0.63]  & [    0.64]  & [    0.63]  & [    0.12]  & [    0.00]  \\ 
\addlinespace[1ex] Control mean  &      0.48 &     18.64 &      8.05 &      5.61 &      9.46 \\ 
Observations  & \multicolumn{1}{c}{     1770}  & \multicolumn{1}{c}{     1770}  & \multicolumn{1}{c}{     1770}  & \multicolumn{1}{c}{     1770}  & \multicolumn{1}{c}{     1737}  \\ 
$ R^2$  &      0.02 &      0.03 &      0.04 &      0.12 &      0.18 \\ 
$ p$-value  &      0.05 &      0.43 &      0.01 &      0.21 &      0.24 \\ 
\addlinespace[1ex] 
\bottomrule
\end{tabular}

\end{center}
\floatfoot{
Notes: Table presents tests for heterogeneity of treatment effects by baseline Household Consumption.  Consumption demeaned before interaction so first four rows give effect of treatment at average value, and next four rows test for differential treatment effect by consumption. Standard errors are (in soft brackets) are clustered at the household level to reflect the design effect, and p-values corrected for False Discovery Rates across all the outcomes in the table are presented in hard brackets.  Stars on coefficient estimates are derived from the FDR-corrected p-values, *=10\%, **=5\%, and ***=1\% significance.  $p$-value in the last row from an F-test on whether treatments have a jointly differential effect by gender.  
}
\end{footnotesize}
\end{table}

\clearpage 

\begin{table}[!h]\caption{Heterogeneity: Baseline local employment rates}\label{t:het_employ}
\begin{footnotesize}
\begin{center}
\begin{tabular}{l *{5}{S}}
\toprule
\multicolumn{1}{c}{\text{ }} & \multicolumn{1}{c}{\text{Employed}} & \multicolumn{1}{c}{\text{\makecell[b]{Productive\\Hours}}} & \multicolumn{1}{c}{\text{\makecell[b]{Monthly\\Income}}} & \multicolumn{1}{c}{\text{\makecell[b]{Productive\\Assets}}} & \multicolumn{1}{c}{\text{Consumption}}\\
\midrule
\multirow[t]{ 2}{0.2\textwidth}{HD} &      0.01 &      2.53 &      0.10 &      1.21\ensuremath{^{***}} &      0.08 \\ 
 & (0.03)  & (1.61)  & (0.27)  & (0.37)  & (0.06)  \\ 
 & [    1.00]  & [    0.37]  & [    1.00]  & [    0.01]  & [    0.60]  \\ 
\addlinespace[1ex] \multirow[t]{ 2}{0.2\textwidth}{GD main} &      0.01 &      4.29\ensuremath{^{**}} &      0.73\ensuremath{^{**}} &      3.68\ensuremath{^{***}} &      0.25\ensuremath{^{***}} \\ 
 & (0.03)  & (1.71)  & (0.26)  & (0.36)  & (0.06)  \\ 
 & [    1.00]  & [    0.05]  & [    0.02]  & [    0.00]  & [    0.00]  \\ 
\addlinespace[1ex] \multirow[t]{ 2}{0.2\textwidth}{GD large} &      0.02 &      0.83 &      0.77 &      3.84\ensuremath{^{***}} &      0.34\ensuremath{^{***}} \\ 
 & (0.05)  & (2.11)  & (0.37)  & (0.49)  & (0.08)  \\ 
 & [    1.00]  & [    1.00]  & [    0.13]  & [    0.00]  & [    0.00]  \\ 
\addlinespace[1ex] \multirow[t]{ 2}{0.2\textwidth}{Combined} &      0.02 &      2.80 &      1.04\ensuremath{^{***}} &      4.30\ensuremath{^{***}} &      0.26\ensuremath{^{***}} \\ 
 & (0.04)  & (2.13)  & (0.31)  & (0.42)  & (0.08)  \\ 
 & [    1.00]  & [    0.60]  & [    0.01]  & [    0.00]  & [    0.01]  \\ 
\addlinespace[1ex] \multirow[t]{ 2}{0.2\textwidth}{HD $\times$ Baseline cell share employed} &      0.35 &     19.81 &      2.94 &      4.69 &      0.15 \\ 
 & (0.32)  & (    16.21)  & (2.56)  & (3.64)  & (0.62)  \\ 
 & [    0.60]  & [    0.60]  & [    0.60]  & [    0.60]  & [    1.00]  \\ 
\addlinespace[1ex] \multirow[t]{ 2}{0.2\textwidth}{GD main $\times$ Baseline cell share employed} &      0.17 &     -1.97 &      3.00 &      1.72 &      0.26 \\ 
 & (0.32)  & (    17.05)  & (2.47)  & (3.50)  & (0.59)  \\ 
 & [    1.00]  & [    1.00]  & [    0.60]  & [    1.00]  & [    1.00]  \\ 
\addlinespace[1ex] \multirow[t]{ 2}{0.2\textwidth}{GD large $\times$ Baseline cell share employed} &      0.34 &      0.26 &      2.38 &      5.25 &     -0.88 \\ 
 & (0.46)  & (    22.84)  & (3.23)  & (4.42)  & (0.79)  \\ 
 & [    1.00]  & [    1.00]  & [    1.00]  & [    0.60]  & [    0.60]  \\ 
\addlinespace[1ex] \multirow[t]{ 2}{0.2\textwidth}{Combined $\times$ Baseline cell share employed} &      0.27 &     -1.71 &      2.68 &      7.59 &      1.51 \\ 
 & (0.43)  & (    21.35)  & (3.11)  & (4.23)  & (0.76)  \\ 
 & [    1.00]  & [    1.00]  & [    1.00]  & [    0.23]  & [    0.16]  \\ 
\addlinespace[1ex] \multirow[t]{ 2}{0.2\textwidth}{Baseline cell share employed} &      0.04 &      0.29 &      1.02 &     -1.41 &     -0.24 \\ 
 & (0.25)  & (    12.35)  & (1.98)  & (2.87)  & (0.50)  \\ 
 & [    1.00]  & [    1.00]  & [    1.00]  & [    1.00]  & [    1.00]  \\ 
\addlinespace[1ex] Control mean  &      0.48 &     18.64 &      8.05 &      5.61 &      9.46 \\ 
Observations  & \multicolumn{1}{c}{     1770}  & \multicolumn{1}{c}{     1770}  & \multicolumn{1}{c}{     1770}  & \multicolumn{1}{c}{     1770}  & \multicolumn{1}{c}{     1737}  \\ 
$ R^2$  &      0.02 &      0.03 &      0.03 &      0.11 &      0.10 \\ 
$ p$-value  &      0.85 &      0.65 &      0.77 &      0.35 &      0.10 \\ 
\addlinespace[1ex] 
\bottomrule
\end{tabular}

\end{center}
\floatfoot{
Notes: Table presents tests for heterogeneity of treatment effects by baseline Employment Rates. Employment demeaned before interaction so first four rows give effect of treatment at average value, and next four rows test for differential treatment effect by employment rates. Standard errors are (in soft brackets) are clustered at the household level to reflect the design effect, and p-values corrected for False Discovery Rates across all the outcomes in the table are presented in hard brackets.  Stars on coefficient estimates are derived from the FDR-corrected p-values, *=10\%, **=5\%, and ***=1\% significance.  $p$-value in the last row from an F-test on whether treatments have a jointly differential effect by gender.  
}
\end{footnotesize}
\end{table}

\clearpage

\begin{table}[!h]\caption{Heterogeneity: Age 23 and over}\label{t:het_older}
\begin{footnotesize}
\begin{center}
\begin{tabular}{l *{5}{S}}
\toprule
\multicolumn{1}{c}{\text{ }} & \multicolumn{1}{c}{\text{Employed}} & \multicolumn{1}{c}{\text{\makecell[b]{Productive\\Hours}}} & \multicolumn{1}{c}{\text{\makecell[b]{Monthly\\Income}}} & \multicolumn{1}{c}{\text{\makecell[b]{Productive\\Assets}}} & \multicolumn{1}{c}{\text{Consumption}}\\
\midrule
\multirow[t]{ 2}{0.2\textwidth}{HD} &      0.06 &      4.32 &      0.26 &      1.17 &      0.09 \\
 & (0.05)  & (2.52)  & (0.43)  & (0.56)  & (0.10)  \\
 & [    0.64]  & [    0.32]  & [    1.00]  & [    0.15]  & [    0.82]  \\
\addlinespace[1ex] \multirow[t]{ 2}{0.2\textwidth}{GD main} &      0.02 &      2.39 &      0.72 &      4.26\ensuremath{^{***}} &      0.28\ensuremath{^{**}} \\
 & (0.05)  & (2.46)  & (0.42)  & (0.53)  & (0.09)  \\
 & [    1.00]  & [    0.82]  & [    0.32]  & [    0.00]  & [    0.02]  \\
\addlinespace[1ex] \multirow[t]{ 2}{0.2\textwidth}{GD large} &      0.00 &      0.03 &      0.88 &      4.27\ensuremath{^{***}} &      0.31\ensuremath{^{*}} \\
 & (0.07)  & (3.23)  & (0.62)  & (0.76)  & (0.12)  \\
 & [    1.00]  & [    1.00]  & [    0.47]  & [    0.00]  & [    0.05]  \\
\addlinespace[1ex] \multirow[t]{ 2}{0.2\textwidth}{Combined} &      0.08 &      2.02 &      1.53\ensuremath{^{**}} &      3.97\ensuremath{^{***}} &      0.21 \\
 & (0.06)  & (2.98)  & (0.47)  & (0.63)  & (0.11)  \\
 & [    0.64]  & [    1.00]  & [    0.01]  & [    0.00]  & [    0.30]  \\
\addlinespace[1ex] \multirow[t]{ 2}{0.2\textwidth}{HD $\times$ Older than 22} &     -0.08 &     -3.01 &     -0.23 &      0.08 &     -0.01 \\
 & (0.07)  & (3.25)  & (0.55)  & (0.74)  & (0.12)  \\
 & [    0.64]  & [    0.82]  & [    1.00]  & [    1.00]  & [    1.00]  \\
\addlinespace[1ex] \multirow[t]{ 2}{0.2\textwidth}{GD main $\times$ Older than 22} &      0.00 &      3.64 &      0.12 &     -1.06 &     -0.05 \\
 & (0.07)  & (3.41)  & (0.53)  & (0.72)  & (0.12)  \\
 & [    1.00]  & [    0.71]  & [    1.00]  & [    0.47]  & [    1.00]  \\
\addlinespace[1ex] \multirow[t]{ 2}{0.2\textwidth}{GD large $\times$ Older than 22} &      0.01 &      1.04 &     -0.29 &     -0.76 &      0.04 \\
 & (0.10)  & (4.23)  & (0.74)  & (0.99)  & (0.15)  \\
 & [    1.00]  & [    1.00]  & [    1.00]  & [    0.97]  & [    1.00]  \\
\addlinespace[1ex] \multirow[t]{ 2}{0.2\textwidth}{Combined $\times$ Older than 22} &     -0.11 &      1.55 &     -0.88 &      0.57 &      0.10 \\
 & (0.09)  & (4.19)  & (0.63)  & (0.86)  & (0.16)  \\
 & [    0.64]  & [    1.00]  & [    0.47]  & [    1.00]  & [    1.00]  \\
\addlinespace[1ex] \multirow[t]{ 2}{0.2\textwidth}{Older than 22} &      0.12\ensuremath{^{*}} &      2.47 &      1.04\ensuremath{^{*}} &      0.28 &      0.22\ensuremath{^{*}} \\
 & (0.05)  & (2.18)  & (0.39)  & (0.53)  & (0.09)  \\
 & [    0.06]  & [    0.64]  & [    0.05]  & [    1.00]  & [    0.06]  \\
\addlinespace[1ex] Control mean  &      0.48 &     18.64 &      8.05 &      5.61 &      9.46 \\
Observations  & \multicolumn{1}{c}{     1770}  & \multicolumn{1}{c}{     1770}  & \multicolumn{1}{c}{     1770}  & \multicolumn{1}{c}{     1770}  & \multicolumn{1}{c}{     1737}  \\
$ R^2$  &      0.02 &      0.04 &      0.04 &      0.11 &      0.11 \\
$ p$-value  &      0.51 &      0.46 &      0.57 &      0.26 &      0.90 \\
\addlinespace[1ex]
\bottomrule
\end{tabular}
\end{center}
\floatfoot{
Notes: Table presents tests for heterogeneity of treatment effects by age.  First four rows give effect of treatment among young, and next four rows test for differential treatment effect for those 23 and over. Standard errors are (in soft brackets) are clustered at the household level to reflect the design effect, and p-values corrected for False Discovery Rates across all the outcomes in the table are presented in hard brackets.  Stars on coefficient estimates are derived from the FDR-corrected p-values, *=10\%, **=5\%, and ***=1\% significance.  $p$-value in the last row from an F-test on whether treatments have a jointly differential effect by gender.  
}
\end{footnotesize}
\end{table}

\clearpage 

\begin{table}[!h]\caption{Spillover effects:  full model, employment outcome}
\label{t:interference_bn_employed}
\begin{footnotesize}
\begin{center}
\begin{tabular}{l *{3}{S}}
\toprule
 & \multicolumn{3}{c}{Treatment} \\ 
\multicolumn{1}{c}{\text{ }} & \multicolumn{1}{c}{\text{HD}} & \multicolumn{1}{c}{\text{GD Main}} & \multicolumn{1}{c}{\text{GD Huge}}\\
\midrule
\multicolumn{4}{l}{\emph{Direct effects of treatment at saturation level of zero}}  \\ 
\addlinespace[1ex] \multirow[t]{ 2}{0.2\textwidth}{Direct effect} &     -0.02 &     -0.02 &      0.05 \\ 
 & (0.07)  & (0.07)  & (0.13)  \\ 
 & [    1.00]  & [    1.00]  & [    1.00]  \\ 
\addlinespace[1ex] \multicolumn{4}{l}{\emph{Spillover effects of treatment onto control individuals}}  \\ 
\addlinespace[1ex] \multirow[t]{ 2}{0.2\textwidth}{Spillover to control} &     -0.01 &     -0.09 &      0.06 \\ 
 & (0.09)  & (0.09)  & (0.16)  \\ 
 & [    1.00]  & [    1.00]  & [    1.00]  \\ 
\addlinespace[1ex] \multicolumn{4}{l}{\emph{Additional effect of treatment onto individuals assigned to\ldots}}  \\ 
\addlinespace[1ex] \multirow[t]{ 2}{0.2\textwidth}{HD} &      0.00 &      0.10 &     -0.10 \\ 
 & (0.11)  & (0.11)  & (0.19)  \\ 
 & [    1.00]  & [    1.00]  & [    1.00]  \\ 
\addlinespace[1ex] \multirow[t]{ 2}{0.2\textwidth}{GD main} &      0.02 &      0.15 &     -0.28 \\ 
 & (0.11)  & (0.12)  & (0.20)  \\ 
 & [    1.00]  & [    1.00]  & [    1.00]  \\ 
\addlinespace[1ex] \multirow[t]{ 2}{0.2\textwidth}{GD large} &      0.03 &     -0.08 &     -0.23 \\ 
 & (0.22)  & (0.17)  & (0.42)  \\ 
 & [    1.00]  & [    1.00]  & [    1.00]  \\ 
\addlinespace[1ex] Saturation mean  &      0.36 &      0.36 &      0.09 \\ 
Saturation SD  &      0.23 &      0.23 &      0.13 \\ 
$ p$-value  &      1.00 &      0.55 &      0.49 \\ 
\addlinespace[1ex] 
\bottomrule
\end{tabular}

\end{center}
\floatfoot{
Notes: Each column describes the direct and spillover effects of a specific treatment on Employment; all results in the table  are from a single estimation. Saturation mean and standard deviation correspond to the distribution of saturation rates for the treatment in question.  Regressions include but do not report the lagged dependent variable, fixed effects for randomization blocks, and a set of LASSO-selected baseline covariates, and are weighted to reflect intensive tracking.  Standard errors are (in soft brackets) are clustered at the household level to reflect the design effect, and p-values corrected for False Discovery Rates across all the outcomes in the table are presented in hard brackets.  Stars on coefficient estimates are derived from the FDR-corrected p-values, *=10\%, **=5\%, and ***=1\% significance.  $p$-value in the last row corresponds to a test for whether the treatment in question has interference effects on any arm, including control.  
}
\end{footnotesize}
\end{table}

\clearpage

\begin{table}[!h]\caption{Spillover effects:  full model, productive hours outcome}
\label{t:interference_bn_productive_hrs}
\begin{footnotesize}
\begin{center}
\begin{tabular}{l *{3}{S}}
\toprule
 & \multicolumn{3}{c}{Treatment} \\ 
\multicolumn{1}{c}{\text{ }} & \multicolumn{1}{c}{\text{HD}} & \multicolumn{1}{c}{\text{GD Main}} & \multicolumn{1}{c}{\text{GD Huge}}\\
\midrule
\multicolumn{4}{l}{\emph{Direct effects of treatment at saturation level of zero}}  \\ 
\addlinespace[1ex] \multirow[t]{ 2}{0.2\textwidth}{Direct effect} &      2.26 &      5.60 &      9.52 \\ 
 & (3.71)  & (3.67)  & (6.27)  \\ 
 & [    1.00]  & [    0.76]  & [    0.76]  \\ 
\addlinespace[1ex] \multicolumn{4}{l}{\emph{Spillover effects of treatment onto control individuals}}  \\ 
\addlinespace[1ex] \multirow[t]{ 2}{0.2\textwidth}{Spillover to control} &      0.81 &      1.74 &     -0.12 \\ 
 & (4.39)  & (4.68)  & (6.56)  \\ 
 & [    1.00]  & [    1.00]  & [    1.00]  \\ 
\addlinespace[1ex] \multicolumn{4}{l}{\emph{Additional effect of treatment onto individuals assigned to\ldots}}  \\ 
\addlinespace[1ex] \multirow[t]{ 2}{0.2\textwidth}{HD} &      0.81 &     -2.50 &     -8.79 \\ 
 & (5.72)  & (5.82)  & (9.63)  \\ 
 & [    1.00]  & [    1.00]  & [    0.87]  \\ 
\addlinespace[1ex] \multirow[t]{ 2}{0.2\textwidth}{GD main} &     -7.55 &      2.76 &    -12.81 \\ 
 & (5.60)  & (6.39)  & (8.73)  \\ 
 & [    0.76]  & [    1.00]  & [    0.76]  \\ 
\addlinespace[1ex] \multirow[t]{ 2}{0.2\textwidth}{GD large} &     -5.79 &    -11.98 &    -31.75 \\ 
 & (    10.27)  & (7.70)  & (    17.74)  \\ 
 & [    1.00]  & [    0.76]  & [    0.76]  \\ 
\addlinespace[1ex] Saturation mean  &      0.36 &      0.36 &      0.09 \\ 
Saturation SD  &      0.23 &      0.23 &      0.13 \\ 
$ p$-value  &      0.65 &      0.49 &      0.09 \\ 
\addlinespace[1ex] 
\bottomrule
\end{tabular}

\end{center}
\floatfoot{ 
Notes: Each column describes the direct and spillover effects of a specific treatment on Productive Hours; all results in the table  are from a single estimation. Saturation mean and standard deviation correspond to the distribution of saturation rates for the treatment in question.  Regressions include but do not report the lagged dependent variable, fixed effects for randomization blocks, and a set of LASSO-selected baseline covariates, and are weighted to reflect intensive tracking. Standard errors are (in soft brackets) are clustered at the household level to reflect the design effect, and p-values corrected for False Discovery Rates across all the outcomes in the table are presented in hard brackets.  Stars on coefficient estimates are derived from the FDR-corrected p-values, *=10\%, **=5\%, and ***=1\% significance.   $p$-value in the last corresponds to a test for whether the treatment in question has interference effects on any arm, including control.
}
\end{footnotesize}
\end{table}

\clearpage

\begin{table}[!h]\caption{Spillover effects:  full model, monthly income outcome}
\label{t:interference_bn_monthly_income}
\begin{footnotesize}
\begin{center}
\begin{tabular}{l *{3}{S}}
\toprule
 & \multicolumn{3}{c}{Treatment} \\ 
\multicolumn{1}{c}{\text{ }} & \multicolumn{1}{c}{\text{HD}} & \multicolumn{1}{c}{\text{GD Main}} & \multicolumn{1}{c}{\text{GD Huge}}\\
\midrule
\multicolumn{4}{l}{\emph{Direct effects of treatment at saturation level of zero}}  \\ 
\addlinespace[1ex] \multirow[t]{ 2}{0.2\textwidth}{Direct effect} &      0.23 &      0.96 &      1.77 \\ 
 & (0.55)  & (0.58)  & (0.88)  \\ 
 & [    1.00]  & [    1.00]  & [    1.00]  \\ 
\addlinespace[1ex] \multicolumn{4}{l}{\emph{Spillover effects of treatment onto control individuals}}  \\ 
\addlinespace[1ex] \multirow[t]{ 2}{0.2\textwidth}{Spillover to control} &      0.48 &     -0.69 &      0.53 \\ 
 & (0.82)  & (0.79)  & (1.13)  \\ 
 & [    1.00]  & [    1.00]  & [    1.00]  \\ 
\addlinespace[1ex] \multicolumn{4}{l}{\emph{Additional effect of treatment onto individuals assigned to\ldots}}  \\ 
\addlinespace[1ex] \multirow[t]{ 2}{0.2\textwidth}{HD} &     -0.92 &      0.91 &      0.05 \\ 
 & (0.87)  & (0.82)  & (1.57)  \\ 
 & [    1.00]  & [    1.00]  & [    1.00]  \\ 
\addlinespace[1ex] \multirow[t]{ 2}{0.2\textwidth}{GD main} &     -0.28 &      0.39 &     -1.10 \\ 
 & (0.91)  & (0.92)  & (1.40)  \\ 
 & [    1.00]  & [    1.00]  & [    1.00]  \\ 
\addlinespace[1ex] \multirow[t]{ 2}{0.2\textwidth}{GD large} &     -1.61 &     -1.75 &      2.08 \\ 
 & (1.41)  & (1.40)  & (3.37)  \\ 
 & [    1.00]  & [    1.00]  & [    1.00]  \\ 
\addlinespace[1ex] Saturation mean  &      0.36 &      0.36 &      0.09 \\ 
Saturation SD  &      0.23 &      0.23 &      0.13 \\ 
$ p$-value  &      0.66 &      0.28 &      0.88 \\ 
\addlinespace[1ex] 
\bottomrule
\end{tabular}

\end{center}
\floatfoot{ 
Notes: Each column describes the direct and spillover effects of a specific treatment on Monthly Income (IHS); all results in the table  are from a single estimation. Saturation mean and standard deviation correspond to the distribution of saturation rates for the treatment in question.  Regressions include but do not report the lagged dependent variable, fixed effects for randomization blocks, and a set of LASSO-selected baseline covariates, and are weighted to reflect intensive tracking. Standard errors are (in soft brackets) are clustered at the household level to reflect the design effect, and p-values corrected for False Discovery Rates across all the outcomes in the table are presented in hard brackets.  Stars on coefficient estimates are derived from the FDR-corrected p-values, *=10\%, **=5\%, and ***=1\% significance.  $p$-value in the last row corresponds to a test for whether the treatment in question has interference effects on any arm, including control.
}
\end{footnotesize}
\end{table}

\clearpage 

\begin{table}[!h]\caption{Spillover effects:  productive assets outcome}
\label{t:interference_bn_tot_prod_assetval}
\begin{footnotesize}
\begin{center}
\begin{tabular}{l *{3}{S}}
\toprule
 & \multicolumn{3}{c}{Treatment} \\ 
\multicolumn{1}{c}{\text{ }} & \multicolumn{1}{c}{\text{HD}} & \multicolumn{1}{c}{\text{GD Main}} & \multicolumn{1}{c}{\text{GD Huge}}\\
\midrule
\multicolumn{4}{l}{\emph{Direct effects of treatment at saturation level of zero}}  \\ 
\addlinespace[1ex] \multirow[t]{ 2}{0.2\textwidth}{Direct effect} &      2.34\ensuremath{^{***}} &      4.35\ensuremath{^{***}} &      5.64\ensuremath{^{***}} \\ 
 & (0.74)  & (0.73)  & (1.09)  \\ 
 & [    0.01]  & [    0.00]  & [    0.00]  \\ 
\addlinespace[1ex] \multicolumn{4}{l}{\emph{Spillover effects of treatment onto control individuals}}  \\ 
\addlinespace[1ex] \multirow[t]{ 2}{0.2\textwidth}{Spillover to control} &      1.05 &      1.69 &      2.18 \\ 
 & (1.00)  & (1.10)  & (1.72)  \\ 
 & [    0.36]  & [    0.26]  & [    0.34]  \\ 
\addlinespace[1ex] \multicolumn{4}{l}{\emph{Additional effect of treatment onto individuals assigned to\ldots}}  \\ 
\addlinespace[1ex] \multirow[t]{ 2}{0.2\textwidth}{HD} &     -0.61 &     -1.95 &     -2.85 \\ 
 & (1.32)  & (1.17)  & (2.10)  \\ 
 & [    0.49]  & [    0.26]  & [    0.32]  \\ 
\addlinespace[1ex] \multirow[t]{ 2}{0.2\textwidth}{GD main} &     -2.17 &      0.23 &     -1.82 \\ 
 & (1.34)  & (1.11)  & (1.88)  \\ 
 & [    0.26]  & [    0.52]  & [    0.36]  \\ 
\addlinespace[1ex] \multirow[t]{ 2}{0.2\textwidth}{GD large} &     -0.88 &     -2.22 &     -7.88 \\ 
 & (1.90)  & (1.87)  & (4.10)  \\ 
 & [    0.49]  & [    0.35]  & [    0.20]  \\ 
\addlinespace[1ex] Saturation mean  &      0.36 &      0.36 &      0.09 \\ 
Saturation SD  &      0.23 &      0.23 &      0.13 \\ 
$ p$-value  &      0.62 &      0.21 &      0.27 \\ 
\addlinespace[1ex] 
\bottomrule
\end{tabular}

\end{center}
\floatfoot{ 
Notes: Each column describes the direct and spillover effects of a specific treatment on Productive Assets (IHS); all results in the table  are from a single estimation. Saturation mean and standard deviation correspond to the distribution of saturation rates for the treatment in question.  Regressions include but do not report the lagged dependent variable, fixed effects for randomization blocks, and a set of LASSO-selected baseline covariates, and are weighted to reflect intensive tracking. Standard errors are (in soft brackets) are clustered at the household level to reflect the design effect, and p-values corrected for False Discovery Rates across all the outcomes in the table are presented in hard brackets.  Stars on coefficient estimates are derived from the FDR-corrected p-values, *=10\%, **=5\%, and ***=1\% significance.  $p$-value in the last row corresponds to a test for whether the treatment in question has interference effects on any arm, including control.
}
\end{footnotesize}
\end{table}

\clearpage

\begin{table}[!h]\caption{Spillover effects:  consumption}
\begin{footnotesize}
\label{t:interference_hh_month_consumption_pc}
\begin{center}
\begin{tabular}{l *{3}{S}}
\toprule
 & \multicolumn{3}{c}{Treatment} \\ 
\multicolumn{1}{c}{\text{ }} & \multicolumn{1}{c}{\text{HD}} & \multicolumn{1}{c}{\text{GD Main}} & \multicolumn{1}{c}{\text{GD Huge}}\\
\midrule
\multicolumn{4}{l}{\emph{Direct effects of treatment at saturation level of zero}}  \\ 
\addlinespace[1ex] \multirow[t]{ 2}{0.2\textwidth}{Direct effect} &      0.01 &      0.29 &      0.20 \\ 
 & (0.12)  & (0.11)  & (0.20)  \\ 
 & [    1.00]  & [    0.23]  & [    1.00]  \\ 
\addlinespace[1ex] \multicolumn{4}{l}{\emph{Spillover effects of treatment onto control individuals}}  \\ 
\addlinespace[1ex] \multirow[t]{ 2}{0.2\textwidth}{Spillover to control} &      0.05 &      0.02 &     -0.36 \\ 
 & (0.16)  & (0.18)  & (0.25)  \\ 
 & [    1.00]  & [    1.00]  & [    0.75]  \\ 
\addlinespace[1ex] \multicolumn{4}{l}{\emph{Additional effect of treatment onto individuals assigned to\ldots}}  \\ 
\addlinespace[1ex] \multirow[t]{ 2}{0.2\textwidth}{HD} &      0.16 &     -0.05 &     -0.01 \\ 
 & (0.20)  & (0.19)  & (0.32)  \\ 
 & [    1.00]  & [    1.00]  & [    1.00]  \\ 
\addlinespace[1ex] \multirow[t]{ 2}{0.2\textwidth}{GD main} &     -0.21 &     -0.08 &      0.53 \\ 
 & (0.17)  & (0.19)  & (0.28)  \\ 
 & [    1.00]  & [    1.00]  & [    0.40]  \\ 
\addlinespace[1ex] \multirow[t]{ 2}{0.2\textwidth}{GD large} &     -0.04 &      0.18 &      1.22 \\ 
 & (0.33)  & (0.30)  & (0.59)  \\ 
 & [    1.00]  & [    1.00]  & [    0.37]  \\ 
\addlinespace[1ex] Saturation mean  &      0.36 &      0.36 &      0.09 \\ 
Saturation SD  &      0.23 &      0.23 &      0.13 \\ 
$ p$-value  &      0.68 &      0.90 &      0.11 \\ 
\addlinespace[1ex] 
\bottomrule
\end{tabular}

\end{center}
\floatfoot{ 
Notes: Each column describes the direct and spillover effects of a specific treatment on Household Consumption (IHS); all results in the table are from a single estimation. Saturation mean and standard deviation correspond to the distribution of saturation rates for the treatment in question. Standard errors are (in soft brackets) are clustered at the household level to reflect the design effect, and p-values corrected for False Discovery Rates across all the outcomes in the table are presented in hard brackets.  Stars on coefficient estimates are derived from the FDR-corrected p-values, *=10\%, **=5\%, and ***=1\% significance.  $p$-value in the last row corresponds to a test for whether the treatment in question has interference effects on any arm, including control.
}
\end{footnotesize}
\end{table}

   \begin{sidewaystable}[!hp]
    \caption{Cost-Effectiveness (benefit per \$100 spent), Primary outcomes}
    \label{t:benefit_cost_ratios_primary}
    \begin{footnotesize}
    \begin{tabular}{l *{6}{S[table-format=03.3]} SSSS}
\toprule
 & & \multicolumn{4}{c}{GiveDirectly} & & \multicolumn{4}{c}{$ p$-values} \\ 
\cmidrule(lr){3-6} \cmidrule(lr){8-11}
\multicolumn{1}{c}{\text{ }} & \multicolumn{1}{c}{\text{HD}} & \multicolumn{1}{c}{\text{Lower}} & \multicolumn{1}{c}{\text{Middle}} & \multicolumn{1}{c}{\text{Upper}} & \multicolumn{1}{c}{\text{Large}} & \multicolumn{1}{c}{\text{Combined}} & \multicolumn{1}{c}{\text{(a)}} & \multicolumn{1}{c}{\text{(b)}} & \multicolumn{1}{c}{\text{(c)}} & \multicolumn{1}{c}{\text{(d)}}\\
\midrule
\multirow[t]{ 2 }{0.15\textwidth}{Employed }  &     0.007 &     0.008 &     0.010 &     0.000 &     0.001 &     0.001 &      0.85 &      0.95 &      0.57 &      0.94 \\ 
 & (    0.010)  & (    0.012)  & (    0.009)  & (    0.008)  & (    0.005)  & (    0.005)  \\ \addlinespace[1ex] 
\multirow[t]{ 2 }{0.15\textwidth}{Productive hours }  &     0.838 &     0.699 &     1.332 &     0.603 &     0.132 &     0.275 &      0.14 &      0.82 &      0.33 &      0.63 \\ 
 & (    0.471)  & (    0.593)  & (    0.489)  & (    0.427)  & (    0.244)  & (    0.241)  \\ \addlinespace[1ex] 
\multirow[t]{ 2 }{0.15\textwidth}{Monthly income }  &     0.094 &     0.192 &     0.220 &     0.194 &     0.086 &     0.124 &      0.16 &      0.31 &      0.24 &      0.41 \\ 
 & (    0.077)  & (    0.092)  & (    0.069)  & (    0.059)  & (    0.041)  & (    0.038)  \\ \addlinespace[1ex] 
\multirow[t]{ 2 }{0.15\textwidth}{Productive assets }  &     0.463 &     0.998 &     0.775 &     0.650 &     0.475 &     0.526 &      0.00 &      0.00 &      0.00 &      0.42 \\ 
 & (    0.107)  & (    0.118)  & (    0.102)  & (    0.078)  & (    0.055)  & (    0.052)  \\ \addlinespace[1ex] 
\multirow[t]{ 2 }{0.15\textwidth}{HH consumption per capita }  &     0.016 &     0.050 &     0.054 &     0.040 &     0.042 &     0.032 &      0.37 &      0.12 &      0.67 &      0.31 \\ 
 & (    0.018)  & (    0.020)  & (    0.018)  & (    0.013)  & (    0.009)  & (    0.008)  \\ \addlinespace[1ex] 
\bottomrule
\end{tabular}

    \floatfoot{
    Note:  Table gives the impact per \$100 spent, which is calculated by dividing the estimated ITT impacts by the cost per arm in hundreds of dollars.  The standard errors in the table are similarly the ITT SEs divided by costs.  Reported $p$-values in final three columns derived from $F$-tests of hypotheses that cost-benefit ratios are equal between: (a) joint test across all arms, (b) GD Lower and HD; (c) GD Lower and GD Large; and (d) GD Large and Combined arms.
    }
    \end{footnotesize}
    \end{sidewaystable}


    \begin{sidewaystable}[!hp]
    \caption{Cost-Effectiveness (benefit per \$100 spent), Secondary outcomes}
    \label{t:benefit_cost_ratios_secondary}
    \begin{footnotesize}
    \begin{tabular}{l *{6}{S[table-format=03.3]} SSSS}
\toprule
 & & \multicolumn{4}{c}{GiveDirectly} & & \multicolumn{4}{c}{$ p$-values} \\ 
\cmidrule(lr){3-6} \cmidrule(lr){8-11}
\multicolumn{1}{c}{\text{ }} & \multicolumn{1}{c}{\text{HD}} & \multicolumn{1}{c}{\text{Lower}} & \multicolumn{1}{c}{\text{Middle}} & \multicolumn{1}{c}{\text{Upper}} & \multicolumn{1}{c}{\text{Large}} & \multicolumn{1}{c}{\text{Combined}} & \multicolumn{1}{c}{\text{(a)}} & \multicolumn{1}{c}{\text{(b)}} & \multicolumn{1}{c}{\text{(c)}} & \multicolumn{1}{c}{\text{(d)}}\\
\midrule
\multicolumn{11}{l}{\emph{Panel A. Beneficiary welfare}}  \\ 
\addlinespace[1ex] \multirow[t]{ 2 }{0.15\textwidth}{Subjective well-being }  &     0.058 &     0.101 &     0.107 &     0.081 &     0.066 &     0.048 &      0.09 &      0.08 &      0.12 &      0.17 \\ 
 & (    0.020)  & (    0.024)  & (    0.020)  & (    0.016)  & (    0.011)  & (    0.010)  \\ \addlinespace[1ex] 
\multirow[t]{ 2 }{0.15\textwidth}{Mental health }  &    -0.013 &    -0.016 &     0.014 &     0.005 &     0.013 &     0.014 &      0.60 &      0.89 &      0.22 &      0.93 \\ 
 & (    0.022)  & (    0.023)  & (    0.019)  & (    0.016)  & (    0.011)  & (    0.011)  \\ \addlinespace[1ex] 
\multirow[t]{ 2 }{0.15\textwidth}{Beneficiary-specific consumption }  &     0.045 &     0.129 &     0.125 &     0.105 &     0.053 &     0.082 &      0.00 &      0.01 &      0.01 &      0.11 \\ 
 & (    0.035)  & (    0.030)  & (    0.027)  & (    0.020)  & (    0.018)  & (    0.014)  \\ \addlinespace[1ex] 
\multicolumn{11}{l}{\emph{Panel B. Household wealth}}  \\ 
\addlinespace[1ex] \multirow[t]{ 2 }{0.15\textwidth}{HH net non-land wealth }  &    -0.054 &     0.041 &     0.244 &     0.226 &     0.131 &     0.106 &      0.05 &      0.56 &      0.53 &      0.67 \\ 
 & (    0.120)  & (    0.150)  & (    0.089)  & (    0.070)  & (    0.049)  & (    0.057)  \\ \addlinespace[1ex] 
\multirow[t]{ 2 }{0.15\textwidth}{HH livestock wealth }  &    -0.002 &     0.445 &     0.374 &     0.447 &     0.257 &     0.263 &      0.00 &      0.00 &      0.12 &      0.92 \\ 
 & (    0.110)  & (    0.123)  & (    0.105)  & (    0.075)  & (    0.056)  & (    0.053)  \\ \addlinespace[1ex] 
\multirow[t]{ 2 }{0.15\textwidth}{Savings }  &     0.311 &     0.264 &     0.262 &     0.265 &     0.169 &     0.202 &      0.14 &      0.56 &      0.22 &      0.39 \\ 
 & (    0.070)  & (    0.080)  & (    0.068)  & (    0.051)  & (    0.036)  & (    0.032)  \\ \addlinespace[1ex] 
\multirow[t]{ 2 }{0.15\textwidth}{Debt }  &     0.122 &    -0.024 &    -0.046 &    -0.095 &    -0.043 &     0.000 &      0.18 &      0.19 &      0.86 &      0.45 \\ 
 & (    0.084)  & (    0.105)  & (    0.088)  & (    0.076)  & (    0.050)  & (    0.045)  \\ \addlinespace[1ex] 
\multicolumn{11}{l}{\emph{Panel C. Beneficiary cognitive and non-cognitive skills}}  \\ 
\addlinespace[1ex] \multirow[t]{ 2 }{0.15\textwidth}{Locus of control }  &     0.019 &     0.032 &     0.005 &     0.000 &     0.010 &     0.027 &      0.65 &      0.59 &      0.30 &      0.12 \\ 
 & (    0.018)  & (    0.021)  & (    0.017)  & (    0.014)  & (    0.010)  & (    0.009)  \\ \addlinespace[1ex] 
\multirow[t]{ 2 }{0.15\textwidth}{Aspirations }  &    -0.002 &     0.020 &    -0.009 &     0.022 &     0.003 &     0.016 &      0.52 &      0.41 &      0.48 &      0.26 \\ 
 & (    0.022)  & (    0.024)  & (    0.019)  & (    0.015)  & (    0.010)  & (    0.009)  \\ \addlinespace[1ex] 
\multirow[t]{ 2 }{0.15\textwidth}{Big Five index }  &     0.035 &     0.019 &     0.022 &     0.003 &    -0.009 &     0.003 &      0.17 &      0.55 &      0.25 &      0.35 \\ 
 & (    0.020)  & (    0.025)  & (    0.019)  & (    0.015)  & (    0.011)  & (    0.010)  \\ \addlinespace[1ex] 
\multirow[t]{ 2 }{0.15\textwidth}{Business knowledge }  &     0.197 &     0.022 &     0.016 &     0.011 &    -0.003 &     0.074 &      0.00 &      0.00 &      0.29 &      0.00 \\ 
 & (    0.022)  & (    0.024)  & (    0.019)  & (    0.016)  & (    0.011)  & (    0.011)  \\ \addlinespace[1ex] 
\multirow[t]{ 2 }{0.15\textwidth}{Business attitudes }  &     0.037 &     0.049 &     0.038 &     0.017 &     0.007 &     0.018 &      0.25 &      0.63 &      0.08 &      0.40 \\ 
 & (    0.020)  & (    0.024)  & (    0.018)  & (    0.015)  & (    0.011)  & (    0.010)  \\ \addlinespace[1ex] 
\bottomrule
\end{tabular}

    \floatfoot{
      Note:  Table gives the impact per \$100 spent, which is calculated by dividing the estimated ITT impacts by the cost per arm in hundreds of dollars.  The standard errors in the table are similarly the ITT SEs divided by costs.  Reported $p$-values in final three columns derived from $F$-tests of hypotheses that cost-benefit ratios are equal between: (a) joint test across all arms, (b) GD Lower and HD; (c) GD Lower and GD Large; and (d) GD Large and Combined arms.
    }
    \end{footnotesize}
    \end{sidewaystable}

    
    \begin{sidewaystable}[!hp]
    \caption{Program impacts on household-to-household transfers}
    \label{t:itt_transfers}
    \begin{footnotesize}

    \begin{center}
    \begin{tabular}{l *{6}{S} ScSSSS}
\toprule
 & & \multicolumn{4}{c}{GiveDirectly} & & \multicolumn{1}{c}{\raisebox{-1ex}[0pt]{Control}} & & & \multicolumn{3}{c}{$ p$-values} \\ 
\cmidrule(lr){3-6} \cmidrule(lr){11-13}
\multicolumn{1}{c}{\text{ }} & \multicolumn{1}{c}{\text{HD}} & \multicolumn{1}{c}{\text{Lower}} & \multicolumn{1}{c}{\text{Middle}} & \multicolumn{1}{c}{\text{Upper}} & \multicolumn{1}{c}{\text{Large}} & \multicolumn{1}{c}{\text{Combined}} & \multicolumn{1}{c}{\text{Mean}} & \multicolumn{1}{c}{\text{Obs.}} & \multicolumn{1}{c}{\text{$ R^2$}} & \multicolumn{1}{c}{\text{(a)}} & \multicolumn{1}{c}{\text{(b)}} & \multicolumn{1}{c}{\text{(c)}}\\
\midrule
\multirow[t]{ 3 }{0.15\textwidth}{HH loans made }  &      0.11 &      0.90\ensuremath{^{**}} &      0.45 &      1.41\ensuremath{^{***}} &      0.68\ensuremath{^{*}} &      1.24\ensuremath{^{***}} &      2.24 &      1705 &      0.15 &      0.10 &      0.18 &      0.27 \\ 
 & (0.30)  & (0.44)  & (0.44)  & (0.45)  & (0.42)  & (0.43)  \\  & [    0.31]  & [    0.05]  & [    0.19]  & [    0.00]  & [    0.09]  & [    0.01]  \\ \addlinespace[0.5ex] 
\multirow[t]{ 3 }{0.15\textwidth}{HH gifts made }  &     -0.33 &      1.75\ensuremath{^{***}} &      2.67\ensuremath{^{***}} &      2.62\ensuremath{^{***}} &      3.11\ensuremath{^{***}} &      2.46\ensuremath{^{***}} &      4.90 &      1704 &      0.17 &      0.00 &      0.60 &      0.33 \\ 
 & (0.36)  & (0.57)  & (0.55)  & (0.59)  & (0.53)  & (0.50)  \\  & [    0.19]  & [    0.00]  & [    0.00]  & [    0.00]  & [    0.00]  & [    0.00]  \\ \addlinespace[0.5ex] 
\multirow[t]{ 3 }{0.15\textwidth}{HH gifts made }  &      0.39 &      0.68\ensuremath{^{*}} &      1.87\ensuremath{^{***}} &      0.56 &      0.32 &      0.81\ensuremath{^{*}} &      3.41 &      1675 &      0.15 &      0.63 &      0.20 &      0.36 \\ 
 & (0.32)  & (0.42)  & (0.48)  & (0.47)  & (0.44)  & (0.45)  \\  & [    0.15]  & [    0.09]  & [    0.00]  & [    0.15]  & [    0.20]  & [    0.08]  \\ \addlinespace[0.5ex] 
\bottomrule
\end{tabular}

    \end{center}

    \floatfoot{
      Note:  Regressions include but do not report the lagged dependent variable, fixed effects for randomization blocks, and a set of LASSO-selected baseline covariates, and are weighted to reflect intensive tracking. Standard errors are (in soft brackets) are clustered at the household level to reflect the design effect, and p-values corrected for False Discovery Rates across all the outcomes in the table are presented in hard brackets.  Stars on coefficient estimates are derived from the FDR-corrected $p$-values, *=10\%, **=5\%, and ***=1\% significance. Reported $p$-values in final three columns derived from $F$-tests of hypotheses that cost-benefit ratios are equal between: (a) GD Lower and HD; (b) GD Lower and GD Large; and (c) GD Large and Combined treatments. 
    }
    \end{footnotesize}
    \end{sidewaystable}

\clearpage
\section{Supplementary figures}\label{app:MoreFigures}

\clearpage
\global\pdfpageattr\expandafter{\the\pdfpageattr/Rotate 90}
\begin{landscape}

    \vspace*{\fill}

\begin{figure}[!h]
\caption{Project timeline}
\label{f:timeline}

\begin{center}
\resizebox{\textwidth}{!}{
	\definecolor{gdgreen}{RGB}{144,238,144}
	\definecolor{hdblue}{RGB}{135,206,250}
	\definecolor{ipagrey}{RGB}{245,245,220}

	\newcommand\tikzunits{1.7}

	\begin{tikzpicture}[scale=\tikzunits,shorten >=1pt,semithick] 

	\draw[thick,->] (0,0) -- (22,0);
	\foreach \x in {0,...,21} 
		\draw[thin] (\x,-0.25) -- (\x,0.25) ;
	\draw (0,0.25) node[draw=none,anchor=south] {2017};
	\draw (0,-0.25)  node[draw=none,anchor=north] {Nov};
	\draw (1,-0.25)  node[draw=none,anchor=north] {Dec};
	\draw (2,0.25) node[draw=none,anchor=south] {2018};
	\draw (2,-0.25)  node[draw=none,anchor=north] {Jan};
	\draw (3,-0.25)  node[draw=none,anchor=north] {Feb};
	\draw (4,-0.25)  node[draw=none,anchor=north] {Mar};
	\draw (5,-0.25)  node[draw=none,anchor=north] {Apr};
	\draw (6,-0.25)  node[draw=none,anchor=north] {May};
	\draw (7,-0.25)  node[draw=none,anchor=north] {Jun};
	\draw (8,-0.25)  node[draw=none,anchor=north] {Jul};
	\draw (9,-0.25)  node[draw=none,anchor=north] {Aug};
	\draw (10,-0.25)  node[draw=none,anchor=north] {Sep};
	\draw (11,-0.25)  node[draw=none,anchor=north] {Oct};
	\draw (12,-0.25)  node[draw=none,anchor=north] {Nov};
	\draw (13,-0.25)  node[draw=none,anchor=north] {Dec};
	\draw (14,0.25) node[draw=none,anchor=south] {2019};
	\draw (14,-0.25)  node[draw=none,anchor=north] {Jan};
	\draw (15,-0.25)  node[draw=none,anchor=north] {Feb};
	\draw (16,-0.25)  node[draw=none,anchor=north] {Mar};
	\draw (17,-0.25)  node[draw=none,anchor=north] {Apr};
	\draw (18,-0.25)  node[draw=none,anchor=north] {May};
	\draw (19,-0.25)  node[draw=none,anchor=north] {Jun};
	\draw (20,-0.25)  node[draw=none,anchor=north] {Jul};
	\draw (21,-0.25)  node[draw=none,anchor=north] {Aug};

	\draw (-2,1.5) node[draw=none,anchor=north west,minimum height = 0.875*\tikzunits cm] {\emph{\large{Measurement}}} ;
	\draw (0,1.5) node[rectangle,draw=black,anchor=north west,minimum width=2,minimum height=0.375*\tikzunits cm,fill=ipagrey] {
		Tranche I baseline
		};
	\draw (1,1) node[rectangle,draw=black,anchor=north west,minimum width=2,minimum height=0.375*\tikzunits cm,fill=ipagrey] {Tranche II baseline};
	\draw (19,1.5) node[rectangle,draw=black,anchor=north west,minimum width=2,minimum height=0.375*\tikzunits cm,fill=ipagrey] {Tranche I follow-up};
	\draw (20,1) node[rectangle,draw=black,anchor=north west,minimum width=2,minimum height=0.375*\tikzunits cm,fill=ipagrey] {Tranche II follow-up};

	\draw (12,1.5) node[rectangle,draw=black,anchor=north west, minimum width=6*\tikzunits cm,minimum height=0.875*\tikzunits cm,fill=ipagrey] {Rolling phone surveys} ; 

	\newcommand\yhdi{-2}  
	\newcommand\yhdii{-3}  
	\draw (-2,-1) node[draw=none,anchor=north west,minimum height=0.75*\tikzunits cm] {\large{\emph{Cash transfers}}};
	\draw (-2,\yhdi) node[draw=none,anchor=north west,minimum height=0.75*\tikzunits cm] {\large{\emph{HD Tranche I}}};
	\draw (-2,\yhdii) node[draw=none,anchor=north west,minimum height=0.75*\tikzunits cm] {\large{\emph{HD Tranche II}}};

	\draw (6,-1) node[rectangle,draw=black,anchor=north west, minimum width=12,minimum height=0.75*\tikzunits cm,fill=gdgreen] {Cash transfer 1}; 
	\draw (8,-1) node[rectangle,draw=black,anchor=north west, minimum width=12,minimum height=0.75*\tikzunits cm,fill=gdgreen] {Cash transfer 2}; 

	\draw (3,\yhdi) node[rectangle,draw=black,anchor=north west, minimum width=2,minimum height=0.75*\tikzunits cm,text width=1.8*\tikzunits cm,align=left,fill=hdblue] {
		Workforce \\ readiness training
		};   
	\draw (4,\yhdii) node[rectangle,draw=black,anchor=north west, minimum width=22,minimum height=0.75*\tikzunits cm,text width=1.8*\tikzunits cm,align=left,fill=hdblue] {
	Workforce  \\ readiness training}; 

	\draw (5,\yhdi) node[rectangle,draw=black,anchor=north west, minimum width=22,minimum height=0.75*\tikzunits cm,text width=3.8*\tikzunits cm,align=left,fill=hdblue] {Skills training/ \\ Be Your Own Boss} ; 
	\draw (6,\yhdii) node[rectangle,draw=black,anchor=north west, minimum width=22,minimum height=0.75*\tikzunits cm,text width=3.8*\tikzunits cm,align=left,fill=hdblue] {Skills training/ \\ Be Your Own Boss} ; 

	\draw (9,\yhdi) node[rectangle,draw=black,anchor=north west, minimum width=22,minimum height=0.75*\tikzunits cm,text width=2.8*\tikzunits cm,align=left,fill=hdblue] {Job placement and coaching} ; 
	\draw (10,\yhdii) node[rectangle,draw=black,anchor=north west, minimum width=22,minimum height=0.75*\tikzunits cm,text width=2.8*\tikzunits cm,align=left,fill=hdblue] {Job placement and coaching} ; 

	\end{tikzpicture}
}
\end{center}

\end{figure}

    \vspace*{\fill}

\end{landscape}
\clearpage 
\global\pdfpageattr\expandafter{\the\pdfpageattr/Rotate 0}

\begin{figure}
\caption{CDF of Savings Stocks (IHS)}
\label{f:saving_CDF}
\begin{center}
\includegraphics[width=0.8\linewidth]{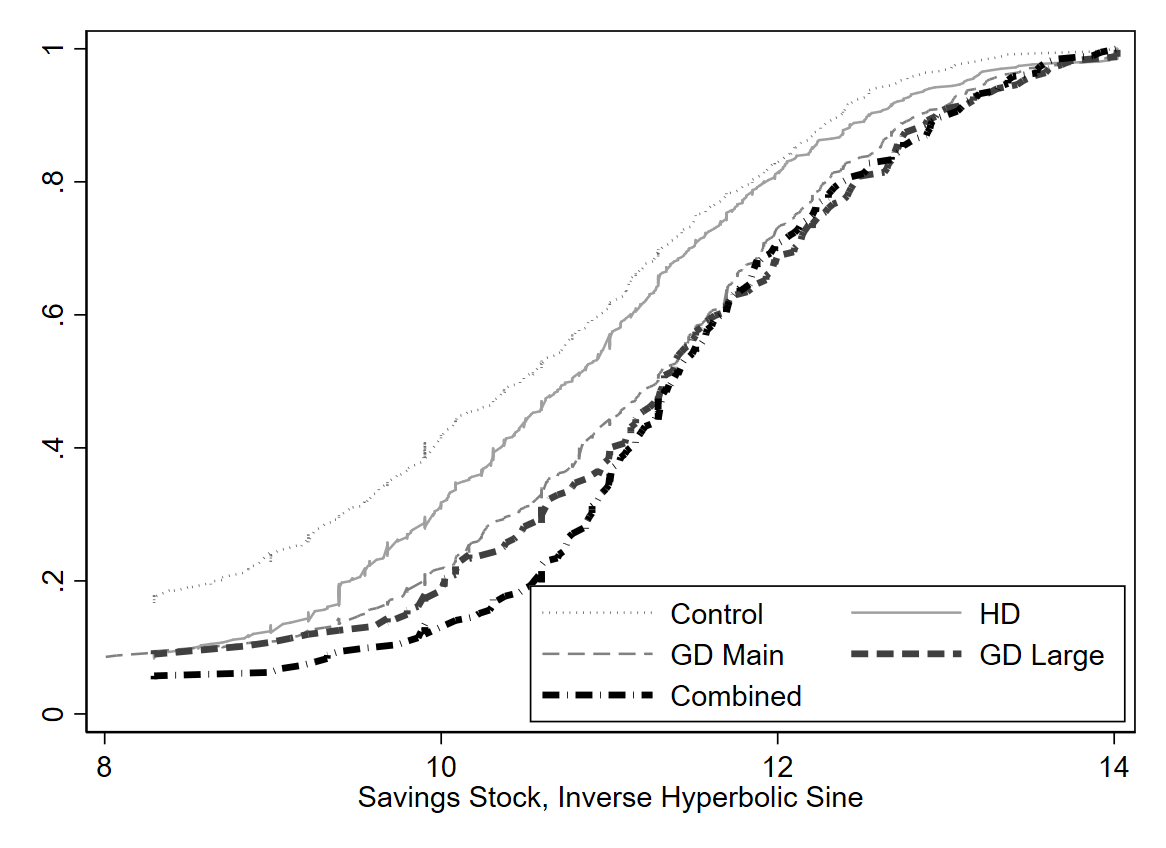}
\end{center}
\end{figure}

\begin{figure}
\caption{CDF of Productive Hours}
\label{f:prod_hours_CDF}
\begin{center}
\includegraphics[width=0.8\linewidth]{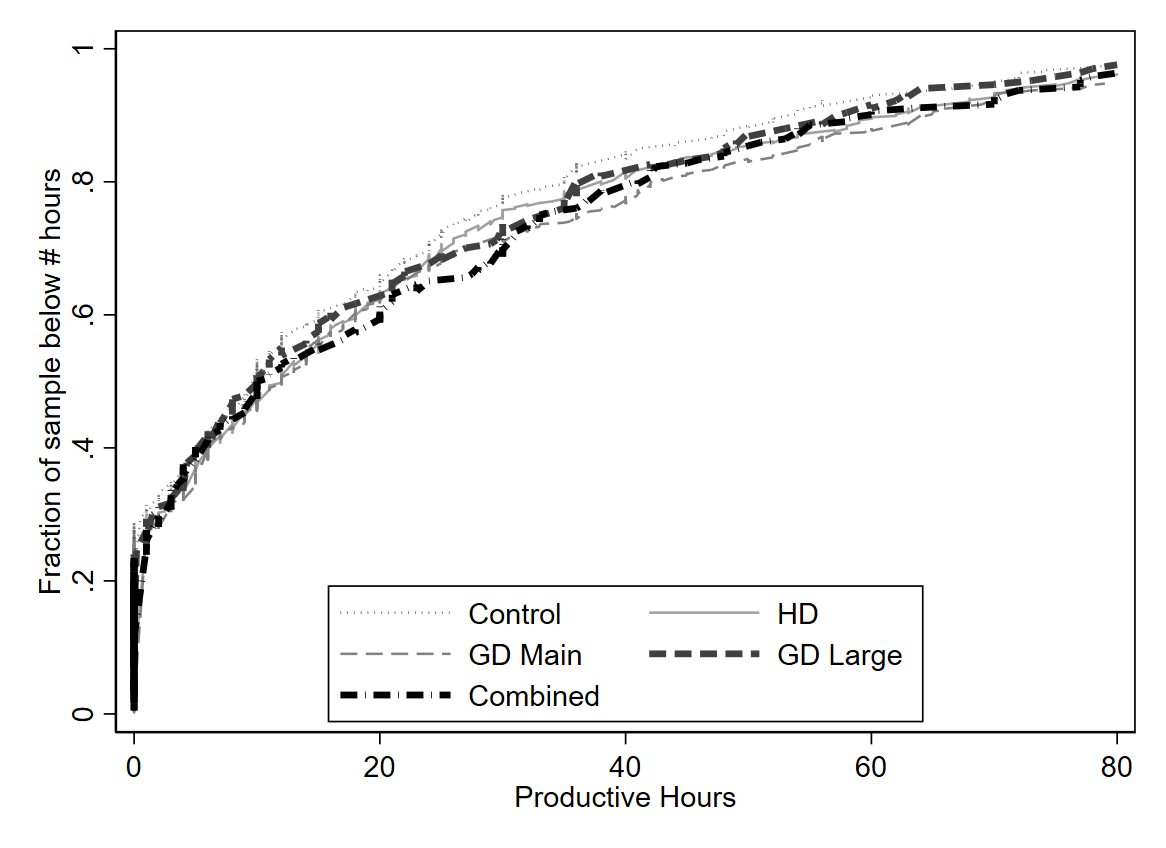}
\end{center}
\end{figure}

\begin{figure}
\caption{Non-Ag Wage Employment, varying hours thresholds}
\label{f:hours_noagric}
\begin{center}
\includegraphics[width=0.8\linewidth]{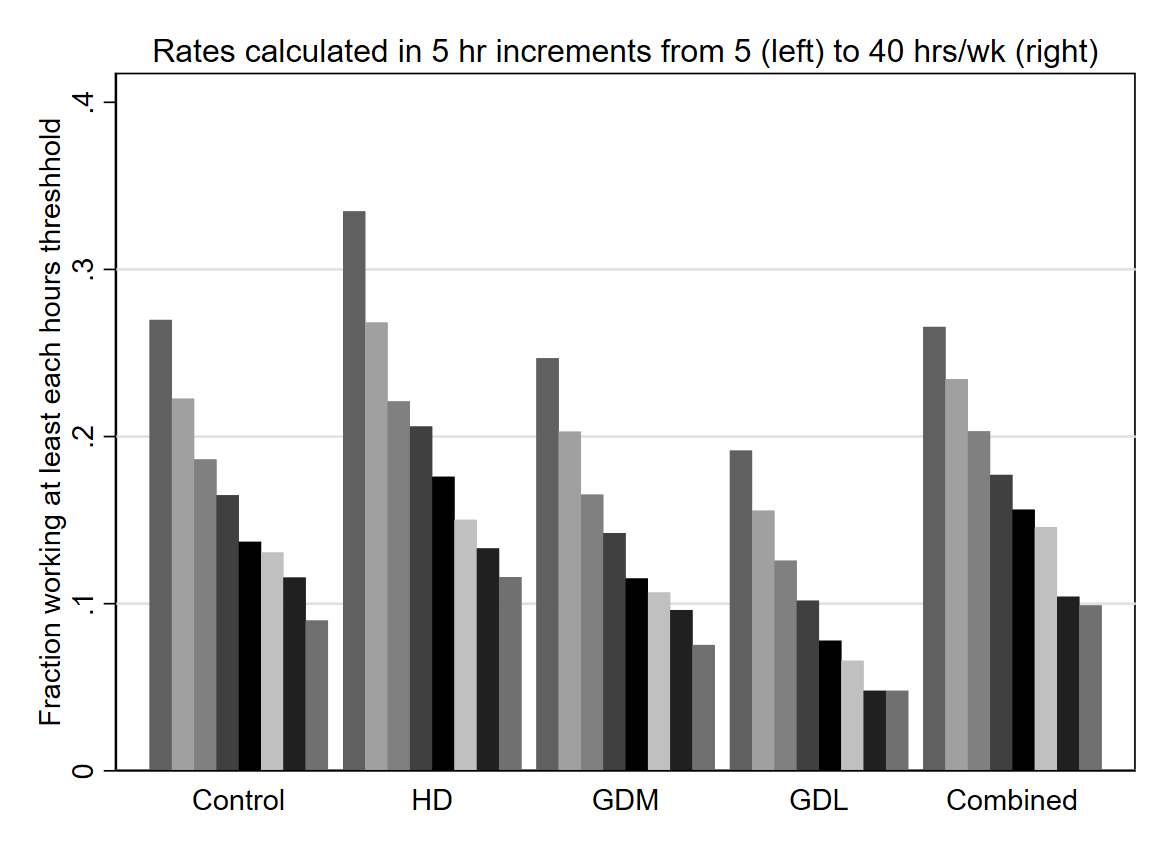}
\end{center}
\end{figure}

\begin{figure}
\caption{Non-Ag Self Employment, varying hours thresholds}
\label{f:hours_enterp}
\begin{center}
\includegraphics[width=0.8\linewidth]{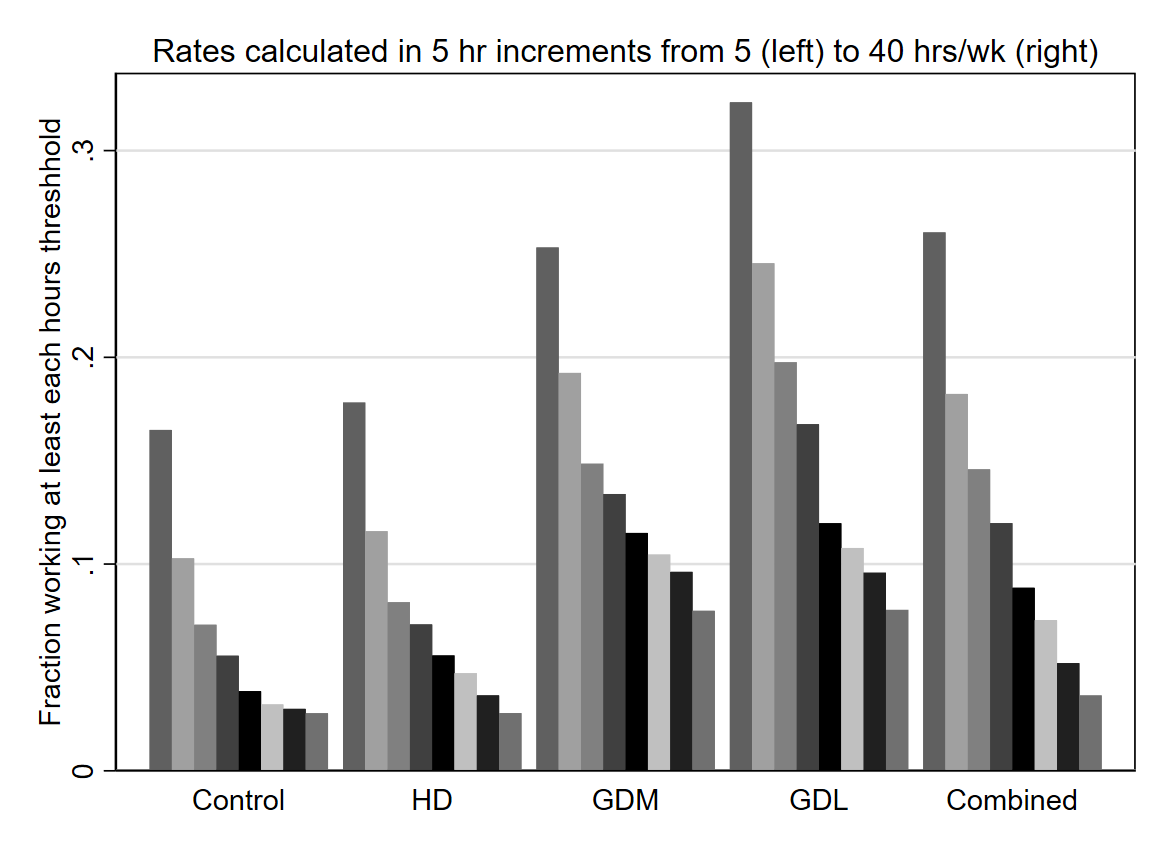}
\end{center}
\end{figure}

\begin{figure}
\caption{Ag Wage Employment, varying hours thresholds}
\label{f:hours_employed_farm}
\begin{center}
\includegraphics[width=0.8\linewidth]{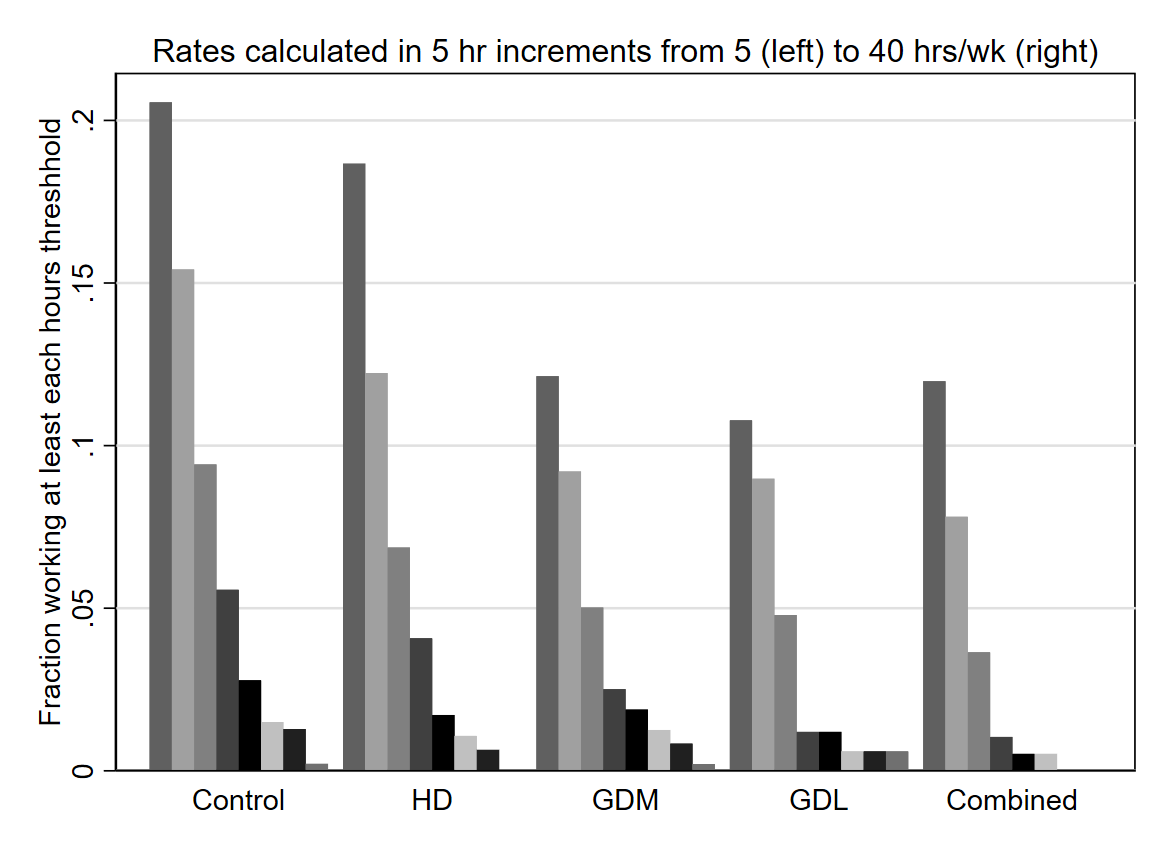}
\end{center}
\end{figure}

\begin{figure}
\caption{Ag Self Employment, varying hours thresholds}
\label{f:hours_semploy}
\begin{center}
\includegraphics[width=0.8\linewidth]{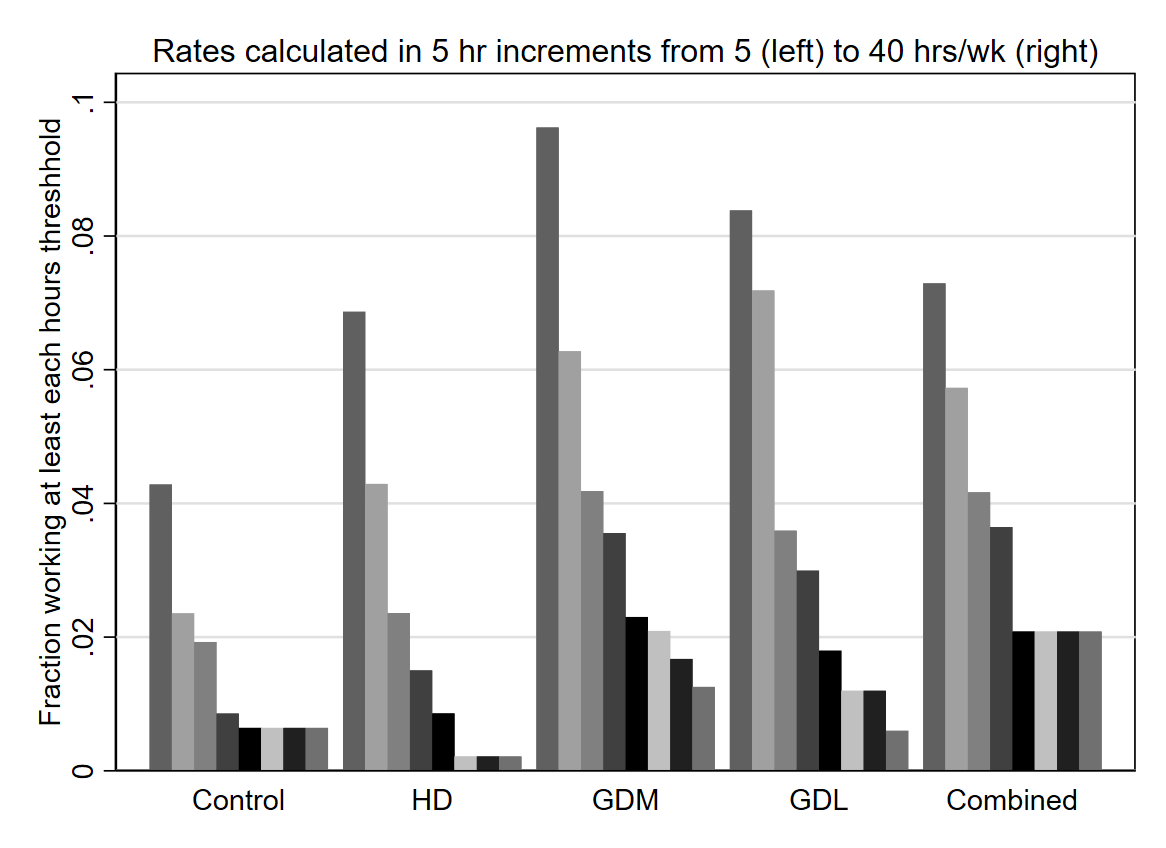}
\end{center}
\end{figure}

 \begin{figure}
\caption{Ag processing or trading Employment, varying hours thresholds}
\label{f:hours_agroprocess}
\begin{center}
\includegraphics[width=0.8\linewidth]{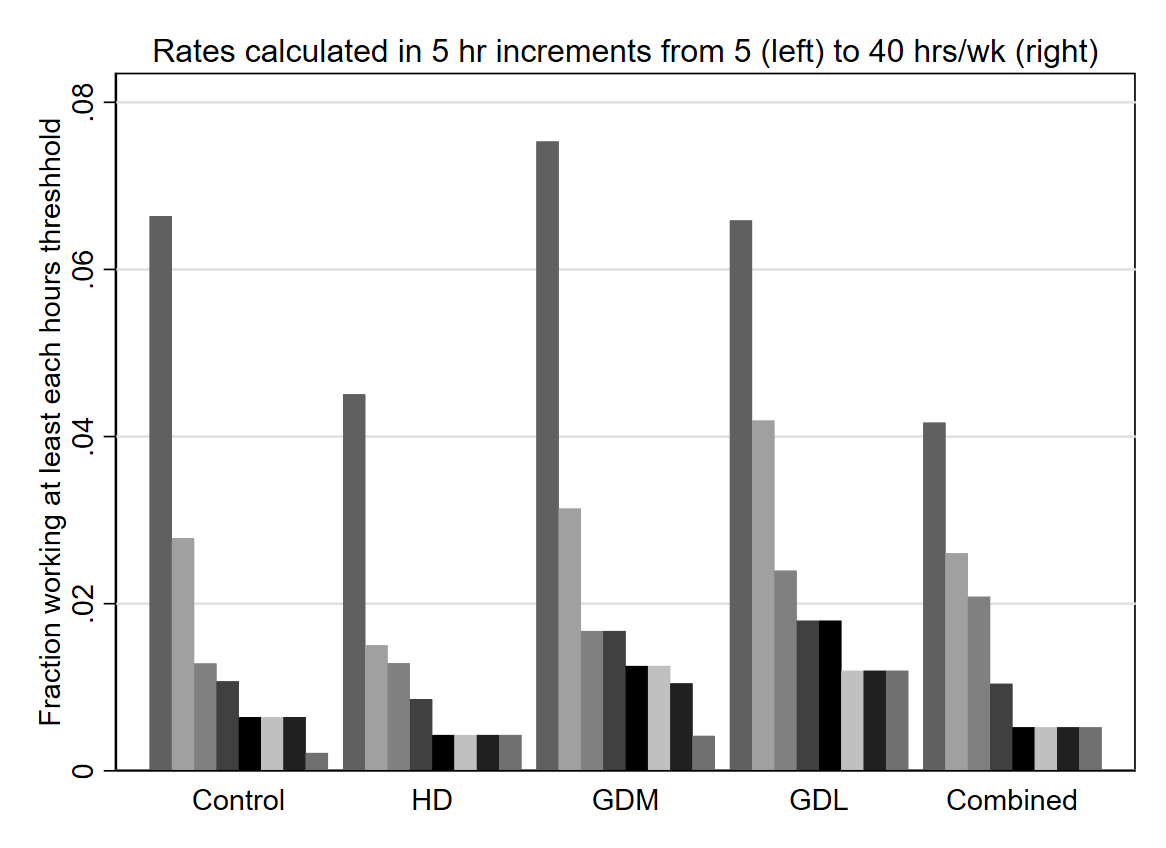}
\end{center}
\end{figure}

\begin{figure}
\caption{Rolling Phone Survey:  Monthly impacts on Productive Hours}
\label{f:phone_survey_productive}
\begin{center}
\includegraphics[width=0.8\linewidth]{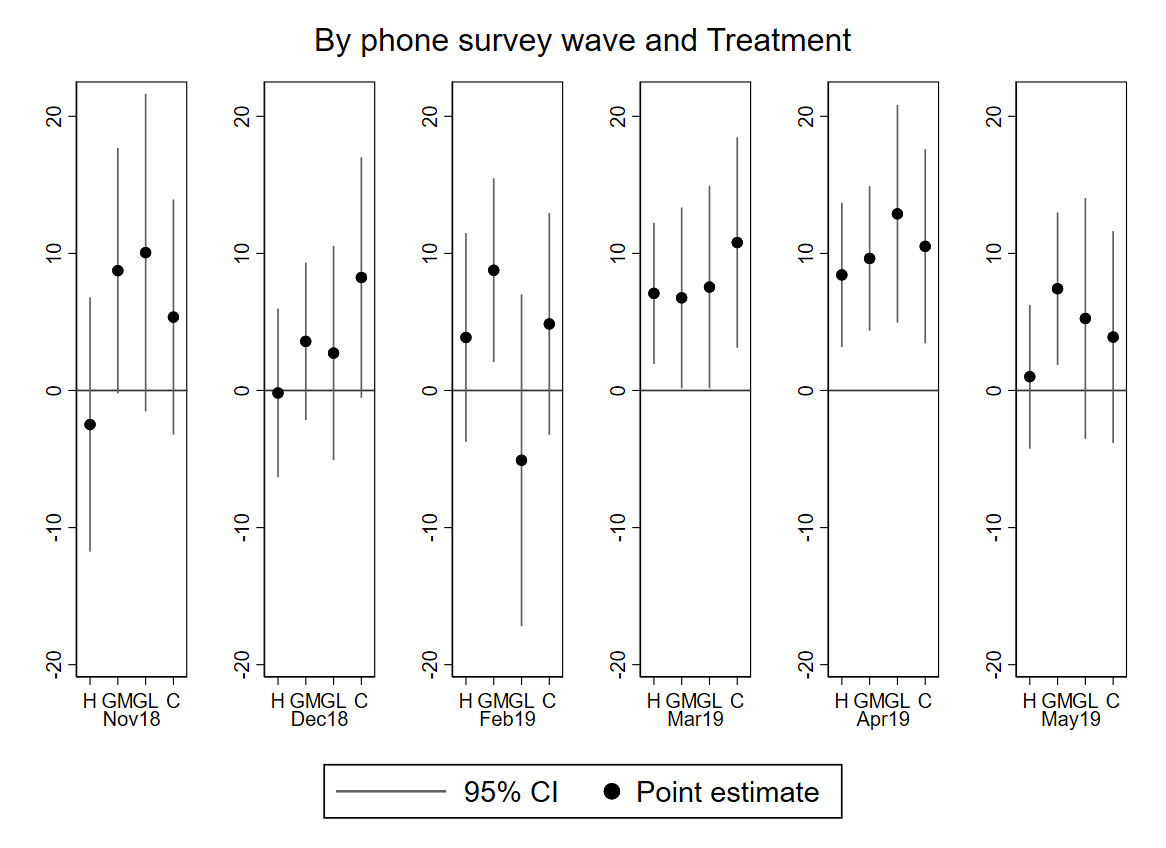}
\end{center}
\end{figure}

\begin{figure}
\caption{Rolling Phone Survey:  Monthly impacts on Apprenticeship Hours}
\label{f:phone_survey_apprenticeship}
\begin{center}
\includegraphics[width=0.8\linewidth]{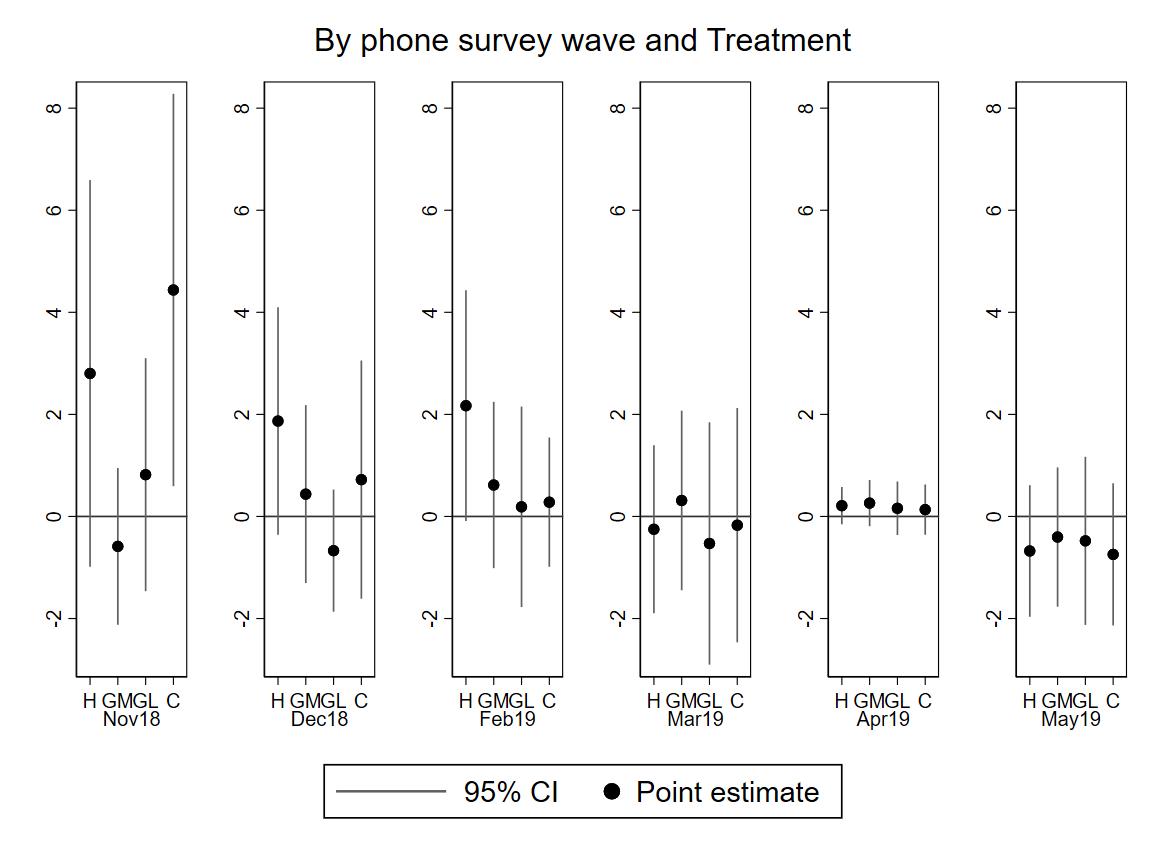}
\end{center}
\end{figure}

\begin{figure}
\caption{Cost Equivalence on Secondary Outcomes}
\label{f:CE_secondary}
\begin{center}
\includegraphics[width=0.8\linewidth]{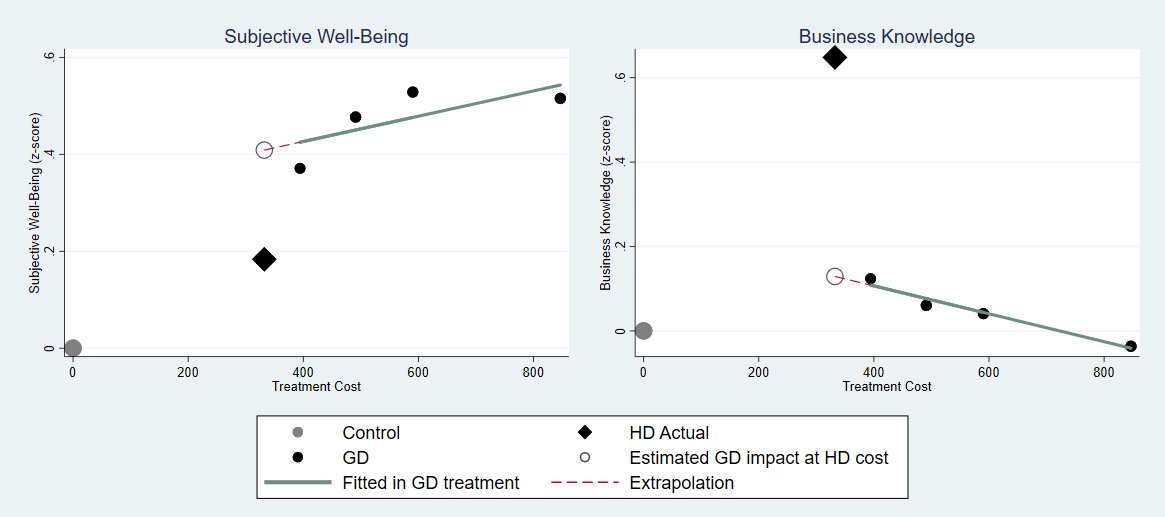}
\end{center}
\end{figure}

\begin{figure}
\caption{Village-Level Treatment Saturations}
\label{f:saturation_densities}
\begin{center}
\includegraphics[width=0.8\linewidth]{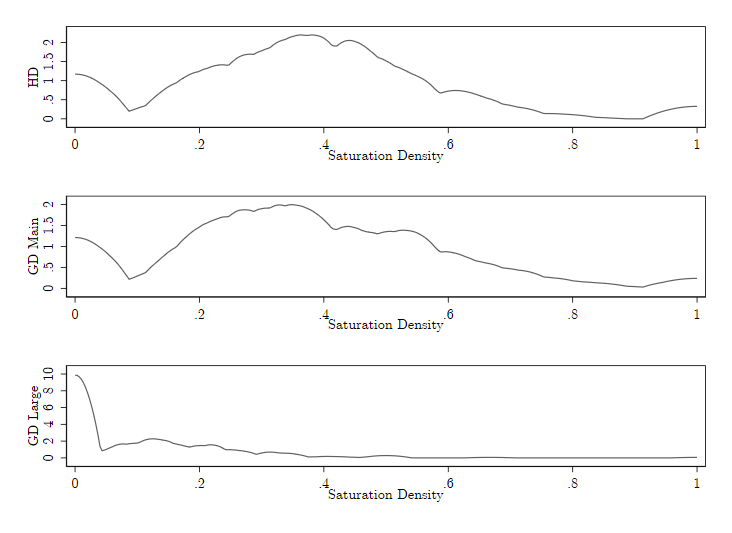}
\end{center}
\end{figure}

\clearpage
\section{Selection of control variables}\label{app:ModelSelection}

In our pre-analysis plan, we state that control variables for the primary specification  ``will be selected on the basis of their ability to predict the primary outcomes''.  In doing so, we seek to build on recent developments that balance the challenge of using baseline data to select variables that will reduce residual variance in equation \eqref{eq:PrimarySpec} with the danger that researcher freedom in the selection of control variables can lead to $p$-hacking, in which right-hand-side variables are selected specifically on the basis of the statistical significance of the coefficient of interest \citep{CarKrue95aer,CasGlenMig12qje}, thereby invalidating inference. 

To balance these concerns, we adapt the \emph{post-double-selection} approach set forth in \citet[henceforth BCH]{BelCheHan14restud}. BCH advocate a two-step procedure in which, first, Lasso is used to automate the selection of control variables, and second, the post-Lasso estimator \citep{BelCheHan12ecta} is used to estimate the coefficients of primary interest in Equation \eqref{eq:PrimarySpec}, effectively using Lasso as a model selection device but \emph{not} imposing the shrunken coefficients that results from the Lasso estimates directly. \citet{BelCheHan14restud} demonstrate that this approach not only reduces bias in estimated treatment effects better than alternative approaches---less a concern given the successful randomization in our experiment---but that it may improve power while retaining uniformly valid inference.  

In the first stage, model selection is undertaken by retaining control variables from the union of those chosen either as predictive of the treatment assignment or of the outcome.  
This model selection stage can be undertaken after residualizing to account for a set of control variables that the authors have a priori determined below in the model, as in \citet{BelCheHan14jep}. 
In our case, we retain block fixed effects, lagged values of the outcome, and lagged values of (the inverse hyperbolic sine of) household wealth in all specifications, per our pre-analysis plan. 

We modify the BCH approach for application to a randomized experiment in three ways. 
First, again following \citep{JonMolRei19qje}, for each outcome we choose the Lasso penalty parameter that minimizes the 10-fold cross-validated mean squared error. 
Second, to ensure that chance differences in the leverage of observations across different covariate sets do not lead to different conclusions about the (relative) impacts of treatment across different outcomes \citep{You19inference}, we take the union of covariate sets selected to be predictive of the five primary outcomes of the study, and use these as controls for all outcomes.  
And third, we modify the heteroskedasticity-robust Lasso estimator of \citet{BelCheHan12ecta} to incorporate sampling weights consistent with our design.\footnote{Specifically, we up-weight observations in our `intensive tracking' endline sample by the inverse of the fraction of not-initially-reached individuals in the follow-up survey who were then assigned to intensive tracking.} 

The set of \emph{potential} covariates is determined as follows:  
\begin{itemize}
    \item Baseline values of all primary outcomes, including the individual components of the employment status, productive time use, monthly income variables outlined in Section \ref{ss:outcomes};
    \item Baseline values of all secondary outcomes, 
    \item Baseline values of all dimensions of heterogeneity pre-specified in Section \ref{ss:heterogeneity}.
    \item The number of study participants (in any arm of the study) in an individual's village, which is defined as the measure of network `degree' for each individual in the spillover analysis of Section \ref{ss:spillovers}.
\end{itemize}
All variables are normalized prior to inclusion in the selection routine, to have mean zero and variance of one in the baseline sample.  
We include squares of all continuous variables and all pairwise interactions among the potential covariates above, and between the potential covariates above and the set of variables that force the routine to include without penalty
To ensure that sample size is not affected by the choice of covariates, we impute values of zero for all variables in the \emph{potential} covariate list, and for each potential covariate we include an indicator for whether such an imputation was undertaken among the list of potential covariates to be fed into the BCH first-stage selection procedure.

\clearpage 
\section{Administrative information}

\subsection{Funding}

All research funding for this project was provided by USAID.

\subsection{Details of Study Participant Selection}

To meet the Huguka Dukore eligiblity criteria, participating youth must meet the following criteria:
\begin{itemize}
    \item 6-12 years of basic education (inclusive).
    \item Age 16-30 at enrollment.
    \item Drawn from Ubudehe poverty groups 1 and 2, per GiveDirectly's remit from the Rwandan Government to treat only the poorest households with cash transfers.   
\end{itemize}
 
Additionally, HD in its outreach specifically targeted the following criteria for inclusion, meaning that such youth will be specially recruited to participate:
\begin{itemize}
    \item Out of school for three consecutive years
    \item Income of less than \$1.75 per day
    \item Youth exhibiting some form of disability (that can be accommodated in HD programming)
    \item Women. 
    \item Youth who have not benefited from related interventions in the past. 
\end{itemize}
Hard eligibility criteria and targeted characteristics were provided to local government leaders, who provided lists of potential candidates to EDC. Those candidates were then invited to the information session and formally screened for eligibility.

All listing and determination of eligibility were conducted by EDC via an ‘over-subscription’ process.  Under this protocol, EDC enrolled more eligible individuals than they were able to treat with HD, in order to generate the samples for the alternate (household grants) arm and the control.   In the end we recruited 1848 study youth from approximately 250 villages in our 13 sectors, for an average of roughly 7.4 study individuals per village.  

Below, we characterize the process for (over)subscription, which delivered the sample of individuals for the baseline. 
\begin{enumerate}
    \item \textbf{Sector-level meeting} to discuss HD with local leaders that introduced the study.  In this meeting, sector officials were fully informed about the scope of the study, emphasizing the separateness of the two interventions and implementers. 
    \item  \textbf{Announcement to the community} in public places (churches, community halls) or a meeting to engage potential beneficiaries.  At this point only the HD program wsa described to beneficiaries, and with only general language about the household grants arm.  Guiding language:  \textit{``We are pleased to be able to bring programming to this community that seeks to improve the livelihoods of vulnerable youth. To this end, we are requesting the names and contact details of youth meeting the following criteria:  <insert eligibility criteria here>.  Participating youth should be willing and interested to join an employment skills program, called HD, that will provide training and work experience to participants.''} 
    \item \textbf{Screening of youth} by the selection committee which produced the final list of potential beneficiaries that was passed to local implementing partners (IPs). 
    \item \textbf{Invitation of potential beneficiaries to an orientation meeting}.  The language of this invitation reflected the fact that potential beneficiaries were not guaranteed places in HD, and might be randomly allocated to a different program or the control. Guiding language for official communication:  \textit{``We have determined that you are eligible for the Huguka Dukore program. There may be more eligible individuals than Huguka Dukore will treat this year, so you are not yet guaranteed a place, though some of those not treated by Huguka Dukore will be supported by another NGO.  To find out more about the Huguka Dukore program and to take the next step toward this opportunity, please attend an orientation meeting at XXX on YYY date.''}
    \item \textbf{Orientation and awareness meeting} with selected youth by local IPs at which they are given further explanation about the program. In HD’s other districts, these orientation meetings convey information about the scope of that program, under a presumption that those who participate in the orientation meeting can have a place in HD should they choose to take it up.  

    \item \textbf{Description of the lottery for program assignment.} The lottery is described during this meeting with reference to another intervention providing livelihoods assistance that will also be determined by the lottery.  Guiding language:  \textit{“Today you have learned more about the Huguka Dukore program.  This is one of two programs that are being delivered by distinct NGOs, in coordination with Sector and District officials, both of which seek to improve livelihoods for vulnerable youth. If you decide that you are interested in participating in one of these programs, there is one more step in the selection process.  To participate, you must attest that you have the time and interest required to participate in Huguka Dukore.  Your name will then be entered into a pool of applicants.  There will be a public meeting in which a lottery will be used to determine which of these applicants receives a place in HD. You may attend this meeting if you wish, but you do not have to do so in order to gain a place.  Not all whose names are entered into the lottery will be placed in HD.  Some of those who participate in the lottery will be passed to a second NGO, which provides assistance to individuals seeking to improve their livelihoods.  Those who receive a place in either program will be contacted directly by the relevant organization after the lottery.  To gain access to either program, you must participate in this lottery.  If you are willing to participate, please provide your name and contact details in writing.   Prior to the lottery, you may be contacted by an independent research organization called Innovations for Poverty Action, who are conducting a survey of potential beneficiaries.  You do not have to participate in this survey in order to gain access to our program, and participation will not affect your chances of enrollment. However, we would be grateful for your willingness to participate in an interview with IPA, which will help us to understand the design and impacts of our work. }
    \item \textbf{Registration for the lottery assignment.} To correctly reflect the lottery process to participants, they were told when asked to enroll in the study that it is  “a lottery in which you will have a chance of receiving HD, a chance of receiving assistance from a different organization that gives household grants, and a chance that you do not receive either program."  Individuals who do not choose to register for the study will not be excluded from receiving HD if they are eligible \& choose to participate.

\end{enumerate}

\subsubsection{Defining Primary outcomes}\label{ss:PrimaryOutcomes}

For each of the outcomes defined below, we provide a definition, followed by an explanation of how that measure will be constructed from survey data.  Survey questions either begin with a `B-' for the beneficiary instrument or a `H-' for the household instrument, followed by the two-digit section number, followed by `q' and the question number.  These refer to the beneficiary and household instruments, respectively.\footnote{In the electronic survey instrument, all variables begin with an `m' prefix, but this notation does not guarantee uniqueness across instruments. Consequently for the purposes of this PAP we adopt the `B-' and `H-' convention above.}

    There are five primary outcomes:
    \begin{enumerate}
    
    \item \emph{Employment status}.  A binary indicator variable taking a value of one if the beneficiary spent 10 hours or more in the prior week working in a wage job or as primary operator of a microenterprise.  The  1 week recall is per ILO definition.   Defined as 'Yes' if beneficiary spent 10 hours or more on any of the following activities: 
      \begin{itemize}
        \item Processing or trading of agricultural goods (\verb|B02qagroprocesshrs|) 
        \item Agricultural (off farm) wage labor (\verb|B02qfarmhours|) 
        \item Non-agricultural wage labor (\verb|B02qnoagrichrs|)
        \item Non-agricultural microenterprise (\verb|B02qenterphrs|)
        \item Microenterprise or other self employment (\verb|B02qsemployhrs|).  
        \end{itemize}
    
    \item \emph{Off-own-farm productive time use}.  Defined as the number of productive hours over the past 7 days.   Sum of hours from questions: 
        \begin{itemize}
        \item Processing or trading of agricultural goods (\verb|B02qagroprocesshrs|)
        \item Agricultural (off farm) wage labor (\verb|B02qfarmhours|) 
        \item  Non-agricultural wage labor (\verb|B02qnoagrichrs|) 
        \item  Non-agricultural microenterprise (\verb|B02qenterphrs|) 
        \item Microenterprise or other self employment (\verb|B02qsemployhrs|)
        \item Apprenticeship (\verb|B02qapprenticehrs|)
        \end{itemize} 
        
    \item \emph{Beneficiary's (monthly) income}. Defined as the sum of the following monthly recall questions:
        \begin{itemize}
        \item Agricultural own-farm income (\verb|B02qagricearn|)
        \item Agricultural wage income (\verb|B02qfarmwage|)
        \item Non-agricultural wage income (\verb|B02qnoagricwage|)
        \item Microenterprise profits (\verb|B02qenterpwage| + \verb|B02qsemploywage|);
        \item Livestock rearing income (\verb|B02qlivestockwage|)
        \item Agricultural processing and trading income (\verb|B02qagroprocessearn|) 
        \item Apprenticeship income (\verb|B02qapprenticewage|) 
        \end{itemize}
        This outcome will be winsorized at the 1st and 99th percentile, and we will take the inverse hyperbolic sine transformation of this as the primary measure.
        
    \item \emph{Productive assets} under beneficiary control. (Sum of asset values from beneficiary enterprise module that are reported as used in the beneficiary's business, Section \verb|B05|: tools, machinery, furniture, inventories, and other physical assets.) This outcome will be winsorized at the 1st and 99th percentile, and we will take the inverse hyperbolic sine transformation of this as the primary measure.

    \item \emph{Household consumption} per capita.  Sum of monthly purchase values of Section \verb|H10|, divided by adult-equivalent household members.  This outcome will be winsorized at the 1st and 99th percentile, and we will take the inverse hyperbolic sine transformation of this as the primary measure.
    \end{enumerate}

The first three of these primary outcomes provide direct measures of the extent to which a study participant is productively employed:  their formal (non-farm) employment categorization, their productive time use, and their earnings.  To the extent that these measures are potentially seasonal in nature, one might worry that interventions could differentially affect the sectoral composition of employment, and that differential seasonality across these would tip the scales in favor of one or the other mode of intervention. More broadly, income may be more fully measured in one sector relative to another.  Such concerns are partly addressed by the inclusion of household consumption as a primary outcome: to the extent that beneficiaries smooth consumption, household consumption will be less susceptible to such concerns.  In addition, we will include as a robustness check an analysis of impacts on a rolling panel of employment status measures, collected over the six months prior to the endline. 

One potential challenge for the analysis of monetary outcomes (income, assets, and consumption) is that, if treatments induce migration, they may cause subjects to face different prices.  Such differences in prices could cause the study to over- (or under-)state the the real value of estimated impacts.  On the other hand, deflating values to control-group prices is not straightforward, for at least two reasons:  study subjects may alter the \emph{quality} of products purchased in ways not captured by the study, therefore giving the appearance of price impacts; and study subjects may earn incomes in more expensive locations but intend for part of that income may be consumed---by the subject themselves, or by family members to which they remit income---in their place of origin.  To address these concerns, we will report as a robustness check an analysis of primary outcomes (3)--(5) that uses control-group prices to deflate these values.  This will be particularly important to the interpretation of the study results if treatments have effects on migration.

\subsubsection{Defining Secondary outcomes}\label{sss:secondary}

We propose to analyze three families of secondary outcome:  one which speaks to alternative measures of beneficiary welfare; a second that speaks to wealth effects that may indicate likely long-term benefits; and a third family that highlights key mechanisms of interest.  

\begin{enumerate}
\item \textbf{Alternative measures of beneficiary welfare}

Within this family, we consider the following alternative measures of beneficiary well-being:
\begin{enumerate}
    \item Subjective well being:  Index of responses to \verb|B10_swb_happiness| and \verb|B10_swb_lifesatisfaction|, constructed as the average of z-scores. 

    \item Mental health:  Index of section \verb|B11| responses.  Z-score of the simple average across all questions for each beneficiary.

    \item Beneficiary-specific consumption expenditures (sum of values from Section B08). This outcome will be winsorized at the 1st and 99th percentile, and we will take the inverse hyperbolic sine transformation of this as the primary measure.
    
\end{enumerate}

\item \textbf{Household net wealth, and its components}

Like productive assets, the accumulation and protection of household wealth.  Conditional on this, households' access to borrowing opportunities---viewed as a measure of their financial access---may be a mechanism through which the interventions studied are multiplied.  Given this welfare ambiguity, we propose to analyze both total household net (non-land) wealth, as well as stocks of savings and debt, taken individually.

\begin{enumerate}
    \item Household net non-land wealth. Sum of values of household assets (\verb|H12|), plus savings value (\verb|H06|), value of loans outstanding that are expected to be repaid (\verb|H08|), less debt value (\verb|H07|). This outcome will be winsorized at the 1st and 99th percentile, and we will take the inverse hyperbolic sine transformation of this as the primary measure.
    \item Total value of all livestock wealth. Sum of values of household livestock assets (\verb|H12|). Specifically, summing over values derived from \verb|H12_oxen| through \verb|H12_ducks| in the household instrument. This outcome will be winsorized at the 1st and 99th percentiles, and we will take the inverse hyperbolic sine transformation of this as the primary measure.
    \item Stock of savings. Beneficiary stock of savings, sum of values in \verb|B06|. Plus household stock of savings from analogous questions (\verb|H06|).  This outcome will be winsorized at the 1st and 99th percentile, and we will take the inverse hyperbolic sine transformation of this as the primary measure.
    \item Stock of debt. Beneficiary sum of borrowed amounts from all (formal and informal) sources (\verb|B07|), plus household borrowings from analogous questions (\verb|H07|). This outcome will be winsorized at the 1st and 99th percentile, and we will take the inverse hyperbolic sine transformation of this as the primary measure.
\end{enumerate}

\item \textbf{Cognitive and non-cognitive skills}

A specific feature of the theory of change that motivates EDC's curriculum is that a focus not just on specific skills, but on non-cognitive attitudes and attitudes, may make that intervention more likely to have persistent effects.  At the same time, cash transfers may also change, inter alia, beneficiaries' sense of control and aspirations.  To test these mechanisms, we define the following family of secondary outcomes:

\begin{enumerate}
    \item Locus of control:  Index of responses to \verb|B09|. Z-score of the simple average across all questions for each beneficiary.
    
    \item Aspirations:  Index of responses to \verb|B13|.  Z-score of the simple average across all questions for each beneficiary.
    
    \item 
    Conscientiousness, agreeableness, and emotional stability from BFI (Section \verb|B12|). Each index is the Z-score of the simple average of the questions related to the corresponding dimension.  Following EDC's analysis of Akaze Kanoze employers,\footnote{Povec Pagel, Olaru, Alcid, and Beauvy-Sany, 2017, ``Identifying cross-cutting non-cognitive skills for positive youth development'', Final report, Education Development Center, Inc.} we will examine program impacts on the three most highly-rated components of the Big-Five Index from employers' perspective: conscientiousness, agreeableness, and emotional stability. 
  
    \item Business knowledge.  Index of \verb|B14|. Z-score of the simple average across all questions for each beneficiary.
    
    \item Business attitudes. Index of \verb|B15|. Z-score of the simple average across all questions for each beneficiary.
    
\end{enumerate}
\end{enumerate}

\subsection{Institutional Board Review (ethics approval)}

Details of the procedures and of the consent process were read aloud, in Kinyarwanda, to each respondent prior to each measurement activity.  

\vspace{11pt}In addition to acquiring ethical approval from RNEC, the research team has acquired approval from the IRB at Innovations for Poverty Action, and from Georgetown University and the University of California, San Diego

\vspace{11pt}\textbf{Informed consent}

Participant consent for inclusion in the study can be divided into two separate components:
\begin{enumerate}
    \item Consent for inclusion in the identification of beneficiaries. Eligible applicants to EDC's HD program were informed that, given oversubscription, there was a chance they would not receive this program, but that they might receive an alternative benefit instead. Determination of eligibility was undertaken by EDC and its HD partner organizations, and required the collection of data regarding the socio-economic status (Ubudehe) of households, and the ages and education levels of youth in the household. The collection of these details was required for enrollment in the study sample. 
    \item Consent for the collection of socio-economic data, via questionnaire, which included details of savings, consumption, and nutritional outcomes.  Households were informed that participation in this questionnaire was not required for inclusion in any of the programs under study.  Households were also informed of the opportunity to decline to answer any specific question within the questionnaire.
\end{enumerate}
 
\subsection{Declaration of interest}
The authors declare that they have no relevant or material financial interests that relate to the research described in this paper.

\subsection{Acknowledgements}
We are grateful to DIV, Google.org, and USAID Rwanda for funding, and to USAID, EDC, GiveDirectly, and IPA for their close collaboration. We thank Leodimir Mfura, Marius Chabi, and Phillip Okull for
overseeing the fieldwork, and Sarait Cardenas-Rodriguez, Aruj Shukla, and Diana Martinez for research assistance.

\end{document}